\documentclass[envcountsame]{llncs}

\usepackage{
        dsfont,
        bbm,
        amsmath,
        theorem,
        epsfig,
        color
}

\usepackage{times,amsmath,amssymb,multicol,stmaryrd,enumerate,graphicx,color,xspace,fancybox,paralist,ifthen,verbatim,morefloats,afterpage,ifdraft}

\spnewtheorem{algorithm}[theorem]{Algorithm}{\bfseries}{\rm}

\begin{document}

\newcommand{\Nat}{{\mathbb N}}
\newcommand{\angl}[1]{\langle #1 \rangle}
\newcommand{\size}{{\sf size}}
\newcommand{\dom}{{\sf dom}}
\newcommand{\diff}{{\sf diff}}
\newcommand{\pref}{{\sf pref}}
\newcommand{\Max}{{\sf max}}
\newcommand{\EL}{{\cal L}}
\newcommand{\B}{{\cal B}}


\newcommand{\sem}[1]{\llbracket #1 \rrbracket}
\newcommand{\rk}{{\sf rk}}
\newcommand{\T}{\mathcal{T}}
\newcommand{\U}{\mathcal{U}}
\newcommand{\C}{\mathcal{C}}
\newcommand{\cS}{\mathcal{S}}
\newcommand{\ot}{\!:\!}
\newcommand{\rhs}{{\sf rhs}}
\newcommand{\maxdiff}{{\sf lubdiff}}
\newcommand{\lub}{{\sf lub}}
\newcommand{\maxrhs}{{\sf maxrhs}}
\newcommand{\can}{{\sf can}}
\newcommand{\orig}{{\sf or}}
\newcommand{\orignode}{{\sf orn}}
\newcommand{\height}{{\sf ht}}
\newcommand{\lab}{{\sf lab}}
\newcommand{\fix}{{\sf fix}}
\newcommand{\maxfix}{{\sf maxfix}}
\newcommand{\sumfix}{{\sf sumfix}}
\newcommand{\rlabs}{{\sf rlabs}}
\newcommand{\td}{{\sf td}}
\newcommand{\lk}{{\sf link}}
\newcommand{\blk}{{\sf blink}}
\newcommand{\nod}{{\sf nod}}
\newcommand{\out}{{\sf out}}
\newcommand{\diftup}{{\sf diftup}}
\newcommand{\maxdiftup}{{\sf lubdiftup}}
\newcommand{\spi}{{\sf spi}}
\newcommand{\red}{{\sf red}}

\title{Look-Ahead Removal for Total Deterministic \\Top-Down Tree Transducers
}


\author{Joost Engelfriet\inst{1} \and Sebastian Maneth\inst{2} \and Helmut Seidl\inst{3}
}


\institute{LIACS, Leiden University, The Netherlands \\
\email{j.engelfriet@liacs.leidenuniv.nl}
\and School of Informatics, University of Edinburgh, United Kingdom \\
\email{smaneth@inf.ed.ac.uk}
\and Institut f\"ur Informatik, Technische Universit\"at M\"unchen, Germany  \\
\email{seidl@in.tum.de}  
}

\maketitle


\begin{abstract}
Top-down tree transducers are a convenient formalism for describing
tree transformations. They can be equipped with regular look-ahead,
which allows them to inspect a subtree before processing it.
In certain cases, such a look-ahead can be avoided and the transformation
can be realized by a transducer without look-ahead.
Removing the look-ahead from a transducer, if possible,
is technically highly challenging.
For a restricted class of transducers with look-ahead, namely those that
are total, deterministic, ultralinear, and bounded erasing, 
we present an algorithm that, for a given transducer from that class,
(1)~decides whether it is equivalent to a total deterministic transducer without look-ahead,
and (2)~constructs such a transducer if the answer is positive.
For the whole class of total deterministic transducers with look-ahead
we present a similar algorithm, which assumes that a so-called difference bound
is known for the given transducer. The designer of a transducer can 
usually also determine a difference bound for it. 
\end{abstract}

\section{Introduction}

Many simple tree transformations can be modeled by top-down tree 
transducers~\cite{DBLP:journals/mst/Rounds70,DBLP:journals/jcss/Thatcher70}. 
They are recently used in XML database theory (e.g., \cite{DBLP:journals/jcss/EngelfrietMS09,haruo,DBLP:conf/pods/LemayMN10,DBLP:conf/dbpl/ManethN99,DBLP:journals/jcss/MartensN07,DBLP:journals/iandc/MartensNG08}), 
in computational linguistics (e.g., \cite{DBLP:conf/cicling/KnightG05,DBLP:conf/mol/Maletti11,DBLP:journals/siamcomp/MalettiGHK09}) 
and in picture generation \cite{drewes}.
A \emph{top-down tree transducer} is a finite-state device 
that scans the input tree in a (parallel)
top-down fashion, simultaneously producing the output tree in
a (parallel) top-down fashion. 
A more expressive (but also more complex) model for specifying tree translations
is the top-down tree transducer 
\emph{with regular look-ahead}~\cite{DBLP:journals/mst/Engelfriet77}.
It consists of a top-down tree transducer and
a finite-state bottom-up tree automaton, called the look-ahead automaton. 
We may think of its execution in two phases: 
In a first phase the input tree is relabeled 
by attaching to each input node the active state of the automaton,
called the look-ahead state at that node. 
In the second phase the top-down tree transducer is
executed over the relabeled tree, thus possibly making
use of the look-ahead information in the new input labels.
As an example, consider a deterministic transducer $M_{\text{ex}}$ of which
the look-ahead automaton checks whether the input tree
contains a leaf labeled $a$. If so, then
$M_{\text{ex}}$ outputs $a$, and otherwise  
it outputs a copy of the input tree.
It should be clear that there is no deterministic top-down tree transducer (without look-ahead)
that realizes the same translation as $M_{\text{ex}}$.
The intuitive reason is that in general the complete
input tree must be read and buffered in memory, 
before the appropriate choice of output can be made. 
How can we formally prove that indeed no deterministic top-down tree
transducer (without look-ahead) can realize this translation?
In general, is there a method to determine 
for a given top-down tree transducer \emph{with look-ahead}, 
whether or not its translation can be realized by a 
top-down tree transducer \emph{without look-ahead}?
And if the answer is yes, can such a transducer be constructed from the given one?

In this paper we give two partial answers to these questions,
where we restrict ourselves to total deterministic transducers
(which will not be mentioned any more in the remainder of this introduction). 
For such transducers we provide a general method as discussed above.
However, part of the method is not automatic, but depends on 
additional knowledge about the given transducer with look-ahead
(which can usually be determined by the designer of the transducer). 
For a restricted type of transducers (where the restrictions 
concern the capability of the transducer to copy and erase) that 
knowledge can also be obtained automatically, which means that 
for a thus restricted transducer with look-ahead it is decidable whether 
its translation can be realized by a (nonrestricted) transducer without look-ahead, 
and if so, such a transducer can be constructed from the given transducer. 

The main notion on which our method is based, is that of a \emph{difference tree}
of a top-down tree transducer with regular look-ahead.
Consider two trees obtained from one input tree by
replacing one of its leaves by two different look-ahead states of the transducer $M$.
Compare now the two output trees of $M$ on these input trees, 
where $M$ treats the look-ahead state as representing an input subtree for which the 
look-ahead automaton of $M$ arrives in that state 
at the root of the subtree. Since $M$ is total and deterministic, 
these output trees exist and are unique, respectively.
By removing the largest common prefix of the two output
trees (i.e., every node of which every ancestor has the same label in each of the two trees), 
we obtain a number of output subtrees that we call difference trees of $M$.
Intuitively, the largest common prefix is the part of the output that does not depend on 
the two possible look-ahead states of the subtree, 
whereas a difference tree is a part of the output that can be 
produced because $M$ knows the look-ahead state of the subtree. 
Thus, the set $\diff(M)$ of all difference trees of $M$ can be viewed as a measure 
of the impact of the look-ahead on the behaviour of $M$. 
For the example transducer $M_{\text{ex}}$ above, 
$\diff(M_{\text{ex}})$ consists of the one-node tree $a$
and all trees of which no leaf is labeled $a$ 
(with one leaf representing a subtree without $a$-labeled leaves); 
thus, $\diff(M_{\text{ex}})$ is infinite. 

Now the idea of our method is as follows, where we 
use \emph{dtla} and \emph{dtop} to abbreviate 
top-down tree transducer with and without look-ahead, respectively.
In~\cite{DBLP:journals/jcss/EngelfrietMS09} it was shown that 
for every dtop an equivalent canonical earliest dtop can be constructed.
Earliest means that each output node is produced as early as possible by the transducer,
and canonical means that different states of the transducer are inequivalent. 
We prove that also for every dtla an equivalent canonical earliest dtla 
(with the same look-ahead automaton) can be constructed, where the earliest and canonical 
properties are relativized with respect to each look-ahead state. 
Thus, to devise our method we may restrict attention to canonical earliest transducers.
Assume there exists a (canonical earliest) dtop $N$ 
equivalent to the (canonical earliest) dtla $M$.
Then the dtla $M$ is at least as early as $N$. In other words, at each moment of the translation, 
$M$ may be ahead of $N$ but not vice versa, i.e., the output of $N$ is a prefix of
that of $M$, which is because $M$ has additional information through its look-ahead.
The output of $N$ is the part of $M$'s output that does not depend on the look-ahead state.
Thus, when removing the output of $N$ from that of $M$, 
the remaining trees are difference trees of $M$. 
Since $N$ must be able to simulate $M$, it has to store these difference trees
in its states. Hence, $\diff(M)$ must be finite. Moreover, it turns out that the above description 
of $N$'s behaviour completely determines $N$, and so, roughly speaking, 
$N$ can be constructed from $M$ and $\diff(M)$.  
Note that since $\diff(M_{\text{ex}})$ is infinite, the translation of $M_{\text{ex}}$
cannot be realized by a dtop.

A natural number $h$ is a \emph{difference bound} for a dtla $M$ if the following holds: 
if $M$ has finitely many difference trees, then $h$ is an upper bound on their height; 
in other words, if a tree in $\diff(M)$ has height $>h$, then $\diff(M)$ is infinite. 
Our first main result is 
that it is decidable for a given dtla $M$ for which a difference bound is also given, whether 
$M$ is equivalent to a dtop $N$, and if so, such a dtop $N$ can be constructed. 
We do not know whether a difference bound can be computed for every dtla $M$, but the designer 
of $M$ will usually be able to determine $\diff(M)$ and hence a difference bound for $M$. 
Our second main result is that a difference bound can be computed for dtlas that are 
ultralinear and bounded erasing.
Ultralinearity means that the transducer cannot copy an input subtree when it is in a cycle
(i.e., in a computation that starts and ends in the same state). Thus it is weaker than the linear
property (which forbids copying) but stronger than the finite-copying property
\cite{DBLP:journals/jcss/EngelfrietRS80,DBLP:journals/iandc/EngelfrietM99}.
The latter implies that the size of the output tree of an ultralinear dtla is linear in the size 
of its input tree.
Bounded erasing means that the transducer has no cycle in which no output is produced. 
The proof that a difference bound can be computed for ultralinear and bounded erasing dtlas, 
is based on pumping arguments that are technically involved. 

The paper is structured as follows. Section~\ref{sec:prelim} contains basic terminology,
in particular concerning prefixes of trees. 
Section~\ref{sec:dtop} defines the dtla (deterministic top-down tree transducer with regular look-ahead)
and discusses some of its basic properties. It also explains the treatment of look-ahead states
that occur in the input tree. 
In Section~\ref{sec:difftree} we define the notions of difference tree and difference bound,
illustrated by some examples. In Section~\ref{sec:normform} we discuss some normal forms 
for dtlas, in particular \emph{look-ahead uniformity} which 
is technically convenient. 
We prove that for every dtla $M$ there is an equivalent canonical earliest dtla $M'$ 
(which is also look-ahead uniform), 
and we show how to compute a difference bound for $M'$ from one of $M$. 
Our first main result is proved in Section~\ref{sec:diff},
which is divided in three subsections.
Section~\ref{sec:difftup} starts with the definition of a \emph{difference tuple} of a dtla $M$, 
which generalizes the notion of difference tree by considering all look-ahead states of $M$
rather than just two. If $N$ is a dtop equivalent to $M$, then its states are in one-to-one 
correspondence with the difference tuples of~$M$ 
(assuming that both $N$ and $M$ are canonical earliest), see Lemma~\ref{lm:Qdiff}, 
and its axiom and rules are completely determined by $M$, 
see Lemmas~\ref{lm:dtladtopprop2a} and~\ref{lm:dtladtopprop2b}. 
In Section~\ref{sec:diffalg} we present the algorithm that computes 
$N$ from $M$ for a given difference bound for $M$ (if such an $N$ exists), 
still assuming that $M$ is canonical earliest. 
Section~\ref{sec:diffdec} states the first main result (for arbitrary $M$) in Theorem~\ref{th:alg}. 
Sections~\ref{sec:origins}--~\ref{sec:graphs} are devoted to the proof of our second main result 
(Theorem~\ref{th:ultra}). 
Section~\ref{sec:origins} continues Section~\ref{sec:dtop} by discussing some basic properties 
of dtlas: the \emph{links} that exist between an input tree and its corresponding output tree, 
and for each node of the output tree, its \emph{origin} in the input tree. 
In Section~\ref{sec:upper} the problem of computing a difference bound for a dtla $M$ is reduced to
that of computing two related upper bounds: an \emph{output bound} for $M$ 
and an \emph{ancestral bound} for $M$. An output bound can be computed for every dtla. 
Finally, in Section~\ref{sec:graphs}, an ancestral bound is computed
for every ultralinear and bounded erasing dtla. 
The computed output bound (in the previous section) 
and ancestral bound for a dtla $M$ are both based on pumping arguments 
(simple for the output bound, complicated for the ancestral bound). 
In both cases a part of the input tree on which $M$ has a cyclic computation, is pumped 
in such a way that the corresponding output tree contains arbitrarily large difference trees.
In the ancestral case the pumping argument is technically based on the fact that $M$ cannot copy and 
must produce output during its cyclic computations. Since the pumping of trees makes it hard 
to address nodes by the usual Dewey notation, a \emph{dependency graph} is defined for $M$ such that 
a cyclic computation of $M$ corresponds to a cycle in its dependency graph; 
pumping the input tree then corresponds to repeating a cycle in the graph. 
At the end of Section~\ref{sec:graphs} we consider two other classes of dtlas
for which equivalence to a dtop is decidable (and if so, such a dtop can be constructed): 
\emph{output-monadic} dtlas and \emph{depth-uniform} dtlas. 
Output-monadic means that every node of an output tree has at most one child. 
Depth-uniform means, in its simplest form, that all states in the right-hand sides of the rules 
of the dtla are at the same depth.

\subsubsection*{Related Work.}

For deterministic string transducers with regular look-ahead, 
look-ahead removal is decidable, i.e., it is decidable whether a given transducer 
with look-ahead is equivalent to a transducer without look-ahead, and if so, 
such a transducer can be constructed.
This was proved in \cite{DBLP:journals/tcs/Choffrut77} 
(see also \cite[Theorem IV.6.1]{berstel}), for so-called subsequential functions.
We extend that result (for the total case) by proving that look-ahead removal 
is decidable for output-monadic dtlas.

Look-ahead has been investigated for other types of tree transducers.
For macro tree transducers \cite{DBLP:journals/jcss/EngelfrietV85,DBLP:journals/iandc/EngelfrietM99} 
and streaming tree transducers \cite{DBLP:conf/icalp/AlurD12}, 
regular look-ahead can always be removed. 
The same is true for nondeterministic visibly pushdown transducers \cite{DBLP:conf/sofsem/FiliotS12}. For  
deterministic visibly pushdown transducers the addition of regular look-ahead
increases their power, but the decidability of look-ahead removal 
for these transducers is not studied in \cite{DBLP:conf/sofsem/FiliotS12}.  

In~\cite{DBLP:journals/ipl/FulopKV04} the deterministic multi bottom-up
tree transducer (dmbot) was introduced and shown
to have (effectively) the same expressive power as the dtla.
Thus, our results can also be viewed 
as partial answers to the question whether it is decidable for a given 
dmbot to be equivalent to a dtop. 

\subsubsection*{Note.}
The results of this paper were first presented at DLT~2014, see \cite{DBLP:conf/dlt/EngelfrietMS14}.

\section{Preliminaries}\label{sec:prelim}

The set of natural numbers is $\Nat=\{0,1,2,\dots\}$, 
and $\Nat_+=\{1,2,\dots\}$.
For $k\in\Nat$ we define $[k]=\{1,\dots,k\}$; in particular, $[0]=\emptyset$.
We assume the usual linear order $\leq$ on~$\Nat$ and on $\Nat\cup\{\infty\}$, 
with $n<\infty$ for all $n\in\Nat$.
For a set $S\subseteq \Nat\cup\{\infty\}$, $\lub \,S$ denotes the least upper bound of the elements of $S$.
If $S$ is finite and nonempty, then $\lub \,S$ is the maximal element of $S$. 
Also, $\lub \,\emptyset=0$.

The domain of a partial function $f$ is denoted $\dom(f)$.
For a set $A$, we denote by $A^*$ the set of sequences, or strings, 
of elements of $A$. A string $(a_1,\dots,a_n)\in A^*$ will be denoted
$a_1\cdots a_n$, unless there is a danger of confusion. 
The concatenation of two strings $u$ and $v$ is denoted $u\cdot v$ or just $uv$,
and the empty string is denoted $\varepsilon$.
A string $u$ is a prefix (postfix) of a string $v$ 
if there exists a string $w$ such that $v=uw$ ($v=wu$);
it is a proper prefix (postfix) if $w\neq\varepsilon$.
The length of a string $u$ is denoted $|u|$.
The cardinality of a set $A$ is denoted $|A|$.

A directed edge-labeled graph $G$ over a set $A$ (of edge labels) consists of a set $V$ of nodes and a set $E\subseteq V\times A\times V$ of edges. An edge $(u,a,v)$ is said to be an edge with label $a$ that starts at $u$ and ends at $v$, or shortly, from $u$ to $v$. 
A (directed) path $\pi$ in $G$ is a sequence $e_1\cdots e_n$ with $n\geq 0$ and $e_i\in E$ 
for every $i\in[n]$, such that for every $m\in[n-1]$ the edge $e_{m+1}$ starts at the node where $e_m$ ends. 
If $n=0$, i.e., $\pi=\varepsilon$, then $\pi$ is a path from $u$ to $u$ for every $u\in V$. 
If $n\geq 1$, then $\pi$ is a path from the start node of $e_1$ to the end node of $e_n$.
If $\pi$ is a path from $u$ to $v$, then it is said to start at $u$ and end at $v$. 
A path is a cycle if it is nonempty and starts and ends at the same node.  

\smallskip
Top-down tree transducers, which will be recalled in Section~\ref{sec:dtop},
work on ranked trees.
This means that the number of children 
of a node of a tree is determined by the symbol at that node.
A ranked alphabet $\Sigma$ is a finite set of symbols 
such that each symbol $a\in\Sigma$
is implicitly equipped with a rank $\rk(a)\in\Nat$.
For $k\in\Nat$ we define $\Sigma^{(k)}=\{a\in\Sigma\mid\rk(a)=k\}$.
To avoid trivialities, we assume that $\Sigma^{(0)}\neq\emptyset$. 
To indicate that $\sigma\in\Sigma$ has rank $k$, we also write it as $\sigma^{(k)}$.

The set $\T_{\Sigma}$ of (finite, ordered, ranked) trees over the ranked
alphabet $\Sigma$ is the smallest set (of terms) such that $a(t_1,\dots,t_k)\in\T_\Sigma$ 
if $k\in\Nat$, $a\in\Sigma^{(k)}$, and $t_1,\dots,t_k\in\T_\Sigma$. 
If $a\in\Sigma^{(0)}$, then we also write $a$ for the tree $a()$.
If $a\in\Sigma^{(1)}$ and $t\in\T_\Sigma$, then we also write $at$ for $a(t)$.
More generally, for a string $w=a_1\cdots a_n$ with $n\in\Nat$ and $a_i\in\Sigma^{(1)}$, 
we write $w(t)=a_1\cdots a_n(t)$ for the tree $a_1(a_2(\cdots a_n(t)\cdots))$;
in particular $\varepsilon(t)=t$. 

We represent the nodes of a tree in Dewey notation, i.e., by
strings of positive natural numbers. 
The empty string $\varepsilon$ represents the root node 
and, for $i\in\Nat_+$, $vi$ represents the $i$th child of the node $v$
(and $v$ is the parent of $vi$). 
Every node $v$ of a tree $t$ has a label in $\Sigma$, denoted $\lab(t,v)$.
Formally, the set $V(t)\subseteq\Nat_+^*$ of nodes (together with their labels) 
of the tree $t$ is inductively defined as:
$V(t)=\{\varepsilon\}\cup\{iv\mid i\in[k], v\in V(t_i)\}$
if $t=a(t_1,\dots,t_k)$, $a\in\Sigma^{(k)}$, and $t_1,\dots,t_k\in\T_\Sigma$; 
moreover, $\lab(t,\varepsilon)=a$ and $\lab(t,iv)=\lab(t_i,v)$. 
A node $u$ is an ancestor of node $v$ 
(and $v$ is a descendant of $u$)
if $u$ is a prefix of $v$;
it is a proper ancestor/descendant if 
it is a proper prefix. 
For $\Delta\subseteq\Sigma$, we define 
$V_\Delta(t)=\{v\in V(t)\mid \lab(t,v)\in\Delta\}$; 
for $a\in\Sigma$, we write $V_a(t)$ instead of $V_{\{a\}}(t)$.
The subtree of $t$ rooted at $v\in V(t)$ is denoted by $t/v$;
formally, $t/\varepsilon=t$ and if $t=a(t_1,\dots,t_k)$ then $t/iv=t_i/v$. 
The size of $t$, denoted $\size(t)$, is its number $|V(t)|$ of nodes.
The height of $t$, denoted $\height(t)$, is the maximal length of its nodes,
i.e., $\Max\{|v|\mid v\in V(t)\}$.
As an example, if $t=\sigma(\sigma(a,b),\tau(b))$, then 
$V(t)=\{\varepsilon,1,(1,1),(1,2),2,(2,1)\}$, $\;\lab(t,(1,2))=b$, $\;V_b(t)=\{(1,2),(2,1)\}$,
$\;t/1=\sigma(a,b)$, $\;\size(t)=6$, and $\height(t)=2$. 
The height of a tuple of trees $\bar{t}=(t_1,\dots,t_k)$, $k\geq 1$, 
is defined as $\height(\bar{t})=\Max\{\height(t_i)\mid i\in[k]\}$.

Let $\Delta$ be a ranked alphabet such that every symbol in $\Sigma\cap\Delta$
has the same rank in $\Sigma$ and $\Delta$. 
For a set of trees $\T\subseteq\T_\Delta$, we define 
$\Sigma(\T)\subseteq\T_{\Sigma\cup\Delta}$ to be the set of trees $a(t_1,\dots,t_k)$ such that 
$k\in\Nat$, $a\in\Sigma^{(k)}$, and $t_1,\dots,t_k\in\T$, and we define
$\T_\Sigma(\T)\subseteq\T_{\Sigma\cup\Delta}$ to be the smallest set of trees $\T'$ such that 
$\T\cup\Sigma(\T')\subseteq \T'$. Note that $\T_\Sigma(\emptyset)=\T_\Sigma$.

A $\Sigma$-\emph{pattern} is an upper portion, or prefix, of a 
tree in $\T_\Sigma$. Formally the set $\mathcal{P}_\Sigma$ of $\Sigma$-patterns 
is defined to be the set of trees $\T_{\Sigma}(\{\bot\})$,
where $\bot$ is a new symbol of rank zero that is not in $\Sigma$.
If $t_0$ is a pattern containing exactly $k$ occurrences of $\bot$,
and $t_1,\dots,t_k$ is a sequence of $k$ patterns, 
then the pattern $t=t_0[t_1,\dots,t_k]$ is obtained from 
$t_0$ by replacing the $i$th occurrence of $\bot$ 
(in left-to-right order) by $t_i$.
A $\Sigma$-\emph{context} is a $\Sigma$-pattern that 
contains exactly one occurrence of $\bot$.
The set of all $\Sigma$-contexts is denoted $\C_\Sigma$.
Thus, for $C\in\C_\Sigma$ and $t\in\T_\Sigma$, the tree $C[t]\in\T_\Sigma$ 
is obtained from the context $C$ by replacing the unique occurrence of $\bot$ in $C$ by $t$. 

On the set $\mathcal{P}_\Sigma$ 
we define a partial order $\sqsubseteq$ as follows:
for patterns $t$ and $t'$ in $\mathcal{P}_\Sigma$,
$t'$ is a \emph{prefix} of $t$, denoted $t'\sqsubseteq t$, if
$t = t'[t_1,\dots,t_k]$ for suitable patterns $t_1,\dots,t_k$;
equivalently, $V_a(t')\subseteq V_a(t)$ for every $a\in\Sigma$.
Obviously, $\bot\sqsubseteq t$ for every pattern $t$.
We note that in~\cite{DBLP:journals/jcss/EngelfrietMS09} the inverse of the partial order 
$\sqsubseteq$ is used.
Every nonempty set $\Pi$ of $\Sigma$-patterns has a 
greatest lower bound $\sqcap \Pi$ in $\mathcal{P}_\Sigma$, 
called the \emph{largest common prefix} of the patterns in $\Pi$; 
it is the unique pattern $t'$ such that 
for every $v\in\Nat_+^*$ and $a\in\Sigma$, 
$v\in V_a(t')$ if and only if 
(1) $v\in V_a(t)$ for every $t\in\Pi$ and
(2)~every proper ancestor of $v$ is in $V(t')$.
This implies the following easy lemma. 

\begin{lemma}\label{lm:pat}
Let $\Pi$ be a nonempty subset of $\T_\Sigma$,
and let $v\in\Nat_+^*$. \\
Then, $v\in V_\bot(\sqcap \Pi)$ if and only if
\begin{enumerate}
\item[(1)] $v\in V(t)$ for every $t\in\Pi$, 
\item[(2)] $\lab(t_1,\hat{v})=\lab(t_2,\hat{v})$ 
for every proper ancestor $\hat{v}$ of $v$ and all $t_1,t_2\in\Pi$, and
\item[(3)] there exist $t_1,t_2\in \Pi$ such that $\lab(t_1,v)\neq\lab(t_2,v)$.
\end{enumerate} 
\end{lemma}
For instance, $\sqcap\{\sigma(\tau(a),b),\sigma(b,b)\}= 
\sigma(\tau(a),b)\sqcap \sigma(b,b)= \sigma(\bot,b)$. 

For $t,t'\in\T_\Sigma$ and $v\in V(t)$, we denote by $t[v\leftarrow t']$
the tree that is obtained from $t$ by replacing its subtree $t/v$ by $t'$.
More precisely, if $C$ is the unique context in $\C_\Sigma$ such that 
$C\sqsubseteq t$ and $C/v=\bot$, 
then $t[v\leftarrow t']=C[t']$. 

Let $\cS$ be a subset of $\T_\Sigma$ such that 
no $s\in\cS$ is a subtree of $s'\in\cS$ with $s\neq s'$.
For a tree $t\in\T_\Sigma$ and 
a partial function $\psi: \cS\to \T_\Sigma$, we define 
$t[s \leftarrow \psi(s)\mid s\in \cS]$ to be the result
of replacing every subtree $s$ of $t$ by $\psi(s)$, for every $s\in\cS$. 
More precisely, $t[s \leftarrow \psi(s)\mid s\in \cS]=
t[v_1\leftarrow \psi(t/v_1)]\cdots[v_k\leftarrow \psi(t/v_k)]$
where $\{v_1,\dots,v_k\}=\{v\in V(t)\mid t/v\in\cS\}$. 
Note that $v_i$ is not an ancestor of $v_j$, for $i\neq j$, 
and hence the order of the substitutions $[v_i\leftarrow \psi(t/v_i)]$ is irrelevant.
Note also that $t[s \leftarrow \psi(s)\mid s\in \cS]$ is defined 
if and only if $\psi(t/v_i)$ is defined for every $i\in[k]$. 

To formulate the rules of top-down tree transducers, 
we use variables $x_i$, with $i\in\Nat$, 
which are assumed to have rank~0.
The set $\{x_0,x_1,x_2,\dots\}$ of all such variables is denoted $X$.
For $k\in\Nat$, we denote 
$\{x_1,\dots,x_k\}$ by $X_k$; note that $X_0=\emptyset$.

\section{Deterministic Top-Down Tree Transducers}
\label{sec:dtop}

A \emph{deterministic top-down tree transducer with regular look-ahead}
(\emph{dtla} for short) is a tuple
$M=(Q,\Sigma,\Delta,R,A,P,\delta)$ where 
$Q$ is a finite set of states of rank $1$, $\Sigma$ and $\Delta$ are the
ranked input and output alphabets, respectively,
and $P$ is a finite nonempty set of look-ahead states.
The function $A$ maps look-ahead states
to trees in $\T_\Delta(Q(\{x_0\}))$; 
for $p\in P$, the tree $A(p)$ is called 
the $p$-axiom of $M$.
The finite set $R$ provides at most one rule 
\[
q(a(x_1\ot p_1,\dots,x_k\ot p_k))\to \zeta
\]
for every state $q$, every input symbol $a$ of rank $k\geq 0$ 
and every sequence $p_1,\dots,p_k$ of look-ahead states. 
The right-hand side $\zeta$ of the rule is a tree in $\T_\Delta(Q(X_k))$,
i.e., $\zeta = t[q_1(x_{i_1}),\ldots, q_r(x_{i_r})]$
for some pattern $t\in\mathcal{P}_{\Delta}$,
$r=|V_\bot(t)|$, $q_j\in Q$, and $x_{i_j}\in\{x_1,\ldots,x_k\}$ for $j\in[r]$;
we will denote $\zeta$ also by $\rhs(q,a,p_1,\dots,p_k)$. 
Finally, $\delta$ is the transition function
of the (total deterministic bottom-up) 
look-ahead automaton $(P,\delta)$. 
That means that $\delta(a,p_1,\dots,p_k)\in P$
for every $k\geq 0$, $a\in\Sigma^{(k)}$, and $p_1,\dots,p_k\in P$.

Examples of dtlas are given in the next section. 
Whenever we consider a dtla with the name $M$, it will be understood
that its components are named $(Q,\Sigma,\Delta,R,A,P,\delta)$.
When necessary we provide the components of a dtla $M$ with the subscript $M$. 
Then we have $Q_M$, $\Sigma_M$, $\Delta_M$, $R_M$, $\rhs_M$, etc.
We denote by $\maxrhs(M)$ the maximal height 
of the axioms and the right-hand sides of the rules of $M$.

We now define the semantics of the dtla $M$, starting with the semantics 
of its look-ahead automaton $(P,\delta)$. 
The transition function $\delta$ gives rise to a function $\delta^*$ that maps
$\T_\Sigma$ to $P$. It is defined by
$\delta^*(a(s_1,\ldots,s_k)) = \delta(a,\delta^*(s_1),\ldots,\delta^*(s_k))$
for $a\in\Sigma^{(k)}$ and $s_1,\dots,s_k\in\T_\Sigma$.
For convenience, we denote the function $\delta^*$ by $\delta$ as well.
For $p\in P$ we denote by $\sem{p}_M$ 
the set of trees $s\in \T_\Sigma$ that have look-ahead state $p$, i.e., $\delta(s)=p$;
we drop the subscript $M$ from $\sem{p}_M$ whenever it is clear from the context. 
Note that $\{\sem{p}\mid p\in P\}$ is a partition of $\T_\Sigma$.
For a node $u$ of an input tree $s\in \T_\Sigma$, 
we also say that $\delta(s/u)$ is the look-ahead state at $u$. 

For $q\in Q$, $s\in\T_\Sigma$, and $u\in V(s)$, we define 
$\rhs(q,s,u)=\rhs(q,a,p_1,\dots,p_k)$ where 
$\lab(s,u)=a\in\Sigma^{(k)}$ and $p_i=\delta(s/ui)$ for every $i\in[k]$.
Intuitively, $\rhs(q,s,u)$ is the right-hand side of the rule that is applied
when $M$ arrives at node $u$ in state $q$ (if that rule exists); 
it is uniquely determined by the label of $u$ and the look-ahead states at its children. 

A \emph{sentential form} of $M$ for $s\in\T_\Sigma$ is a tree in $\T_\Delta(Q(V(s)))$, 
where the nodes in $V(s)$ are viewed as symbols of rank~0. 
For sentential forms $\xi,\xi'$ we write $\xi \Rightarrow_s \xi'$ 
if there exist $v\in V(\xi)$, $q\in Q$, and $u\in V(s)$ such that 
$$\xi/v=q(u) \mbox{ and } \xi'=\xi[v\leftarrow \rhs(q,s,u)[x_i\leftarrow ui\mid i\in\Nat_+]].$$
This will be called a computation step of $M$ in state $q$ at nodes $u$ and $v$. 
It is easy to see that the rewriting in computations is confluent
(i.e., if $\xi\Rightarrow^*_s\xi_1$ and $\xi\Rightarrow^*_s\xi_2$, 
then there exists a sentential form $\bar{\xi}$ such that 
$\xi_1\Rightarrow^*_s\bar{\xi}$ and $\xi_2\Rightarrow^*_s\bar{\xi}$).
Hence, if $\xi\Rightarrow^*_s t\in\T_\Delta$ and $\xi\Rightarrow^*_s\xi'$, 
then $\xi'\Rightarrow^*_s t$; thus, computations that start with a given sentential form 
lead to a unique tree in $\T_\Delta$ (if it exists). 

The dtla $M$ \emph{realizes} a partial function $\sem{M}: \T_\Sigma\to \T_\Delta$, 
called its \emph{translation}. 
Let $s\in \T_\Sigma$ and $\delta(s)=p\in P$.  
The output tree $\sem{M}(s)$ of the transducer $M$ for the input tree $s$ is the unique tree 
$t\in\T_\Delta$ such that $A(p)[x_0\leftarrow\varepsilon] \Rightarrow_s^* t$ (if it exists). 
For readability, we will write $M(s)$ instead of $\sem{M}(s)$.

Two dtlas $M_1$ and $M_2$ are \emph{equivalent} if they realize the same translation, 
i.e., if $\Sigma_{M_1}=\Sigma_{M_2}$, $\Delta_{M_1}=\Delta_{M_2}$, and $\sem{M_1}=\sem{M_2}$.

Intuitively, a sentential form $\xi$ consists of output that has already been produced by $M$;  
moreover, $\xi/v=q(u)$ means that $M$ has arrived at node $u$ of the input tree $s$
in state $q$ and, starting in that state, will translate the input subtree $s/u$ into the output 
subtree $M(s)/v$. Note that several parallel copies of $M$ can arrive at $u$ for different nodes 
of $\xi$, i.e., there may exist nodes $v'\neq v$ such that $\xi/v'=q'(u)$, 
where $q'$ may also be equal to $q$. 

A sentential form $\xi$ for $s$ is \emph{reachable} 
if $A(p)[x_0\leftarrow\varepsilon] \Rightarrow_s^* \xi$ where $p=\delta(s)$. 
Thus, if $M(s)$ is defined and $\xi$ is a reachable sentential form for $s$, 
then $\xi\Rightarrow_s^* M(s)$.

We also define the semantics of every state $q$ of $M$ as a partial function $\sem{q}_M: \T_\Sigma\to \T_\Delta$
as follows. For $s\in\T_\Sigma$, $\sem{q}_M(s)$ is the unique tree $t\in\T_\Delta$ such that 
$q(\varepsilon)\Rightarrow_s^* t$ (if it exists). 
For readability, we will write $q_M(s)$ instead of $\sem{q}_M(s)$.

The following lemma is easy to prove. 

\begin{lemma}\label{lm:elem}
 Let $s\in\T_\Sigma$ and $t\in\T_\Delta$. 
\begin{enumerate}
\item[(1)] For every $q\in Q$ and $u\in V(s)$, 
$$q(u)\Rightarrow_s^* t \;\mbox{ if and only if }\; q_M(s/u)=t.$$
\item[(2)] For every sentential form $\xi$ for $s$, 
$$\xi\Rightarrow_s^* t \;\mbox{ if and only if }\; 
t=\xi[q(u)\leftarrow q_M(s/u)\mid q\in Q, u\in V(s)].$$
\item[(3)] If $\delta(s)=p$, then 
$$M(s)=A(p)[q(x_0)\leftarrow q_M(s)\mid q\in Q].$$
\item[(4)] For every $\bar{q}\in Q$, if $s=a(s_1,\dots,s_k)$, then 
$$\bar{q}_M(s)=\rhs(\bar{q},a,\delta(s_1),\dots,\delta(s_k))
[q(x_i)\leftarrow q_M(s_i)\mid q\in Q,\,i\in[k]].$$
\end{enumerate}
\end{lemma}
\begin{proof}
(1) follows from the obvious bijection between the nodes of $s/u$ and the nodes of $s$ with prefix $u$ 
(i.e., the descendants of $u$ in $s$).
 
(2) is obvious from (1) and the fact that the computation steps $\xi \Rightarrow_s \xi'$ of $M$ are context-free. In fact, the computations of $M$ on $s$ can be viewed as derivations of a context-free grammar with the set of nonterminals $Q(V(s))$ and with rules $q(u)\to \rhs(q,s,u)[x_i\leftarrow ui\mid i\in\Nat_+]$.
  
(3) and (4) follow from (2), 
taking $\xi=A(p)[x_0\leftarrow\varepsilon]$ and 
$\xi=\rhs(\bar{q},s,\varepsilon)[x_i\leftarrow i\mid i\in[k]]$, respectively.
\qed
\end{proof}
Note that (3) and (4) of Lemma~\ref{lm:elem} form an alternative way of defining the semantics of~$M$ (recursively).

\medskip
{\bf Convention.} 
For a given dtla $M$ it can be assumed that it is \emph{reduced}, i.e., that all its states and look-ahead states are \emph{reachable} in the following sense. A look-ahead state $p$ is reachable if $\sem{p}_M\neq\emptyset$. A state $q$ is reachable if $q$ occurs in an axiom, or if $q$ occurs in the right-hand side of a rule of which the left-hand side starts with a reachable state. 
In what follows we will assume all dtlas to be reduced. However, all properties that we will define, 
also apply to non-reduced dtlas. Whenever we construct a dtla and show that it has certain properties,
we allow the dtla to be non-reduced, but we will ensure that those properties are preserved under reduction (i.e., removing unreachable states and look-ahead states and the rules in which they occur). However, that will not be mentioned explicitly. 
\qed

\smallskip
A \emph{deterministic top-down tree transducer} (\emph{dtop} for short) 
is a dtla $M$ with trivial look-ahead automaton $(P,\delta)$, i.e., $P$ is a singleton. 
Whenever convenient, we drop $(P,\delta)$ from the tuple defining $M$, 
we identify $A$ with the unique axiom $A(p)$, 
we write a rule as $q(a(x_1,\dots,x_k))\to \zeta$ rather than $q(a(x_1\ot p,\dots,x_k\ot p))\to \zeta$ 
(where $p$ is the unique look-ahead state of $M$) and we denote $\zeta$ by $\rhs(q,a)$. 

A dtla $M$ is \emph{proper} (a \emph{dtpla} for short) if it is not a dtop, i.e., if $|P|\geq 2$. 
Obviously, to decide whether $M$ is equivalent to a dtop, we may assume that $M$ is proper. 

A dtla $M$ is \emph{total} if $\dom(\sem{M})=\T_\Sigma$, 
i.e., if its translation $\sem{M}$ is a total function. 
Note that it is decidable whether $M$ is total, 
because $\dom(\sem{M})$ is effectively a regular tree language
(cf.~\cite[Corollary~2.7]{DBLP:journals/mst/Engelfriet77}).
From now on we mostly consider total dtlas.

A dtla $M$  is \emph{complete} if 
$\rhs(q,a,p_1,\dots,p_k)$ is defined for every 
$q\in Q$, $k\in\Nat$, $a\in\Sigma^{(k)}$, and $p_1,\dots,p_k \in P$.
By Lemma~\ref{lm:elem}(4), this means that $q_M(s)$ is defined for every $q\in Q$ and $s\in\T_\Sigma$. 
Thus, by Lemma~\ref{lm:elem}(3), if $M$ is complete, then $M$ is total.

A dtla $M$ is \emph{linear}
if (1) for every $p\in P$, the variable $x_0$ occurs at most once in $A(p)$, and 
(2) for every rule $q(a(x_1\ot p_1,\ldots,x_k\ot p_k)) \to \zeta$, 
each variable $x_i$ occurs at most once in $\zeta$.

A dtla $M$ is \emph{ultralinear}
if there is a mapping $\mu: Q\to\Nat$ 
such that for every rule
$q(a(x_1\ot p_1,\ldots,x_k\ot p_k)) \to \zeta$
the following two properties hold 
for every $\bar{q}(x_i)$ that occurs in $\zeta$: 
(1) $\mu(\bar{q})\geq \mu(q)$, and 
(2) if $\mu(\bar{q})= \mu(q)$, then 
$x_i$ occurs only once in $\zeta$.
Obviously, every linear dtla is ultralinear.
We note that ultra-linearity was first defined for context-free grammars, in~\cite{GinsSpan}.

A dtla $M$ is \emph{nonerasing} if it does not have erasing rules. 
A rule of $M$ is an \emph{erasing rule} 
if its right-hand side is in $Q(X)$, i.e.,
contains no symbols from $\Delta$.

A dtla $M$ is \emph{bounded erasing} (for short, \emph{b-erasing})
if there is no cycle in the directed graph $E_M$ with the set of nodes
$Q$ and an edge from $q$ to $q'$ if there is an erasing rule of the form
$q(a(x_1\ot p_1,\ldots,x_k\ot p_k)) \to q'(x_j)$.
Obviously, every nonerasing dtla is b-erasing. 

Note that all the above properties, except properness, are preserved under reduction.

\subsubsection{Look-ahead states in input trees.}
Let $M=(Q,\Sigma,\Delta,R,A,P,\delta)$ be a total dtla.
To analyze the behaviour of $M$ for different look-ahead states, 
we will consider input trees $\bar{s}$ with occurrences of $p\in P$, 
viewed as input symbol of rank zero, 
representing an absent subtree $s$ with $\delta(s)=p$. 
If $M$ arrives in state $q$ at a $p$-labeled leaf of $\bar{s}$, 
then $M$ will output the new symbol $\angl{q,p}$ of rank zero,
representing the absent output tree $q_M(s)$. 
In this way, $M$ translates input trees $\bar{s}\in\T_\Sigma(P)$ to output trees
in $\T_\Delta(Q\times P)$. Without loss of generality we assume that $P$ and $\Sigma$ are disjoint, 
and so are $Q\times P$ and~$\Delta$.

Formally, we extend $M$ to a dtla $M^\circ=(Q,\Sigma^\circ,\Delta^\circ,R^\circ,A,P,\delta^\circ)$
where $\Sigma^\circ=\Sigma\cup P$ 
such that every element of $P$ has rank zero, 
$\Delta^\circ=\Delta\cup (Q\times P)$ 
such that every element of $Q\times P$ has rank zero, 
$R^\circ$ is obtained from $R$ by adding the rules $q(p)\to \angl{q,p}$  
for all $q\in Q$ and $p\in P$ such that $q_M(s)$ is defined for some $s\in\sem{p}_M$, 
and $\delta^\circ$ is the extension of $\delta$ such that  
$\delta^\circ(p)=p$ for every $p\in P$.

For notational simplicity, we will denote $\delta^\circ(\bar{s})$, $M^\circ(\bar{s})$, and 
$q_{M^\circ}(\bar{s})$ by $\delta(\bar{s})$, $M(\bar{s})$, and $q_M(\bar{s})$, respectively, 
for every input tree $\bar{s}\in\T_\Sigma(P)$.  
But note that we do \emph{not} drop $^\circ$ from 
$\sem{p}_{M^\circ}$, $\sem{M^\circ}$, and $\sem{q}_{M^\circ}$, 
i.e., $\sem{p}_M$, $\sem{M}$, and $\sem{q}_M$ keep their meaning. 

We will use the following elementary lemma, which expresses the above intuition. 

\begin{lemma}\label{lm:MM}
Let $M$ be a total dtla.
Let $\bar{s}$ be a tree in $\T_\Sigma(P)$, and for every $p\in P$,
let $s_p$ be a tree in $\T_\Sigma(P)$ such that $\delta(s_p)=p$. 
Then $\delta(\bar{s}[p\leftarrow s_p\mid p\in P])=\delta(\bar{s})$ and 
\begin{equation}\label{eq:MM}
M(\bar{s}[p\leftarrow s_p\mid p\in P])=
M(\bar{s})\big[\angl{q,p}\leftarrow q_M(s_p)\mid q\in Q, p\in P \big].
\end{equation}
\end{lemma}
\begin{proof}
Let $s'=\bar{s}[p\leftarrow s_p\mid p\in P]$. 
It should be clear that $\delta(s'/v)=\delta(\bar{s}/v)$ for every node $v$ of $\bar{s}$. 
Let $p'=\delta(s')=\delta(\bar{s})$.

We first assume that $s_p\in\T_\Sigma$ for every $p\in P$. Thus $s'\in\T_\Sigma$, 
and so $M(s')$ is defined. 
Let $U$ be the set of nodes $u$ of $\bar{s}$ such that $\lab(\bar{s},u)\in P$, and
for $u\in U$, let $p_u=\lab(\bar{s},u)$.
Now consider a computation 
$A(p')[x_0\leftarrow\varepsilon]\Rightarrow_{s'}^*\xi$ of $M$ 
such that $\xi\in\T_\Delta(Q(U))$ and none of the computation steps is at a node $u\in U$
(and hence not at a descendant of $u$).
Then, by the observation above, also $A(p')[x_0\leftarrow\varepsilon]\Rightarrow_{\bar{s}}^*\xi$. 
By Lemma~\ref{lm:elem}(2), 
$M(s')=\xi[q(u)\leftarrow q_M(s_{p_u})\mid q\in Q, u\in U]$.
Thus, $q_M(s_{p_u})$ is defined for every $q(u)$ that occurs in $\xi$, and so 
$q(p_u)\to \angl{q,p_u}$ is a rule of $M^\circ$. 
Hence $\xi\Rightarrow_{\bar{s}}^*\xi[q(u)\leftarrow \angl{q,p_u}\mid q\in Q, u\in U]$
and so $M(\bar{s})=\xi[q(u)\leftarrow \angl{q,p_u}\mid q\in Q, u\in U]$. 
This proves Equation~(\ref{eq:MM}) for the case where $s_p\in\T_\Sigma$.
It also implies that $M^\circ$ is total, because for every $\bar{s}\in\T_\Sigma(P)$
one can choose some $s_p\in\sem{p}_M$ for each $p\in P$, 
and then $M(\bar{s}[p\leftarrow s_p\mid p\in P])$ is defined and hence so is $M(\bar{s})$. 
Since $M^\circ$ is total, the previous argument 
also proves Equation~(\ref{eq:MM}) for the general case where $s_p\in\T_\Sigma(P)$. 
\qed
\end{proof}

Note that the proof of Lemma~\ref{lm:MM} shows that if $M$ is total, then so is $M^\circ$. 
From Lemma~\ref{lm:MM} we immediately obtain the next lemma for $\Sigma$-contexts. 
Note that for every $C\in\C_\Sigma$ and $p\in P$,
the tree $M(C[p])$ is in $\T_\Delta(Q\times\{p\})$.

\begin{lemma}\label{lm:semantics}
Let $M$ be a total dtla. 
Let $C\in\C_\Sigma$, $s\in\T_\Sigma(P)$, 
and $p\in P$ such that $\delta(s)=p$. Then $\delta(C[s])=\delta(C[p])$ and 
$M(C[s])=M(C[p])\big[\angl{q,p}\leftarrow q_M(s)\mid q\in Q \big]$.
\end{lemma}
\begin{proof}
Apply Lemma~\ref{lm:MM} with $\bar{s}=C[p]$ and $s_p=s$.
\qed
\end{proof}

In Section~\ref{sec:diff} the next lemma will be needed. 

\begin{lemma}\label{lm:reach}
Let $M$ be a total dtop. Then $M$ is complete, and for every $q\in Q$ 
there exists $C\in\C_\Sigma$ such that $\angl{q,p}$ occurs in $M(C[p])$, where $P=\{p\}$. 
\end{lemma}

\begin{proof}
It is convenient to assume that $p=\bot$ (and so $C[p]=C$). 
We first prove the second statement. 
Since we assume, by convention, that every state $q$ of $M$ is reachable, 
we proceed by induction on the definition of reachability. 
If $q(x_0)$ occurs in the axiom $A$ of $M$, then $C=\bot$ satisfies the requirement
because $M(\bot)=A[\bar{q}(x_0)\leftarrow\angl{\bar{q},\bot}\mid \bar{q}\in Q]$. 
Now let $q(a(x_1,\dots,x_k))\to \zeta$ be a rule of $M$ such that $q$ is reachable, 
and let $\zeta/z=q'(x_j)$. 
By induction there exist $C\in\C_\Sigma$ and $v\in V(M(C))$ 
such that $M(C)/v=\angl{q,\bot}$. 
Let $s_j=\bot$ and choose $s_m\in\T_\Sigma$ for $m\in[k]-\{j\}$. 
Let $C'$ be the $\Sigma$-context $C[a(s_1,\dots,s_k)]$.  
By Lemma~\ref{lm:semantics}, 
$M(C')=M(C)[\angl{\bar{q},\bot}\leftarrow \bar{q}_M(a(s_1,\dots,s_k))\mid \bar{q}\in Q]$
and by Lemma~\ref{lm:elem}(4), 
$q_M(a(s_1,\dots,s_k))=\zeta[\bar{q}(x_i)\leftarrow \bar{q}_M(s_i)\mid \bar{q}\in Q, i\in[k]]$.
Then $M(C')/vz=q_M(a(s_1,\dots,s_k))/z=q'_M(s_j)=\angl{q',\bot}$.
So, $\angl{q',\bot}$ occurs in $M(C')$. 

To show that $M$ is complete, let $q\in Q$ and $a\in\Sigma^{(k)}$. 
We have shown that there exist $C\in\C_\Sigma$ and $v\in V(M(C))$ 
such that $M(C)/v=\angl{q,\bot}$.
Let $C'$ be as above. Then $M(C')/v=q_M(a(s_1,\dots,s_k))$. 
Hence $\rhs(q,a)$ is defined by Lemma~\ref{lm:elem}(4). 
\qed
\end{proof}

\section{Difference Trees}
\label{sec:difftree}

Let $M$ be a total dtla. 
We wish to decide whether $M$ is equivalent to a dtop. 
Let $C$ be a $\Sigma$-context and let $p,p'\in P$.
As explained in the Introduction, 
we are interested in the difference between the output of $M$ on input 
$C[p]$ and its output on input $C[p']$, see also Lemma~\ref{lm:semantics}. 
Intuitively, a dtop $N$ that is equivalent to $M$ 
does not know whether the subtree $s$ of an input tree $C[s]$ has look-ahead state $p$ or $p'$, 
and hence, when reading the context $C$, it can output at most the largest common prefix 
$M(C[p])\sqcap M(C[p'])$ of the output trees $M(C[p])$ and $M(C[p'])$. Recall
that $M(C[p])$ denotes $M^\circ(C[p])$, which is defined because $M^\circ$ is total (as shown 
in the proof of Lemma~\ref{lm:MM}).
Let $v$ be a node of $M(C[p])\sqcap M(C[p'])$ with label $\bot$. 
Then we say that $M(C[p])/v$ is a \emph{difference tree} of $M$ (and hence, by symmetry,
so is $M(C[p'])/v$). Thus, a difference tree is a part of the output that can be produced by $M$
because it knows that $s$ has look-ahead state $p$ (or $p'$). 
Intuitively, to simulate $M$, the dtop $N$ must store the difference trees in its state. 
Hence, for $N$ to exist, there should be finitely many difference trees (as will be proved 
in Corollary~\ref{co:difffin}). 
We denote the set of all difference trees of $M$ by $\diff(M)$, 
for varying $C$, $p$, $p'$, and $v$. Thus we define 
\begin{eqnarray*}
\lefteqn{\diff(M)=} \\
& & \hspace{0.4cm}
\{M(C[p])/v \mid C\in\C_\Sigma,\,p\in P,\,\exists p'\in P: v\in V_\bot(M(C[p]) \sqcap M(C[p']))\},
\end{eqnarray*}
which is a subset of $\T_\Delta(Q\times P)$. 
We define the number $\maxdiff(M)\in\Nat\cup\{\infty\}$ to be the least upper bound of the 
heights of all difference trees of $M$, i.e.,
$$\maxdiff(M)=\lub\{\height(t)\mid t\in\diff(M)\}$$
(for the definition of $\lub$ see Section~\ref{sec:prelim}). 
Intuitively, $\maxdiff(M)$ gives a measure of how much the transducer
$M$ makes use of its look-ahead information. 
Clearly, $\maxdiff(M)$ is finite (i.e., is in $\Nat$) if and only if $\diff(M)$ is finite. 
We will say that a number $h(M)\in\Nat$ is a \emph{difference bound} for $M$ if the following holds:
if $\diff(M)$ is finite, then $\maxdiff(M)\leq h(M)$.
Thus, if $\diff(M)$ is infinite, then any natural number is a difference bound for $M$;
if $\diff(M)$ is finite, then any upper bound of the heights of the (finitely many) difference trees 
of $M$ is a difference bound for $M$.
 
Our first main result (Theorem~\ref{th:alg}) is that if a difference bound for $M$ is known, 
then we can decide whether $M$ is equivalent to a dtop, and if so, construct such a dtop from~$M$.
Our second main result (Theorem~\ref{th:ultra}) is that a difference bound can be computed 
for every total dtla $M$ that is ultralinear and b-erasing. 

A node $v\in V_\bot(M(C[p]) \sqcap M(C[p']))$ will be called a \emph{difference node}
of $M(C[p])$ and $M(C[p'])$. It is characterized in the next lemma, 
which follows immediately from Lemma~\ref{lm:pat}. 

\begin{lemma}\label{lm:diff_char}
Let $C\in\C_\Sigma$, $p,p'\in P$, and $v\in\Nat_+^*$. Then, \\
$v$ is a difference node of $M(C[p])$ and $M(C[p'])$ if and only if 
\begin{enumerate}
\item[(1)] $v\in V(M(C[p]))\cap V(M(C[p']))$, 
\item[(2)] $\lab(M(C[p]),\hat{v})=\lab(M(C[p']),\hat{v})$ 
for every proper ancestor $\hat{v}$ of $v$, and
\item[(3)] $\lab(M(C[p]),v)\neq\lab(M(C[p']),v)$.
\end{enumerate}
\end{lemma}

Note that if $v$ is a difference node of $M(C[p])$ and $M(C[p'])$, then $p\neq p'$. 
Thus, in the definition of $\diff(M)$ we can assume that $p\neq p'$.
Hence, if $M$ is a dtop then $\diff(M)=\emptyset$ and so $\maxdiff(M)=0$.
Note also that, to compute $\maxdiff(M)$ for a dtla $M$, it suffices to consider difference trees 
of non-zero height, i.e., difference nodes $v$ that are not leaves of $M(C[p])$.  

We now give some examples of dtlas with their sets of difference trees. 

\begin{example}\label{ex:Mex}
Let $\Sigma=\Delta=\{\sigma^{(1)},a^{(0)},b^{(0)}\}$, which means that 
$\Sigma$ and $\Delta$ are the ranked alphabet $\{\sigma,a,b\}$ with $\rk(\sigma)=1$
and $\rk(a)=\rk(b)=0$. 
We consider the following total dtla $M=(Q,\Sigma,\Delta,R,A,P,\delta)$ with $M(\sigma^na)=a$ and 
$M(\sigma^nb)=\sigma^nb$ for every $n\in\Nat$. It is, in fact, the dtla $M_{\rm ex}$ 
of the Introduction, for this particular input alphabet.  
Its set of look-ahead states is $P=\{p_a,p_b\}$ with transition function $\delta$ 
defined by $\delta(a)=p_a$, $\delta(b)=p_b$,
$\delta(\sigma,p_a)=p_a$, and $\delta(\sigma,p_b)=p_b$. Its set of states is $Q=\{q\}$, 
its two axioms are $A(p_a)=a$ and $A(p_b)=q(x_0)$, 
and its set $R$ of rules contains the two rules 
$q(\sigma(x_1\ot p_b))\to \sigma(q(x_1))$ and $q(b)\to b$. 

Clearly, $\C_\Sigma=\{\sigma^n\bot\mid n\in\Nat\}$ and for $C=\sigma^n\bot$ we have 
$M(C[p_a])=a$ and $M(C[p_b])=\sigma^n\angl{q,p_b}$. 
Since $M(C[p_a])\sqcap M(C[p_b])=\bot$, the only difference node of $M(C[p_a])$ and $M(C[p_b])$
is $\varepsilon$, and we obtain the difference trees $M(C[p_a])$ and $M(C[p_b])$. 
Hence, $\diff(M)=\{a\}\cup\{\sigma^n\angl{q,p_b}\mid n\in\Nat\}$ and $\maxdiff(M)=\infty$.
Since $\diff(M)$ is infinite, $M$ is not equivalent to a dtop, 
as will be shown in Corollary~\ref{co:difffin}.
\qed
\end{example}

\begin{example}\label{ex:Mthree}
Let $\Sigma=\Delta=\{\sigma^{(1)},\tau^{(1)},a^{(0)},b^{(0)}\}$ and consider the following total dtla $M$. 
For an input tree $s$ with leaf $a$, $M$ outputs the top-most 80 unary symbols of $s$ and the leaf $a$ 
if $\size(s)> 80$, and it outputs $s$ if $\size(s)\leq 80$. 
The same is true for an input tree with leaf $b$, with 30 unary symbols instead of 80.
The look-ahead automaton of $M$ is similar to the one of the previous example: 
$P=\{p_a,p_b\}$ with $\delta(a)=p_a$, $\delta(b)=p_b$, and 
$\delta(\gamma,p)=p$ for $\gamma\in\{\sigma,\tau\}$ and $p\in P$. 
The set of states is $Q=\{q^a_i\mid i\in[80]\}\cup\{q^b_i\mid i\in[30]\}$. 
The axioms are $A(p_a)=q^a_{80}(x_0)$ and $A(p_b)=q^b_{30}(x_0)$. 
For the states with superscript $a$, $M$ has the rules $q^a_i(a)\to a$ for $i\in[80]$, 
and for $\gamma\in\{\sigma,\tau\}$, the rules
$q^a_i(\gamma(x_1\ot p_a))\to \gamma(q^a_{i-1}(x_1))$ for $1<i\leq 80$, and
$q^a_1(\gamma(x_1\ot p_a))\to \gamma(a)$. The rules for the states with superscript $b$ are similar
(in the obvious way, with $b$ instead of $a$ and $30$ instead of $80$). 

Every $\Sigma$-context $C$ is of the form $C=w(\bot)$ with $w\in\{\sigma,\tau\}^*$, 
cf. Section~\ref{sec:prelim} for this notation. For such a context we have 
$M(C[p_a])=w(\angl{q^a_{80-|w|},p_a})$ if $|w|<80$, and $M(C[p_a])=w_{[80]}(a)$ if $|w|\geq 80$,
where $w_{[80]}$ is the prefix of $w$ of length $80$. A similar statement holds for $M(C[p_b])$. 
Thus, for $|w|<30$ we obtain that $M(C[p_a])=w(\angl{q^a_{80-|w|},p_a})$ and 
$M(C[p_b])=w(\angl{q^b_{30-|w|},p_b})$, hence 
$M(C[p_a])\sqcap M(C[p_b])=w(\bot)$, which gives the 
difference trees $\angl{q^a_{i+50},p_a}$ and $\angl{q^b_i,p_b}$, where $i=30-|w|$. 
For $w=w_1z$ with $|w_1|=30$ and $|z|<50$, we get that 
$M(C[p_a])=w_1z(\angl{q^a_{80-|w_1z|},p_a})$ and $M(C[p_b])=w_1(b)$, hence
$M(C[p_a])\sqcap M(C[p_b])=w_1(\bot)$,
with the difference trees $z(\angl{q^a_{50-|z|},p_a})$ and $b$. Finally, 
if $w=w_1zw_2$ with $|w_1|=30$ and $|z|=50$, then 
$M(C[p_a])=w_1z(a)$ and $M(C[p_b])=w_1(b)$, hence again $M(C[p_a])\sqcap M(C[p_b])=w_1(\bot)$,
now with the difference trees $z(a)$ and (again) $b$. 
Hence, $\maxdiff(M)=50$. 

It is not difficult to see that there exists a dtop $N$ equivalent to $M$. 
It outputs the top-most $30$ symbols, and then it 
stores in its state the next $50$ (or less) unary symbols of the input tree $s$, and
depending on the leaf label of $s$, it outputs these symbols and $a$ or it outputs $b$. 
Formally, its set of states is 
$Q_N= Q_{N1}\cup Q_{N2}$ with 
$Q_{N1}=\{q_i\mid i\in[30]\}$ and $Q_{N2}=\{q_z\mid z\in\{\sigma,\tau\}^*, \,|z|\leq 50\}$, and
its axiom is $q_{30}(x_0)$. The rules of $N$ are the following, 
for every $q_i\in Q_{N1}$, $q_z\in Q_{N2}$, and $\gamma\in\{\sigma,\tau\}$. 
First, the rules $q_i(\gamma(x_1))\to \gamma(q_{i-1}(x_1))$ for $i\neq 1$, and
$q_1(\gamma(x_1))\to \gamma(q_{\varepsilon}(x_1))$. 
Second, $q_z(\gamma(x_1))\to q_{z\gamma}(x_1)$ for $|z|<50$, and
$q_z(\gamma(x_1))\to q_z(x_1)$ for $|z|=50$.
Third, and finally, $q_i(a)\to a$, $\,q_i(b)\to b$,
$\,q_z(a)\to z(a)$, and $q_z(b)\to b$.
\qed
\end{example}

\begin{example}\label{ex:Mleaves}
Let $\Sigma=\{\sigma^{(2)},aa^{(0)},ab^{(0)},ba^{(0)},bb^{(0)}\}$ 
where we view $aa$, $ab$, $ba$, and $bb$ as symbols, and let $\Delta=\{\sigma^{(3)},\#^{(2)},a^{(0)},b^{(0)}\}\cup\Sigma^{(0)}$.
We consider the following total dtla $M$ such that 
$M(aa)=aa$, $M(ab)=ab$, $M(ba)=ba$, $M(bb)=bb$, and 
for every $s_1,s_2\in\T_\Sigma$, $M(\sigma(s_1,s_2))=\sigma(M(s_1),M(s_2),\#(y,z))$ where 
$y\in\{a,b\}$ is the first letter of the label of the left-most leaf of $\sigma(s_1,s_2)$
and $z\in\{a,b\}$ is the second letter of the label of its right-most leaf.
Its look-ahead automaton has four states $p_{yz}$ with $y,z\in\{a,b\}$, such that 
$\delta(yz)=p_{yz}$ and 
$\delta(\sigma,p_{wx},p_{yz})=p_{wz}$ for all $w,x,y,z\in\{a,b\}$. 
It has one state $q$, its axioms are $A(p_{yz})=q(x_0)$,
and its rules are $q(yz)\to yz$ and 
$$q(\sigma(x_1\ot p_{wx},x_2\ot p_{yz}))\to \sigma(q(x_1),q(x_2),\#(w,z))$$
for all $w,x,y,z\in\{a,b\}$.

Consider a $\Sigma$-context $C$ and the trees $M(C[p_{aa}])$ and $M(C[p_{ba}])$. 
Let $u$ be the node of $C$ with $C/u=\bot$. It is easy to see that 
the difference nodes of $M(C[p_{aa}])$ and $M(C[p_{ba}])$ are the node $u$ and 
all nodes $v\cdot(3,1)$ such that $v\neq u$ is a node of $C$ 
and $u$ is the left-most leaf of $C/v$. 
That gives the difference trees $M(C[p_{aa}])/u=\angl{q,p_{aa}}$, $M(C[p_{ba}])/u=\angl{q,p_{ba}}$,
$M(C[p_{aa}])/v\cdot(3,1)=a$ and $M(C[p_{ba}])/v\cdot(3,1)=b$. 
In this way we obtain that $\diff(M)=\{a,b\}\cup\{\angl{q,p_{yz}}\mid y,z\in\{a,b\}\}$. 
Thus, $\maxdiff(M)=0$. 

Clearly, there is a dtop $N$ equivalent to $M$. It has three states $q,q_1,q_2$, axiom $q(x_0)$,
and rules $q(yz)\to yz$, 
$$q(\sigma(x_1,x_2))\to \sigma(q(x_1),q(x_2),\#(q_1(x_1),q_2(x_2))),$$
$q_i(\sigma(x_1,x_2))\to q_i(x_i)$ for $i=1,2$, $\;q_1(yz)\to y$, and $q_2(yz)\to z$ for $y,z\in\{a,b\}$. 
\qed
\end{example}

\begin{example}\label{ex:Mcounter}
Let $\Sigma=\{\sigma^{(2)},a^{(0)}\}$ and $\Delta=\{e^{(0)},o^{(0)}\}$,
and consider the following total dtla $M$ that translates 
every tree $s\in\T_\Sigma$ into $e$ if $\size(s)$ is even and into $o$ if it is odd. 
Its look-ahead automaton has two states $p_e$ and $p_o$ with $\delta(a)=p_o$, 
$\delta(\sigma,p_e,p_e)=\delta(\sigma,p_o,p_o)=p_e$, and 
$\delta(\sigma,p_e,p_o)=\delta(\sigma,p_o,p_e)=p_o$.
Its set of states is empty, and its axioms are $A(p_e)=e$ and $A(p_o)=o$. 

For every $\Sigma$-context $C$, $\{M(C[p_e]),M(C[p_o])\}=\{e,o\}$. 
Hence $\diff(M)=\{e,o\}$ and $\maxdiff(M)=0$. 
Although $\diff(M)$ is finite, there is obviously no dtop equivalent to $M$. 
\qed
\end{example}

\section{Normal Forms}\label{sec:normform}

In this section we prove normal forms for total dtlas. 
For each of these normal forms we consider its effect on $\maxdiff(M)$. 
We start with a simple normal form in which each axiom consists of one state, 
more precisely, is in $Q(\{x_0\})$. 

A dtla $M$ is \emph{initialized} if for every $p\in P$ there is a state $q_{0,p}$
such that $A(p)=q_{0,p}(x_0)$.
The states $q_{0,p}$ are called initial states; they are not necessarily distinct. 
Note that for an initialized dtla, $M(s)=q_M(s)$ where $q=q_{0,p}$ and $p=\delta(s)$, 
for every $s\in\T_\Sigma$. 
The dtlas and dtops in Examples~\ref{ex:Mthree} and~\ref{ex:Mleaves} are initialized. 
The dtlas in Examples~\ref{ex:Mex} and~\ref{ex:Mcounter} are not. 

Recall that $\maxrhs(M)$ is the maximal height 
of the axioms and the right-hand sides of the rules of $M$.

\begin{lemma}\label{lm:make_initialized}
For every total dtla $M$ an equivalent initialized dtla $M'$ can be constructed,
with the same look-ahead automaton as $M$, such that \\
$|Q_{M'}|=|Q_M|+1$, 
$\;\maxrhs(M')\leq 2\cdot\maxrhs(M)$, and 
$$\maxdiff(M')\leq\maxdiff(M)\leq\Max\{\maxdiff(M'),\maxrhs(M)\}.$$
If $M$ is ultralinear or b-erasing, then so is $M'$.
\end{lemma}

\begin{proof}
To construct $M'$ from $M$, we introduce a new state $q_0$. 
For every $a\in\Sigma^{(k)}$ and $p_1,\dots,p_k\in P$
we add the rule 
$$q_0(a(x_1\ot p_1,\dots,x_k\ot p_k))\to 
A(p)[q(x_0)\leftarrow\rhs(q,a,p_1,\dots,p_k)\mid q\in Q],$$
where $p=\delta(a,p_1,\dots,p_k)$. 
The right-hand side of this rule is defined: since every look-ahead state is reachable, 
there exist trees $s_i$ with $\delta(s_i)=p_i$, and so $\delta(a(s_1,\dots,s_k))=p$; 
since $M$ is total, $M(a(s_1,\dots,s_k))$ is defined and hence  
$\rhs(q,a,p_1,\dots,p_k)$ is defined for every $q$ that occurs in $A(p)$.
After adding the above rules, we change $A(p)$ into $q_0(x_0)$ for every $p\in P$.

Then $M'$ is initialized, with $q_{0,p}=q_0$ for all $p\in P$.
It should be clear from Lemma~\ref{lm:elem}(3/4) that $M'$ is equivalent to $M$. 
It should also be clear that for every $C\in\C_\Sigma$ and $p\in P$,
if $C\neq\bot$ then $M'(C[p])=M(C[p])$. Moreover, for $C=\bot$ we have 
$M'(p)= \angl{q_0,p}$ and $M(p)=A(p)[q(x_0)\leftarrow\angl{q,p}\mid q\in Q]$. 
Since $\height(M'(p)/\varepsilon)=0$ and $\height(M(p)/v)\leq \height(A(p))\leq \maxrhs(M)$
for every $v\in V(M(p))$,
this implies the required inequalities for $\maxdiff(M)$ and $\maxdiff(M')$. 

If $M$ is ultralinear with mapping $\mu$, then we can assume that 
$\mu(q)>0$ for all $q$, and then $M'$ is ultralinear
by extending $\mu$ with $\mu(q_0)=0$.
If $M$ is b-erasing then so is $M'$,
because a cycle in $E_{M'}$ does not contain $q_0$ 
and hence is also a cycle in $E_M$.
\qed
\end{proof}

Note that it follows from the inequalities for $\maxdiff(M)$ and $\maxdiff(M')$ that 
if $h(M')$ is a difference bound for $M'$, 
then $\Max\{h(M'),\maxrhs(M)\}$ is a difference bound for $M$. 
In fact, if $\diff(M)$ is finite, then $\diff(M')$ is finite because $\maxdiff(M')\leq\maxdiff(M)$,
hence $\maxdiff(M')\leq h(M')$, from which it follows that  
$\maxdiff(M)\leq\Max\{\maxdiff(M'),\maxrhs(M)\}\leq\Max\{h(M'),\maxrhs(M)\}$.

We continue with a basic and technically convenient normal form 
in which every state of the dtla only translates 
input trees that have the same look-ahead state; moreover, 
the rules satisfy a generalized completeness condition. 
It is closely related to the uniform i-transducer 
in~\cite{DBLP:journals/jcss/EngelfrietMS09}. 

A dtla $M$ is \emph{look-ahead uniform} (for short, \emph{la-uniform}) 
if there is a mapping $\rho: Q\to P$ (called \emph{la-map})
satisfying the following conditions:
\begin{enumerate}
\item[(1)] For every $p\in P$ and $q\in Q$,
if $q(x_0)$ occurs in $A(p)$, then $\rho(q)=p$.
\item[(2)] For every rule $q(a(x_1\ot p_1,\ldots,x_k\ot p_k))\to \zeta$ in $R$,
\begin{enumerate}
\item[(a)] $\rho(q)=\delta(a,p_1,\ldots,p_k)$ and 
\item[(b)] for every $\bar{q}\in Q$ and $i\in[k]$, 
if $\bar{q}(x_i)$ occurs in $\zeta$, then $\rho(\bar{q})=p_i$.
\end{enumerate}
\item[(3)] For every $q\in Q$, $a\in \Sigma^{(k)}$, and $p_1,\dots,p_k\in P$ 
   such that $\delta(a,p_1,\dots,p_k) = \rho(q)$, \\
   there is a rule
   $q(a(x_1\ot p_1,\dots,x_k\ot p_k))\to \zeta$ in $R$.
\end{enumerate}
Clearly, the dtla $M$ of Example~\ref{ex:Mex} is la-uniform with $\rho(q)=p_b$, and similarly, 
the one of Example~\ref{ex:Mthree} is la-uniform with $\rho(q_i^y)=p_y$ for $y\in\{a,b\}$. 
Note that a dtop is la-uniform if and only if it is complete 
(if and only if it is total, by Lemma~\ref{lm:reach}). 
In general, an la-uniform dtla is not complete; 
in fact, it is easy to see that every complete la-uniform dtla is a dtop. 

We will need the following straightforward properties of an la-uniform dtla.

\begin{lemma}\label{lm:prop_la-uniform}
Let $M$ be an la-uniform dtla with la-map $\rho$. 
\begin{enumerate}
\item[(1)] $\dom(\sem{q}_M)=\sem{\rho(q)}_M$ for every $q\in Q$.
\item[(2)] $M$ is total.
\item[(3)] $M^\circ$ is la-uniform with the same la-map $\rho$ as $M$.
\item[(4)] Let $\xi$ be a reachable sentential form for $s\in\T_\Sigma$. \\
For all $v\in V(\xi)$, $q\in Q$, and $u\in V(s)$, 
if $\xi/v=q(u)$ then $\rho(q)=\delta(s/u)$.
\end{enumerate}
\end{lemma}
\begin{proof}
(1) We prove by structural induction on $s\in\T_\Sigma$ that 
$q_M(s)$ is defined if and only if $\delta(s)=\rho(q)$. 
Let $s=a(s_1,\dots,s_k)$ and $\delta(s_i)=p_i$ for $i\in[k]$.
Then $\delta(s)=\delta(a,p_1,\dots,p_k)$. 
Thus, by conditions~(2)(a) and~(3) above, 
$\delta(s)=\rho(q)$ if and only if 
there is a rule $q(a(x_1\ot p_1,\dots,x_k\ot p_k))\to \zeta$ in $R$.
For such a rule, by condition (2)(b) above, if $\bar{q}(x_i)$ occurs in $\zeta$,
then $\rho(\bar{q})=p_i$ and hence, by induction, $\bar{q}_M(s_i)$ is defined. 
It now follows from Lemma~\ref{lm:elem}(4) that 
$\delta(s)=\rho(q)$ if and only if $q_M(s)$ is defined.

(2) This is immediate from (1) and Lemma~\ref{lm:elem}(3), by condition~(1) above. 

(3) By (1), the set $R^\circ$ of rules of $M^\circ$ is obtained from $R$ by adding 
all the rules $q(p)\to \angl{q,p}$ such that $\rho(q)=p$. Hence $\rho$ also satisfies 
conditions~(2) and~(3) above for $M^\circ$ (and condition~(1) above, because $M^\circ$
has the same axioms as $M$).

(4) The easy proof is by induction on the length of the computation 
$A(\delta(s))[x_0\leftarrow\varepsilon] \Rightarrow_s^* \xi$, 
using conditions~(1) and~(2) above.
\qed
\end{proof}

Note that by (3) of this lemma, for every $C\in\C_\Sigma$ and $p\in P$,
the tree $M(C[p])$ is in $\T_\Delta(Q_p\times\{p\})$ where $Q_p=\{q\in Q\mid \rho(q)=p\}$.

We now prove that la-uniformity is a normal form for total dtlas.

\begin{lemma}\label{lm:make_uniform}
For every total dtla $M$ an equivalent la-uniform dtla $M'$ can be constructed,
with the same look-ahead automaton as $M$, 
such that $|Q_{M'}|=|Q_M|\cdot|P_M|$, $\maxrhs(M')=\maxrhs(M)$, and $\maxdiff(M')=\maxdiff(M)$. 
If $M$ is initialized, ultralinear or b-erasing, then so is $M'$.
\end{lemma}

\begin{proof}
We observe that it may be assumed that $M$ is complete: 
if $\rhs(q,a,p_1,\dots,p_k)$ is undefined, then we add the (dummy) rule 
$q(a(x_1\ot p_1,\dots,x_k\ot p_k))\to d$ where $d$ is any element of $\Delta^{(0)}$. 

We construct $M'$ as follows.
The state set of $M'$ is $Q_{M'}=Q\times P$.
Every axiom $A(p)$ of $M$ is changed into 
$A(p)[q(x_0)\leftarrow\angl{q,p}(x_0)\mid q\in Q]$, and  
every rule 
$$q(a(x_1\ot p_1,\dots,x_k\ot p_k))\to\zeta$$
is changed into the rule
$$\angl{q,p}(a(x_1\ot p_1,\dots,x_k\ot p_k))\to \zeta
[\bar{q}(x_i)\leftarrow \angl{\bar{q},p_i}(x_i)\mid\bar{q}\in Q,\,i\in[k]]$$
where $p=\delta(a,p_1,\dots,p_k)$.

It should be clear that 
$M'$ satisfies conditions~(1) and~(2) of the definition of la-uniformity with la-map $\rho$ 
such that $\rho(\angl{q,p})=p$; since $M$ is complete, condition~(3) is also satisfied.
Hence $M'$ is la-uniform. 

Since $M$ and $M'$ are total, so are $M^\circ$ and $(M')^\circ$
(see the proof of Lemma~\ref{lm:MM}). 
Obviously, for every computation of $(M')^\circ$ 
on an input tree $\bar{s}\in \T_\Sigma(P)$, 
one obtains a computation of $M^\circ$ on that input tree 
by changing every $\angl{q,p}(u)$ that occurs in a sentential form 
into $q(u)$, and every $\angl{\angl{q,p},p}$ into $\angl{q,p}$. 
Hence $M'(\bar{s})=M(\bar{s})[\angl{q,p}\leftarrow \angl{\angl{q,p},p}\mid q\in Q, p\in P]$. 
This implies that $M'$ is equivalent to $M$. 
It also implies that $\maxdiff(M')=\maxdiff(M)$, as can easily be verified.

Obviously, if $M$ is ultralinear
with mapping $\mu_M: Q\to\Nat$,
then so is $M'$ with the mapping $\mu$ such that 
$\mu(\angl{q,p})=\mu_M(q)$. 
Moreover, if there is an edge from $\angl{q,p}$ to $\angl{q',p_j}$ in $E_{M'}$,
then there is an edge from $q$ to $q'$ in $E_M$. 
Hence, if $M$ is b-erasing, then so is $M'$. 
\qed
\end{proof}

Note that since $\maxdiff(M')=\maxdiff(M)$, 
the la-uniform dtla $M'$ has the same difference bounds as $M$. 

\begin{example}\label{ex:la-uniform}
The dtla $M$ of Example~\ref{ex:Mleaves} is not la-uniform. We change it into 
an la-uniform dtla by the construction in the proof of Lemma~\ref{lm:make_uniform} 
(but keep calling it $M$). 
Then it has set of states $Q=\{q_{yz}\mid y,z\in\{a,b\}\}$ 
where $q_{yz}$ abbreviates $\angl{q,p_{yz}}$, so $\rho(q_{yz})=p_{yz}$. 
Its axioms are $A(p_{yz})=q_{yz}(x_0)$, and its rules are $q_{yz}(yz)\to yz$ and 
$$q_{wz}(\sigma(x_1\ot p_{wx},x_2\ot p_{yz}))\to \sigma(q_{wx}(x_1),q_{yz}(x_2),\#(w,z))$$
for all $w,x,y,z\in\{a,b\}$. 
\qed
\end{example}

From now on we mainly consider la-uniform dtlas.
For an la-uniform dtla $M$, its la-map will be denoted $\rho$ (or $\rho_M$ when necessary).

\smallskip
Finally we generalize the normal form for dtops in~\cite{DBLP:journals/jcss/EngelfrietMS09} 
to total dtlas. 
For this normal form it is essential that dtlas need not be initialized, 
i.e., that arbitrary axioms are allowed. 

A dtla $M$ is \emph{earliest} if it is la-uniform and, for every state $q$ of $M$, the set 
$$\rlabs_M(q):=\{\lab(q_M(s),\varepsilon)\mid s\in\dom(\sem{q}_M)\}\subseteq\Delta$$ 
is not a singleton.
This is equivalent with requiring that 
$\sqcap\{q_M(s)\mid s\in\dom(\sem{q}_M)\}=\bot$, cf. the definition of earliest 
in~\cite{DBLP:journals/jcss/EngelfrietMS09}.
In other words, $M$ is \emph{not} earliest if it has a state $q$ 
for which the roots of all output trees $q_M(s)$ have the same label; 
intuitively, the node with that label could be produced earlier in the computation of $M$.  

A dtla $M$ is \emph{canonical} if it is earliest and 
$\sem{q}_M\neq\sem{q'}_M$ for all distinct states $q,q'$ of $M$.
Since it is required that $M$ is la-uniform, the earliest and canonical properties are
appropriately relativized with respect to each look-ahead state, 
see Lemma~\ref{lm:prop_la-uniform}(1). 

It is easy to see that the dtla $M$ of Example~\ref{ex:Mthree} is canonical.  
It is earliest because $\rlabs_M(q_i^a)=\{\sigma,\tau,a\}$ and $\rlabs_M(q_i^b)=\{\sigma,\tau,b\}$. 
For an input tree $w(a)$, 
$\sem{q_i^a}_M$ outputs the first $i$ symbols of $w$ and the leaf $a$ if $|w|\geq i$, 
and it outputs $w(a)$ if $|w|<i$. For an input tree $w(b)$, $\sem{q_i^a}_M$ is undefined. 
And the analogous statement holds for $q_i^b$. 

Similarly, the dtla $M$ of Example~\ref{ex:la-uniform} 
(which is the la-uniform version of the dtla of Example~\ref{ex:Mleaves}) is canonical: 
for all $y,z\in\{a,b\}$, $\rlabs_M(q_{yz})=\{yz,\sigma\}$ and 
$\sem{q_{yz}}_M$ is the restriction of $\sem{M}$ to $\sem{p_{yz}}_M$. 

The term `canonical' suggests that any two equivalent canonical dtlas $M_1$ and $M_2$ are the same, 
modulo a renaming of their states and look-ahead states. That is indeed true for dtops, 
as shown in~\cite[Theorem~15]{DBLP:journals/jcss/EngelfrietMS09}, 
but it does \emph{not} hold for arbitrary dtlas: for instance,  
the dtla $M$ of Example~\ref{ex:la-uniform} and the dtop $N$ of Example~\ref{ex:Mleaves}
are both canonical, and they are equivalent but not the same. It is, however, true 
if $M_1$ and $M_2$ have the same look-ahead automaton 
(by a proof similar to the one of~\cite[Theorem~15]{DBLP:journals/jcss/EngelfrietMS09}),
but that fact will not be needed in what follows. 

For an la-uniform dtla $M$ the sets $\rlabs_M(q)$ can be computed in a standard way. 
In fact, consider the 
directed graph with set of nodes $Q\cup\Delta$ and with the following edges: 
for every rule $q(a(x_1\ot p_1,\dots,x_k\ot p_k))\to\zeta$ of $M$, 
if $\lab(\zeta,\varepsilon)=d\in\Delta$ then there is an edge $q\to d$, and 
if $\zeta=q'(x_j)$ then there is an edge $q\to q'$. Note that the subgraph induced
by $Q$ is the graph $E_M$, as in the definition of a b-erasing dtla in Section~\ref{sec:dtop}.
It is straightforward to show that $\rlabs_M(q)=\{d\in\Delta\mid q\to^* d\}$, as follows: 

\smallskip
$(\subseteq)$ Structural induction on $s=a(s_1,\dots,s_k)$, such that $q_M(s)$ has root label $d$.
By Lemma~\ref{lm:elem}(4),
$q_M(s)=\zeta[\bar{q}(x_i)\leftarrow \bar{q}_M(s_i)\mid \bar{q}\in Q,\,i\in[k]]$ where
$\zeta=\rhs(q,s,\varepsilon)$.
If $\zeta$ has root label $d\in\Delta$, then there is an edge $q\to d$.
If $\zeta=q'(x_j)$, then $q_M(s)=q'_M(s_j)$ and so $q\to q'\to^*d$ by induction. 

$(\supseteq)$ Induction on the length of $q\to^* d$. 
If $q\to d$ then 
$q_M(a(s_1,\dots,s_k))=\zeta[\bar{q}(x_i)\leftarrow \bar{q}_M(s_i)\mid \bar{q}\in Q,\,i\in[k]]$ 
by Lemma~\ref{lm:elem}(4), for any $s_i\in\sem{p_i}$, and so $q_M(a(s_1,\dots,s_k))$ 
has root label $d$. 
We have used that $M$ is la-uniform: if $\bar{q}(x_i)$ occurs in $\zeta$, 
then $\rho(\bar{q})=p_i$ and hence $\bar{q}_M(s_i)$ is defined by Lemma~\ref{lm:prop_la-uniform}(1). 
If $q\to q'\to^*d$ then $q_M(a(s_1,\dots,s_k))=q'_M(s_j)$ has root label $d$, 
because by induction there exists $s_j$ such that $q'_M(s_j)$ has root label $d$.

\smallskip
We now prove that canonicalness is a normal form for total dtlas.
For an la-uniform dtla $M$, let $\fix(M)$ be a fixed subset of $\T_\Sigma$ 
such that for every $p\in P$ there is a unique $s\in\fix(M)$ with $\delta(s)=p$. 
Thus, $\fix(M)$ is a set of representatives of the equivalence classes $\sem{p}$, $p\in P$. 
Since every $\sem{p}$ is a regular tree language, a particular $\fix(M)$
can be computed from $M$. 
For every $p\in P$, let $s_p$ be the unique tree in $\fix(M)$ with $\delta(s_p)=p$.
We define 
$$\sumfix(M)=\sum_{q\in Q}\size(q_M(s_{\rho(q)})).$$
Note that $\sumfix(M)$ is in $\Nat$ and can be computed from $M$.

\begin{theorem}\label{th:uniform_earliest}
For every la-uniform dtla $M$ an equivalent canonical dtla $\can(M)$ can be constructed,
with the same look-ahead automaton as $M$, 
such that 
$$\maxdiff(M)-\sumfix(M)\leq\maxdiff(\can(M))\leq\maxdiff(M)+\sumfix(M).$$
\end{theorem}

\begin{proof}
We first prove the statement of this theorem for the case where $M$ is earliest. 
Since the equivalence of two dtlas is decidable (see~\cite{DBLP:journals/actaC/Esik81} 
and~\cite[Corollary~19]{DBLP:journals/jcss/EngelfrietMS09}), 
it is decidable for two states $q,q'$ of $M$ whether or not $\sem{q}_M=\sem{q'}_M$. 
If this holds, then $q'$ can be replaced by $q$ in every axiom
and every right-hand side of a rule, thus making $q'$
unreachable and hence superfluous. 
Since in $M(C[p])$ every $\angl{q',p}$ is replaced by $\angl{q,p}$, $\maxdiff(M)$ does not change. 
Thus, repeating this procedure
one obtains a canonical dtla $\can(M)$ equivalent to $M$, with $\maxdiff(\can(M))=\maxdiff(M)$. 

For the interested reader we observe that for an earliest dtla $M$ 
the equivalence relation $\equiv$ on $Q$ defined by  
$q\equiv q'$ if and only if $\sem{q}_M=\sem{q'}_M$, 
can in fact easily be computed by fixpoint iteration,
because it is the largest equivalence relation on $Q$
such that if $q\equiv q'$ then 
(a) $\rho(q)=\rho(q')$ and 
(b) if $\rhs(q,a,p_1,\dots,p_k)= t[q_1(x_{i_1}),\dots,q_r(x_{i_r})]$ 
where $t\in\mathcal{P}_\Sigma$ and $r=|V_\bot(t)|$,
then $\rhs(q',a,p_1,\dots,p_k)= t[q'_1(x_{i_1}),\dots,q'_r(x_{i_r})]$
with $q_j\equiv q'_j$ for every $j\in[r]$. 
The straightforward proof of this is left to the reader, 
cf. the proof of~\cite[Theorem 13]{DBLP:journals/jcss/EngelfrietMS09}.
Thus, the full dtla equivalence test of 
\cite{DBLP:journals/actaC/Esik81,DBLP:journals/jcss/EngelfrietMS09}
is not needed.

It remains to be proved that every la-uniform dtla $M$ can be 
transformed into an equivalent earliest dtla~$M'$, 
with the same look-ahead automaton, 
such that the distance between $\maxdiff(M')$ and $\maxdiff(M)$ is at most $\sumfix(M)$.
If $M$ is not earliest, then we obtain $M'$ by repeatedly applying the 
following transformation step.

\emph{Transformation}. We transform $M$ into a dtla $N$ with the same look-ahead automaton.
Let $Q_1$ be the (nonempty) set of states $q\in Q$ such that $\rlabs_M(q)$ is a singleton,
and for every $q\in Q_1$ let $\rlabs_M(q)=\{d_q\}$ and $m_q=\rk(d_q)$. The set of states of $N$ is 
$$Q_N:=(Q-Q_1)\cup \{\angl{q,i}\mid q\in Q_1, i\in[m_q]\}.$$ 
When $M$ arrives in state $q\in Q_1$ at a node $u$ of an input tree $s$, 
$N$ will first output the symbol $d_q$ 
and then arrive at node $u$ in the states $\angl{q,1},\dots,\angl{q,m_q}$,  
to compute the direct subtrees of the tree $q_M(s/u)$,
where $\angl{q,i}$ computes the $i$th direct subtree $q_M(s/u)/i$. 
So, to describe $N$, we define for every tree $\zeta\in\T_\Delta(Q(\Omega))$ 
where $\Omega$ is any set of symbols of rank~0, the tree 
$\zeta\Phi_\Omega=\zeta[q(\omega)\leftarrow d_q(\angl{q,1}(\omega),\dots,\angl{q,m_q}(\omega))
\mid q\in Q_1, \omega\in\Omega]$.
For every $p\in P$, the $p$-axiom of $N$ is $A(p)\Phi_{\{x_0\}}$. 
Every rule $q(a(x_1\ot p_1,\dots,x_k\ot p_k))\to \zeta$ is changed into the rule 
$q(a(x_1\ot p_1,\dots,x_k\ot p_k))\to \zeta\Phi_{X_k}$ if $q\in Q-Q_1$, and into the $m_q$ rules 
$\angl{q,i}(a(x_1\ot p_1,\dots,x_k\ot p_k))\to \zeta\Phi_{X_k}/i$ (with $i\in[m_q]$) if $q\in Q_1$. 
Note that in the latter case the root of $\zeta\Phi_{X_k}$ has label $d_q$ 
and so its $i$th direct subtree is well defined. That is clear if $\lab(\zeta,\varepsilon)=d_q$.
If $\zeta=q'(x_j)$, then $q'\in Q_1$ and $d_{q'}=d_q$; 
so $\zeta\Phi_{X_k}=d_q(\angl{q',1}(x_j),\dots,\angl{q',m_q}(x_j))$ and
one obtains the rules $\angl{q,i}(a(x_1\ot p_1,\dots,x_k\ot p_k))\to\angl{q',i}(x_j)$ for $i\in[m_q]$.

If $M$ has la-map $\rho$, then $N$ is la-uniform with la-map $\rho_N$ 
such that $\rho_N(q)=\rho(q)$ for $q\in Q-Q_1$ and 
$\rho_N(\angl{q,i})=\rho(q)$ for $q\in Q_1$ and $i\in[m_q]$.

It should be clear intuitively that $N$ is equivalent to $M$. 
Formally it can easily be shown for every $s\in\T_\Sigma$ 
and every reachable sentential form $\xi$ of $M$ for $s$, 
that $\xi\Phi_{V(s)}$ is a reachable sentential form of $N$ for $s$
(where each computation step of $M$ is simulated by one or $m_q$ 
computation steps of $N$), and hence $N(s)=M(s)$. 
We will compute $\maxdiff(N)$ below; to do that we need to extend the previous statement 
to trees $\bar{s}\in\T_\Sigma(P)$.
Let $\Psi$ be the substitution 
$[\angl{q,p}\leftarrow d_q(\angl{\angl{q,1},p},\dots,\angl{\angl{q,m_q},p})\mid q\in Q_1,p\in P]$.
Then, for every reachable sentential form $\xi$ of $M^\circ$ for $\bar{s}$, 
the tree $\xi\Phi_{V(s)}\Psi$ is a reachable sentential form of $N^\circ$ for $\bar{s}$,
and hence $N(\bar{s})=M(\bar{s})\Psi$.

\emph{Repetition}. 
Using Lemma~\ref{lm:elem}(3/4), it can easily be shown 
for every $q\in Q$ and $s\in\sem{\rho(q)}$, 
that $q_N(s)=q_M(s)$ if $q\notin Q_1$, and that 
$\angl{q,i}_N(s)=q_M(s)/i$ for every $i\in[m_q]$ if $q\in Q_1$. 
From this (and assuming that $\fix(N)=\fix(M)$), 
it should be clear that $\sumfix(N)<\sumfix(M)$, because, for $q\in Q_1$ and $s\in\sem{\rho(q)}$, 
$$\sum_{i\in[m_q]}\size(\angl{q,i}_N(s))=\sum_{i\in[m_q]}\size(q_M(s)/i)=\size(q_M(s))-1.$$ 
Hence, the repetition of the above transformation stops after at most $\sumfix(M)$ steps, 
with an earliest dtla $M'$ equivalent to $M$. 

\emph{Difference trees}. 
It now remains to prove that the distance between $\maxdiff(N)$ and $\maxdiff(M)$ is at most~1,
i.e., that $\maxdiff(N)\leq \maxdiff(M)+1$ and $\maxdiff(M)\leq \maxdiff(N)+1$.
Consider $C\in\C_\Sigma$ and $p,p'\in P$ with $p\neq p'$.
Recall from above that $N(C[p])=M(C[p])\Psi=
M(C[p])[\angl{q,p}\leftarrow d_q(\angl{\angl{q,1},p},\dots,\angl{\angl{q,m_q},p})\mid q\in Q_1]$
and similarly for $p'$.
We observe that if $v$ is a node of $M(C[p])$, then each proper ancestor of $v$ has the same label in 
$M(C[p])$ and $N(C[p])$, and similarly for $p'$.

Let $v$ be a difference node of $N(C[p])$ and $N(C[p'])$ that is not a leaf of $N(C[p])$.
Then $v\in V(M(C[p]))$. If also $v\in V(M(C[p']))$, then Lemma~\ref{lm:diff_char} implies that
$v$ is also a difference node of $M(C[p])$ and $M(C[p'])$
(in fact, by the above observation every proper ancestor of $v$ 
has the same label in $M(C[p])$ and $M(C[p'])$; 
if $v$ would have the same label in $M(C[p])$ and $M(C[p'])$,
then that label would be in $\Delta$ because $p\neq p'$, and hence 
$v$ would have the same label in $N(C[p])$ and $N(C[p'])$). 
Consequently, $\height(N(C[p])/v)=\height(M(C[p])\Psi/v)\leq \height(M(C[p])/v)+1\leq\maxdiff(M)+1$.
If $v\notin V(M(C[p']))$, then the parent $\hat{v}$ of $v$ 
is a difference node of $M(C[p])$ and $M(C[p'])$
(in fact, the label of $\hat{v}$ in $M(C[p'])$ is $\angl{q,p'}$ for some $q\in Q_1$, 
and so $\hat{v}$ has different labels in $M(C[p])$ and $M(C[p'])$ because $p\neq p'$). Hence, in this case, 
$\height(N(C[p])/v)\leq \height(N(C[p])/\hat{v})\leq \height(M(C[p])/\hat{v})+1\leq\maxdiff(M)+1$.
This proves that $\maxdiff(N)\leq \maxdiff(M)+1$. 

Now let $v$ be a difference node of $M(C[p])$ and $M(C[p'])$ that is not a leaf of $M(C[p])$;
note that $\lab(N(C[p]),v)=\lab(M(C[p]),v)\in\Delta$.
If $v$ is also a difference node of $N(C[p])$ and $N(C[p'])$, 
then $\height(M(C[p])/v)\leq\height(M(C[p])\Psi/v)=\height(N(C[p])/v)\leq\maxdiff(N)$.
Now assume that $v$ is not a difference node of $N(C[p])$ and $N(C[p'])$.
Then, by Lemma~\ref{lm:diff_char}, the label of $v$ in $M(C[p'])$ is $\angl{q,p'}$ for some $q\in Q_1$
and $\lab(N(C[p]),v)=\lab(N(C[p']),v)=d_q$.
This implies, again by Lemma~\ref{lm:diff_char}, that 
the children of $v$ are difference nodes of $N(C[p])$ and $N(C[p'])$.
Let $vi$ be a child of $v$ for which $\height(M(C[p])/vi)$ is maximal. 
Then we have $\height(M(C[p])/v)=\height(M(C[p])/vi)+1\leq\height(N(C[p])/vi)+1\leq\maxdiff(N)+1$.
This proves that $\maxdiff(M)\leq \maxdiff(N)+1$.
\qed
\end{proof}

Note that it follows from the inequalities for $\maxdiff(\can(M))$ that 
if $h(M)$ is a difference bound for $M$, 
then $h(M)+\sumfix(M)$ is a difference bound for $\can(M)$.
In fact (similar to the argument after Lemma~\ref{lm:make_initialized}), 
if $\diff(\can(M))$ is finite, then $\diff(M)$ is finite  
because $\maxdiff(M)\leq\maxdiff(\can(M))+\sumfix(M)$, 
hence $\maxdiff(M)\leq h(M)$ and hence 
$\maxdiff(\can(M))\leq \maxdiff(M)+\sumfix(M)\leq h(M)+\sumfix(M)$. 

Note also that the transformation in the above proof does not preserve the ultralinear property
(as can be seen in the next example). 

\begin{example}
In this example we denote by $Y$ the nonempty subsets of $\{a,b\}$, i.e., $Y=\{\{a\},\{b\},\{a,b\}\}$. 
Let $\Sigma=\{\sigma^{(2)},a^{(0)},b^{(0)}\}$ and 
$\Delta=\{\sigma_y^{(2)}\mid y\in Y\}\cup\{a^{(0)},b^{(0)}\}$.
We consider an la-uniform dtla $M$ such that $M(a)=a$, $M(b)=b$, and 
$M(\sigma(s_1,s_2))=\sigma_y(M(s_1),M(s_2))$ for $s_1,s_2\in\T_\Sigma$,  
where $y$ is the set of labels of the leaves of $\sigma(s_1,s_2)$.
Its set of look-ahead states is $P=\{p_y\mid y\in Y\}$ and $\delta$ is defined in the obvious way:
$\delta(a)=\{a\}$, $\delta(b)=\{b\}$, and $\delta(\sigma,p_y,p_z)=p_{y\cup z}$ for $y,z\in Y$. 
Its set of states is $Q=\{q_y\mid y\in Y\}$ with $\rho(q_y)=p_y$, 
its axioms are $A(p_y)=q_y(x_0)$ for every $y\in Y$, 
and its set $R$ consists of the rules 
$q_{\{a\}}(a)\to a$, $\;q_{\{b\}}(b)\to b$, and 
$$q_{y\cup z}(\sigma(x_1\ot p_y,x_2\ot p_z))\to \sigma_{y\cup z}(q_y(x_1),q_z(x_2))$$ for $y,z\in Y$.

The dtla $M_1$ that is obtained from $M$ by identifying all its states into one state $q$, 
is equivalent to $M$; it has $\rlabs_{M_1}(q)=\Delta$, but it is not la-uniform. 
However, $M$ is not earliest: in fact, 
$\rlabs_M(q_{\{a\}})=\{\sigma_{\{a\}},a\}$ and $\rlabs_M(q_{\{b\}})=\{\sigma_{\{b\}},b\}$, but
$\rlabs_M(q_{\{a,b\}})=\{\sigma_{\{a,b\}}\}$. Let $N$ be the dtla obtained from $M$ 
by applying the transformation in the proof of Theorem~\ref{th:uniform_earliest} once.
Then $Q_1=\{q_{\{a,b\}}\}$. 
We will write the states $\angl{q_{\{a,b\}},1}$ and $\angl{q_{\{a,b\}},2}$ as
$q_{1\{a,b\}}$ and $q_{2\{a,b\}}$, respectively. 
So, $N$ has states $q_{\{a\}}$, $q_{\{b\}}$, $q_{1\{a,b\}}$, and $q_{2\{a,b\}}$. 
Its axioms are $A_N(p_y)=q_y(x_0)$ for $y=\{a\}$ or $y=\{b\}$ (just as in $M$),
and $A_N(p_y)=\sigma_y(q_{1y}(x_0),q_{2y}(x_0))$ for $y=\{a,b\}$. 
For $y=\{a\}$ or $y=\{b\}$, its set $R_N$ of rules contains the rule
$$q_y(\sigma(x_1\ot p_y,x_2\ot p_y))\to \sigma_y(q_y(x_1),q_y(x_2))$$ 
plus the rules $q_{\{a\}}(a)\to a$ and $\;q_{\{b\}}(b)\to b$ (just as in $M$).
Moreover, for $y=\{a,b\}$, $R_N$ contains the rules 
\[
\begin{array}{rll}
q_{1y}(\sigma(x_1\ot p_y,x_2\ot p_z)) & \to & \sigma_y(q_{1y}(x_1),q_{2y}(x_1)) \\
q_{2y}(\sigma(x_1\ot p_z,x_2\ot p_y)) & \to & \sigma_y(q_{1y}(x_2),q_{2y}(x_2))
\end{array}
\]
for all $z\in Y$, plus the rules 
\[
\begin{array}{rll} 
q_{1y}(\sigma(x_1\ot p_w,x_2\ot p_z)) & \to & q_w(x_1) \\
q_{2y}(\sigma(x_1\ot p_z,x_2\ot p_w)) & \to & q_w(x_2)
\end{array}
\]
for all $w,z\in Y$ with $w\neq y$ and $w\cup z=y$. 
For $N$ we have $\rlabs_N(q_{\{a\}})=\{\sigma_{\{a\}},a\}$ and 
$\rlabs_N(q_{\{b\}})=\{\sigma_{\{b\}},b\}$ as for $M$, and we have 
$\rlabs_N(q_{1\{a,b\}})=\rlabs_N(q_{2\{a,b\}})=\Delta$. Hence $N$ is earliest.
Obviously, $N$ is also canonical, and so $N=\can(M)$.  

Clearly, $N$ is not ultralinear, because condition~(2) of the definition of ultralinearity 
cannot be satisfied for the rule 
$q_{1y}(\sigma(x_1\ot p_y,x_2\ot p_z)) \to \sigma_y(q_{1y}(x_1),q_{2y}(x_1))$.
Since $M$ is linear, this shows that ultralinearity is not preserved. 
After the definition of canonicalness we observed (without proof) that the dtla $N=\can(M)$ is 
unique, modulo a renaming of states, and so the nonpreservation of ultralinearity
in the proof of Theorem~\ref{th:uniform_earliest} is, in fact, unavoidable. 
\qed
\end{example}

\section{Difference Tuples and the Algorithm}
\label{sec:diff} 

In this section we introduce the notion of difference tuple, 
generalizing the notion of difference trees by considering 
all look-ahead states of a dtla simultaneously. Based on that notion, 
we present an algorithm which, for a given total dtla $M$ 
and a difference bound for $M$, decides whether $M$ is equivalent to a dtop, 
and if so, constructs such a dtop. By the results of the previous section, 
we may assume $M$ to be canonical. 

\subsection{Difference Tuples}
\label{sec:difftup}

Let $M$ be a total dtpla and let $P=\{\hat{p}_1,\dots,\hat{p}_n\}$, 
where the order of the look-ahead states is fixed as indicated.
Recall that a dtpla is a proper dtla, i.e., a dtla that is not a dtop, hence $n\geq 2$.
For a given context $C$ consider the trees $M(C[\hat{p}_1]),\dots,M(C[\hat{p}_n])$.
Intuitively, the largest common prefix of all these trees
does \emph{not} depend on the look-ahead.
In contrast, the subtrees of the above trees that are not
part of the largest common prefix, \emph{do} depend on the
look-ahead information.

For trees $t_1,\dots,t_n\in\T_\Delta(Q\times P)$ we define 
\[
\diftup(t_1,\dots,t_n):=\{(t_1/v,\dots,t_n/v)\mid v\in V_\bot(\sqcap\{t_1,\dots,t_n\})\},
\]
which is a set of $n$-tuples in $\T_\Delta(Q\times P)^n$.
We will say that $(t_1/v,\dots,t_n/v)$ is the difference tuple of $t_1,\dots,t_n$ at such a node $v$. 
We define the \emph{set of difference tuples} of~$M$ as
\[
\diftup(M):=\bigcup_{C\in \C_\Sigma}\diftup(M(C[\hat{p}_1]),\dots,M(C[\hat{p}_n])).
\]
For a $\Sigma$-context $C$, 
we define $\pref(M,C)\in \mathcal{P}_\Delta$ as 
\[
\pref(M,C):=\sqcap\{M(C[p])\mid p\in P\}=\sqcap\{M(C[\hat{p}_1]),\dots,M(C[\hat{p}_n])\}.
\]
Note that $\pref(M,C)$ is a $\Delta$-pattern because a node with label $\angl{q,\hat{p}_i}$
in $M(C[\hat{p}_i])$ cannot have the same label in $M(C[\hat{p}_j])$ for $i\neq j$.
Note also that 
\[
\diftup(M)=\{(M(C[\hat{p}_1])/v,\dots,M(C[\hat{p}_n])/v)\mid
C\in \C_\Sigma,\,v\in V_\bot(\pref(M,C))\}.
\]
We define the number $\maxdiftup(M)\in\Nat\cup\{\infty\}$ to be the least upper bound of the 
heights of the components of all difference tuples of $M$, i.e.,
\[
\maxdiftup(M)=\lub\{\height(\bar{t})\mid \bar{t}\in \diftup(M)\}.
\]

Difference tuples are introduced for the following reason, cf. Lemma~\ref{lm:semantics}.
We wish to decide whether $M$ is equivalent to a dtop.
If there exists a dtop $N$ that is equivalent to $M$, 
then we expect intuitively for any $s\in\T_\Sigma$ and $C\in\C_\Sigma$, that 
$N(C[s])=t[q_{1N}(s),\dots,q_{rN}(s)]$ 
where $t=\pref(M,C)=\sqcap\{M(C[\hat{p}_1]),\dots,M(C[\hat{p}_n])\}$
and $r=|V_\bot(t)|$. 
Thus, since $N$ does not know the look-ahead state $\delta_M(s)$ of $s$, it translates $C$ into the largest common prefix of the output trees $M(C[\hat{p}_1]),\dots,M(C[\hat{p}_n])$. 
Moreover, if the $i$th occurrence of $\bot$ is at node $v_i$ of $t$ for $i\in[r]$, 
then we expect the difference tuple $(M(C[\hat{p}_1])/v_i,\dots,M(C[\hat{p}_n])/v_i)$ 
of these output trees at $v_i$ to be stored in the
state $q_i$ of $N$; in this way $N$ is prepared to continue its simulation of $M$ on the subtree $s$.  
This will be proved in Lemma~\ref{lm:dtladtopprop2}, 
under the condition that $M$ is canonical and $N$ is earliest. 
If $N$ is also canonical, then its states are in one-to-one correspondence 
with the difference tuples of $M$, as will be proved in Lemma~\ref{lm:Qdiff}.

Before giving some examples, we show that $\maxdiftup(M)$ equals $\maxdiff(M)$, 
defined in Section~\ref{sec:difftree}. 
This implies that $\diftup(M)$ is finite if and only if $\diff(M)$ is finite. 

\begin{lemma}\label{lm:twodiff}
For every total dtpla $M$, $\maxdiff(M)=\maxdiftup(M)$.
\end{lemma}
\begin{proof}
($\leq$) We show that every difference tree is a subtree of a component of a difference tuple.
Consider a difference tree $M(C[p])/v$ with $C\in\C_\Sigma$, $p\in P$, and 
$v$ a difference node of $M(C[p])$ and $M(C[p'])$ where $p'\in P$, 
i.e., $v\in V_\bot(M(C[p]) \sqcap M(C[p']))$.
Since $\pref(M,C) \sqsubseteq M(C[p]) \sqcap M(C[p'])$, 
there is an ancestor $\hat{v}$ of $v$ such that $\hat{v}\in V_\bot(\pref(M,C))$.
Thus, $M(C[p])/\hat{v}$ is a component of a difference tuple, and 
$M(C[p])/v$ is one of its subtrees.

($\geq$) We show that every component of a difference tuple is a difference tree. 
Consider $M(C[p])/v$ with $C\in\C_\Sigma$, $p\in P$, and $v\in V_\bot(\pref(M,C))$.
By Lemma~\ref{lm:pat}, each proper ancestor of $v$ has the same label in all $M(C[\bar{p}])$,
$\bar{p}\in P$, but $v$ does not have the same label in all $M(C[\bar{p}])$. 
Thus, there exists $p'\in P$ such that $v$ has different labels in 
$M(C[p])$ and $M(C[p'])$. 
Then $v$ is a difference node of $M(C[p])$ and $M(C[p'])$ by Lemma~\ref{lm:diff_char}. 
\qed
\end{proof}

We now give examples of $\diftup(M)$ for several total dtplas $M$. 
In the remainder of this subsection, 
we will use the dtla of Example~\ref{ex:Mthree} as a running example.

\begin{example}\label{ex:diftup}
For the dtla $M$ of Example~\ref{ex:Mex},  
with the order $P=\{p_a,p_b\}$, we obtain that 
$\diftup(M)=\{(a,\sigma^n\angl{q,p_b})\mid n\in\Nat\}$.
 
For the dtla $M$ of Example~\ref{ex:Mthree}, 
also with the order $P=\{p_a,p_b\}$, we obtain that 
$\diftup(M)$ consists of all pairs 
\[
\begin{array}{ll}
(\angl{q^a_{i+50},p_a},\,\angl{q^b_i,p_b}) & \textrm{for } i\in[30], 
\\[2mm]
(z(\angl{q^a_{50-|z|},p_a}),\,b) & \textrm{for } z\in\{\sigma,\tau\}^*  \textrm{ with } |z|<50, \textrm{and} 
\\[2mm]
(z(a),\,b) & \textrm{for } z\in\{\sigma,\tau\}^* \textrm{ with } |z|=50.
\end{array}
\]

For the dtla $M$ of Example~\ref{ex:la-uniform} 
(which is the la-uniform version of the dtla of Example~\ref{ex:Mleaves})
it is not difficult to see that 
$\diff(M)=\{a,b\}\cup\{\angl{q_{yz},p_{yz}}\mid y,z\in\{a,b\}\}$,
and that the set 
$\diftup(M)$ consists of the three 4-tuples $(a,a,b,b)$, $\;(a,b,a,b)$, and
$(\angl{q_{aa},p_{aa}},\angl{q_{ab},p_{ab}},\angl{q_{ba},p_{ba}},\angl{q_{bb},p_{bb}})$,
where we have taken the order $P=\{p_{aa},p_{ab},p_{ba},p_{bb}\}$.

For the dtla $M$ of Example~\ref{ex:Mcounter}, $\diftup(M)=\{(e,o),(o,e)\}$.

In the above examples, the components of the difference tuples are exactly the difference trees.
As another example, let $\Sigma=\Delta=\{\sigma^{(1)},a^{(0)},b^{(0)},c^{(0)}\}$ and
consider the dtla $M$ with $P=\{p_a,p_b,p_c\}$, 
$\delta(y)=p_y$ and $\delta(\sigma,p_y)=p_y$ for $y\in\{a,b,c\}$,
$Q=\emptyset$, $A(p_a)=a$, $A(p_b)=\sigma(b)$, and $A(p_c)=\sigma(\sigma(c))$.
Thus, for every $n\in\Nat$, $M$ translates $\sigma^na$ into $a$, $\sigma^nb$ into $\sigma b$, 
and $\sigma^nc$ into $\sigma\sigma c$. 
Since $a\sqcap\sigma b=\bot$, $\;a\sqcap\sigma\sigma c=\bot$, and 
$\sigma b\sqcap\sigma\sigma c=\sigma\bot$, we obtain that 
$\diff(M)=\{a,\sigma b,\sigma\sigma c,b,\sigma c\}$.
Since $a\sqcap\sigma b\sqcap\sigma\sigma c=\bot$, we obtain that 
$\diftup(M)=\{(a,\sigma b,\sigma\sigma c)\}$. 
Thus, $b$ and $\sigma c$ are difference trees that are not components of a difference tuple
(but are subtrees of such components). 
Note that there is a dtop with one state that is equivalent to $M$. 
\qed
\end{example}

In the next lemmas $N$ is a total dtop, equivalent to $M$. 
We assume that the unique look-ahead state of $N$ is $\bot$.
So, $N^\circ$ translates input trees in $\T_\Sigma(\{\bot\})$, 
in particular $\Sigma$-contexts, into output trees in $\T_\Delta(Q_N\times\{\bot\})$;
for a $\Sigma$-context $C$ we of course write $C$ instead of $C[\bot]$.
The unique axiom $A_N(\bot)$ is denoted by $A_N$, 
a rule $q(a(x_1\ot\bot,\dots,x_k\ot\bot))\to\zeta$ is written  
$q(a(x_1,\dots,x_k))\to\zeta$, and $\zeta$ is denoted $\rhs_N(q,a)$.
For a tree $t\in\T_\Delta(Q_N\times \{\bot\})$ 
we define the pattern $t\Phi\in \mathcal{P}_\Delta$
by $t\Phi=t[\angl{q,\bot}\leftarrow\bot\mid q\in Q_N]$;
similarly, for $t\in\T_\Delta(Q_N(X))$ we define 
$t\Phi=t[q(x_i)\leftarrow\bot\mid q\in Q_N,i\in\Nat]$.

Let $M$ be a canonical dtpla and $N$ a dtop such that $\sem{M}=\sem{N}$.
We first show that the translation of an input tree by $M$ is always ahead of
its translation by $N$, in a uniform way. 
An \emph{aheadness mapping} from $N$ to $M$ is 
a function $\varphi: Q_N\times P_M\to \T_\Delta(Q_M\times P_M)$ 
such that for every $C\in\C_\Sigma$ and $p\in P_M$,
\begin{equation}\label{eq:ahead}
M(C[p]) = N(C)[\angl{q,\bot}\leftarrow\varphi(q,p)\mid q\in Q_N].
\end{equation}
Note that $\varphi(q,p)$ must be in 
$\T_\Delta(\{\angl{\bar{q},p}\mid\bar{q}\in Q_M,\rho_M(\bar{q})=p\})$.
Intuitively, $\varphi$ defines the exact amount 
in which $M$ is ahead of $N$, which is independent of $C$. 
This amount is stored in each state $q$ of $N$, for every look-ahead state $p$ of $M$. 

The next lemma provides an obvious equivalent formulation of Equation~(\ref{eq:ahead}),
using the substitution $\Phi$ defined above.  

\begin{lemma}\label{lm:altahead}
For every $C\in\C_\Sigma$ and $p\in P_M$, 
Equation~(\ref{eq:ahead}) is equivalent to the following two conditions:
\begin{enumerate}
\item[(a)] $N(C)\Phi \sqsubseteq M(C[p])$, and 
\item[(b)] for every $v\in V_\bot(N(C)\Phi)$ and $q\in Q_N$, \\
if $N(C)/v=\angl{q,\bot}$, then $\varphi(q,p)=M(C[p])/v$.
\end{enumerate}
\end{lemma}

\begin{example}\label{ex:ahead}
Consider $M$ and $N$ as in Example~\ref{ex:Mthree}.
As observed after Example~\ref{ex:la-uniform}, the dtpla $M$ is canonical. 
It is easy to check that $N$ is also canonical. 
We will show that there is an aheadness mapping $\varphi$ from $N$ to $M$.
To this aim, let $w\in\{\sigma,\tau\}^*$ and consider the context $C=w(\bot)$. 
In Example~\ref{ex:Mthree} we have computed $M(C[p_a])$ and $M(C[p_b])$.
Let us now compute $N(C)$. 
For $|w|<30$ we obtain that $N(C)=w(\angl{q_{30-|w|},\bot})$. 
Since in this case $M(C[p_a])=w(\angl{q^a_{80-|w|},p_a})$ and 
$M(C[p_b])=w(\angl{q^b_{30-|w|},p_b})$, Equation~(\ref{eq:ahead}) requires for $i=30-[w]$, 
and hence for every $i\in[30]$, that 
$$\varphi(q_i,p_a)=\angl{q^a_{i+50},p_a} \textrm{ }\textrm{ and }\textrm{ } \varphi(q_i,p_b)=\angl{q^b_i,p_b}.$$ 
Note that in this case $M$ is not properly ahead of $N$. 
For $w=w_1z$ with $|w_1|=30$ and $|z|<50$, we get that $N(C)=w_1(\angl{q_z,\bot})$. 
Since $M(C[p_a])=w_1z(\angl{q^a_{80-|w_1z|},p_a})$ and $M(C[p_b])=w_1(b)$, 
the mapping $\varphi$ should satisfy 
$$\varphi(q_z,p_a)= z(\angl{q^a_{50-|z|},p_a})\textrm{ }\textrm{ and }\textrm{ } \varphi(q_z,p_b)=b,
\textrm{ }\textrm{ for } \textrm{ }|z|<50.$$
Finally, if $w=w_1zw_2$ with $|w_1|=30$ and $|z|=50$, then $N(C)=w_1(\angl{q_z,\bot})$.
Since $M(C[p_a])=w_1z(a)$ and $M(C[p_b])=w_1(b)$, we get that 
$$\varphi(q_z,p_a)= z(a)\textrm{ }\textrm{ and }\textrm{ } \varphi(q_z,p_b)=b,
\textrm{ }\textrm{ for } \textrm{ }|z|=50.$$
Clearly, the above requirements define $\varphi$ uniquely, 
and hence $\varphi$ is an aheadness mapping from $N$ to $M$ (and, in fact, the unique one).
\qed
\end{example}

\begin{lemma}\label{lm:aheadness-mapping}
Let $M$ be a canonical dtpla and $N$ a dtop such that
$\sem{M}=\sem{N}$. 
Then there is a unique aheadness mapping from $N$ to~$M$.
\end{lemma}
\begin{proof}
We first show that $M$ is ahead of $N$, 
i.e., that all output symbols produced by $N$
on a given input context are also produced by $M$.

\smallskip
{\bf Claim~1.}\quad Let $p\in P_M$ and let $C$ be a $\Sigma$-context. \\
For every $d\in\Delta$, $V_d(N(C))\subseteq V_d(M(C[p]))$. 
Equivalently, $N(C)\Phi\sqsubseteq M(C[p])$.

\smallskip
Proof. We show that every node $v$ of $N(C)$ with label $d\in\Delta$ 
is also a node of $M(C[p]$, with the same label. 
The proof is by induction on the length of $v$, as follows.
Since $v$'s proper ancestors are in $V_\Delta(N(C))$,
the induction hypothesis implies that $v$ is a node of $M(C[p])$.
Consider an arbitrary $s\in \sem{p}_M$. 
By Lemma~\ref{lm:semantics}, $v$~has label $d$ in $N(C[s])$. 
Since $\sem{M}=\sem{N}$, $M(C[s])=N(C[s])$ and so $v$ has label $d$ in $M(C[s])$.
Suppose that $v$ does not have label $d$ in $M(C[p])$. 
Then, again by Lemma~\ref{lm:semantics},
$v$ must have some label $\angl{q,p}$ in $M(C[p])$ such that  
$q_M(s)$ has root label $d$. 
Since this holds for every $s\in\sem{p}_M$, we obtain that 
$\rlabs_M(q)=\{d\}$ contradicting the fact that $M$ is earliest. 
Note that, since $M$ is la-uniform, $\rho_M(q)=p$ by Lemma~\ref{lm:prop_la-uniform}(3)
and hence $\sem{p}_M=\dom(\sem{q}_M)$ by Lemma~\ref{lm:prop_la-uniform}(1). 
This proves the claim.

\smallskip
Next we show that the amount in which $M$ is ahead of $N$, is independent of $C$. 

\smallskip
{\bf Claim~2.}
Let $C_1,C_2$ be $\Sigma$-contexts, $v_1,v_2\in\Nat_+^*$, $q\in Q_N$, and $p\in P_M$. \\
If $N(C_1)/v_1=N(C_2)/v_2=\angl{q,\bot}$, then $M(C_1[p])/v_1=M(C_2[p])/v_2$.

\smallskip
Proof. 
By Claim~1, $v_i$ is a node of $M(C_i[p])$.
Let $t_i\in \T_\Delta(Q_M\times\{p\})$ denote the tree $M(C_i[p])/v_i$.
For every $s\in\sem{p}_M$, 
$N(C_1[s])/v_1=N(C_2[s])/v_2=q_N(s)$ by Lemma~\ref{lm:semantics}, and so 
$M(C_1[s])/v_1=M(C_2[s])/v_2$. Hence, again by Lemma~\ref{lm:semantics},
$t_1\Psi_s=t_2\Psi_s$ for all $s\in \sem{p}_M$, where
$\Psi_s=[\angl{q,p}\leftarrow q_M(s)\mid q\in Q_M]$.
Suppose that $t_1\not=t_2$. Then there is a leaf $v$ of, e.g., $t_1$
with label $\angl{q_1,p}$ such that $v$ is a node of $t_2$ with 
$t_2/v\not= \angl{q_1,p}$. 
If the root label of $t_2/v$ is $d\in\Delta$, 
then $q_{1M}(s)$ has root label $d$ for all $s\in\sem{p}_M$, 
contradicting the fact that $M$ is earliest.
If $t_2/v$ equals $\angl{q_2,p}$ with $q_1\neq q_2$, then 
$q_{1M}(s)=q_{2M}(s)$ for all $s\in\sem{p}_M$.
Since, as observed in the proof of Claim~1, 
$\sem{p}_M$ is the domain of both $\sem{q_1}_M$ and $\sem{q_2}_M$
by Lemma~\ref{lm:prop_la-uniform},
we obtain that $\sem{q_1}_M=\sem{q_2}_M$, 
contradicting the fact that $M$ is canonical.
This proves the claim.

\smallskip
An aheadness mapping from $N$ to $M$ can now be defined as follows. 
Let $q\in Q_N$ and $p\in P_M$. 
By Lemma~\ref{lm:reach}, there is a $\Sigma$-context $C$
such that $N(C)$ has a node $v$ labeled $\angl{q,\bot}$.
By Claim~1, $v$ is a node of $M(C[p])$
and we define $\varphi(q,p)=M(C[p])/v$.
By Claim~2, the definition of $\varphi$ does not depend on $C$ and $v$.
It follows from Claim~1 that if $N(C)=t[\angl{q_1,\bot},\dots,\angl{q_r,\bot}]$ 
with $t\in\mathcal{P}_\Delta$ and $r=|V_\bot(t)|$, then
$M(C[p])=t[M(C[p])/v_1,\dots,M(C[p])/v_r]$ where $v_i$ is the $i$th occurrence of $\bot$ in $t$, 
and hence $M(C[p])=t[\varphi(q_1,p),\dots,\varphi(q_r,p)]$. 
Thus, $\varphi$ satisfies Equation~(\ref{eq:ahead}): the requirement for an aheadness mapping. 
Obviously, if $\varphi'$ is an aheadness mapping from $N$ to $M$, and $C$ is a $\Sigma$-context
such that $N(C)$ has a node $v$ labeled $\angl{q,\bot}$, then $M(C[p])/v=\varphi'(q,p)$
for every $p\in P_M$, by Lemma~\ref{lm:altahead}(b). Thus $\varphi'=\varphi$, which shows 
that $\varphi$ is the unique aheadness mapping from $N$ to $M$. 
\qed
\end{proof}

Let $M$ and $N$ be as in Lemma~\ref{lm:aheadness-mapping}, with aheadness mapping $\varphi$. 
Let $\bar{s}$ be an input tree and $v$ a node of $\bar{s}$ with $\bar{s}/v=s$. 
When $N$ arrives at $v$ in state $q$,
we expect that, when processing $s$, it will first output $\varphi(q,\delta_M(s))$ 
and then produce the output of $M$ on $s$. 
This is formalized in the following lemma. 

\begin{lemma}\label{lm:dtladtopprop1}
Let $M$ be a canonical dtpla and $N$ a dtop such that
$\sem{M}=\sem{N}$. Let $\varphi$ be the aheadness mapping from $N$ to $M$. 
For every $s\in\T_\Sigma$ and $q\in Q_N$, if $\delta_M(s)=p$, then 
$$q_N(s)=\varphi(q,p)[\angl{\bar{q},p}\leftarrow {\bar{q}}_M(s)\mid \bar{q}\in Q_M].$$
\end{lemma}
\begin{proof}
By Lemma~\ref{lm:reach}, 
there are $C,v$ such that $N(C)/v=\angl{q,\bot}$.
It follows from Lemma~\ref{lm:altahead}(b) that $M(C[p])/v=\varphi(q,p)$.
Since $M$ and $N$ are equivalent, $N(C[s])=M(C[s])$.
Applying Lemma~\ref{lm:semantics} twice, we obtain  
$q_N(s)=N(C[s])/v=M(C[s])/v=
(M(C[p])/v)[\angl{\bar{q},p}\leftarrow {\bar{q}}_M(s)\mid \bar{q}\in Q_M]$, 
which proves the equation.
\qed
\end{proof}

\begin{example}
We continue Example~\ref{ex:ahead}. 
Consider a state $q_z$ of $N$ with $|z|<50$, and let $s=w_1w_2(a)$ with $|zw_1|=50$.
Then $q_{zN}(s)=zw_1(a)$.  
Now $\varphi(q_z,p_a)= z(\angl{q^a_{50-|z|},p_a})=z(\angl{q^a_{|w_1|},p_a})$ by Example~\ref{ex:ahead},
and $q_{|w_1|M}^a(s)=w_1(a)$.
\qed
\end{example}

In the next lemma we prove our basic intuition that the output of $N$ on input $C$
is the largest common prefix of the outputs of $M$ on all inputs $C[p]$, $p\in P$, 
such that the difference tuples of $M$ are stored in the states of $N$. 

\begin{lemma}\label{lm:dtladtopprop2}
Let $M$ be a canonical dtpla and $N$ an earliest dtop such that $\sem{M}=\sem{N}$. 
Let $C\in\C_\Sigma$ and let $\varphi$ be the aheadness mapping from $N$ to $M$. 
Then
\begin{enumerate}
\item[(1)] $N(C)\Phi = \pref(M,C)$, and 
\item[(2)] for every $v\in V_\bot(N(C)\Phi)$, $q\in Q_N$, and $p\in P_M$, \\
if $N(C)/v=\angl{q,\bot}$, then $\varphi(q,p)=M(C[p])/v$.
\end{enumerate}
\end{lemma}
\begin{proof}
By Lemma~\ref{lm:altahead}(b) it suffices to prove statement~(1). 
By Lemma~\ref{lm:altahead}(a), 
$N(C)\Phi\sqsubseteq M(C[p])$ for every $p\in P_M$
and so $N(C)\Phi\sqsubseteq\pref(M,C)$ by the definition of greatest lower bound.  
To show equality, we prove for every node $v$ of $N(C)\Phi$ with label $\bot$
that $v$ has label $\bot$ in $\pref(M,C)$ too. 
Let $N(C)/v=\angl{q,\bot}$ for $q\in Q_N$. Then, 
by Lemma~\ref{lm:altahead}(b),
$M(C[p])/v=\varphi(q,p)$ for every $p\in P_M$. 
Suppose now that $v$ has label $d\in\Delta$ in $\pref(M,C)$.
Then $v$ has label $d$ in every $M(C[p])$, and so the root symbol of 
$\varphi(q,p)$ is $d$ for every $p\in P_M$. 
It then follows from Lemma~\ref{lm:dtladtopprop1} that $q_N(s)$ has root label $d$
for every $s\in\T_\Sigma$, contradicting the fact that $N$ is earliest.
Hence $v$ has label $\bot$ in $\pref(M,C)$.
\qed
\end{proof}

\begin{example}
For $M$ and $N$ of Example~\ref{ex:Mthree}, the statements of Lemma~\ref{lm:dtladtopprop2}
are easy to check from Examples~\ref{ex:Mthree} and~\ref{ex:ahead}. 
As an example, let $C=w_1z(\bot)$ with $|w_1|=30$ and $|z|<50$. 
Then $N(C)=w_1(\angl{q_z,\bot})$, $M(C[p_a])=w_1z(\angl{q^a_{80-|w_1z|},p_a})$ and $M(C[p_b])=w_1(b)$.
Thus $N(C)\Phi=w_1(\bot)=M(C[p_a])\sqcap M(C[p_b])$, and 
for the node $v$ with $N(C)/v=\angl{q_z,\bot}$ we have, 
using the definition of $\varphi$ in Example~\ref{ex:ahead}, that 
$\varphi(q_z,p_a)= z(\angl{q^a_{50-|z|},p_a})=M(C[p_a])/v$ 
and $\varphi(q_z,p_b)=b=M(C[p_b])/v$.
\qed
\end{example}

If a dtpla $M$ is equivalent to a dtop, then it is equivalent to a canonical dtop 
by Theorem~\ref{th:uniform_earliest}. 
By~\cite[Theorem~15]{DBLP:journals/jcss/EngelfrietMS09}, two (reduced) canonical dtops are equivalent 
if and only if they are the same (modulo a renaming of their states). 
Thus, if a dtpla $M$ is equivalent to a dtop, 
then it is equivalent to a \emph{unique} canonical dtop $\td(M)$.
In the next three lemmas we give another proof of this, and 
we show that the dtop $\td(M)$ can be constructed from $M$ and $\diftup(M)$.

We start by showing that $Q_{\td(M)}$ can be identified with $\diftup(M)$, 
assuming $N$ to be canonical too. 

\begin{lemma}\label{lm:Qdiff}
Let $M,N$ be equivalent canonical dtlas such that $M$ is a dtpla and $N$ is a dtop. 
Let $\varphi: Q_N\times P_M\to T_\Delta(Q_M\times P_M)$ be
the aheadness mapping from $N$ to $M$, and let for $q\in Q_N$
$$\psi(q)=(\varphi(q,\hat{p}_1),\dots,\varphi(q,\hat{p}_n))$$
where $P_M=\{\hat{p}_1,\dots,\hat{p}_n\}$.
Then $\psi$ is a bijection between $Q_N$ and $\diftup(M)$.
\end{lemma}
\begin{proof}
The proof is in three parts.

\smallskip
(i) For all $q\in Q_N$, $\psi(q)\in\diftup(M)$. 

\smallskip
\noindent
Proof. Let $C\in\C_\Sigma$ and $v\in\Nat^*_+$ be such that $N(C)/v=\angl{q,\bot}$, by Lemma~\ref{lm:reach}.
By Lemma~\ref{lm:dtladtopprop2}, $v\in V_\bot(\pref(M,C))$ and
$M(C[\hat{p}_i])/v=\varphi(q,\hat{p}_i)$ for every $i\in[n]$.
That shows that $\psi(q)\in\diftup(M)$.

\smallskip
(ii) For every $(t_1,\dots,t_n)\in\diftup(M)$ there exists $q\in Q_N$ such 
that $\psi(q)=(t_1,\dots,t_n)$.

\smallskip
\noindent
Proof. 
If $(t_1,\dots,t_n)\in\diftup(M)$ then there are $C\in\C_\Sigma$ and $v\in\Nat^*_+$ such 
that $\pref(M,C)/v=\bot$ and $M(C[\hat{p}_i])/v=t_i$ for $i\in[n]$.
By Lemma~\ref{lm:dtladtopprop2},
$N(C)/v=\angl{q,\bot}$ for some $q\in Q_N$, and
$M(C[\hat{p}_i])/v=\varphi(q,\hat{p}_i)$. 
Thus, $t_i=\varphi(q,\hat{p}_i)$ for $i\in[n]$.

\smallskip
(iii) For all $q_1,q_2\in Q_N$, if $\psi(q_1)=\psi(q_2)$ then $q_1=q_2$.

\smallskip
\noindent
Proof. 
Let $\psi(q_1)=\psi(q_2)$. By Lemma~\ref{lm:dtladtopprop1}, 
$q_{1N}(s)=q_{2N}(s)$ for all $s\in\T_\Sigma$.
In other words, $\sem{q_1}_N=\sem{q_2}_N$ and hence
$q_1=q_2$ because $N$ is canonical.  
\qed
\end{proof}

Thus, by renaming the states of $N$ according to $\psi$, 
we obtain a canonical dtop $N'$, equivalent to $M$, such that $Q_{N'}=\diftup(M)$.  
Moreover, the aheadness mapping $\varphi'$ from $N'$ to $M$
satisfies $\varphi'((t_1,\dots,t_n),\hat{p}_i)=t_i$ for every $(t_1,\dots,t_n)\in\diftup(M)$ and $i\in[n]$,
in other words, $\varphi'$ consists of the projections of difference tuples on their components. 

\begin{example}\label{ex:psi}
For $M$ and $N$ of Example~\ref{ex:Mthree}, 
we obtain from Example~\ref{ex:ahead} that 
\[
\begin{array}{ll}
\psi(q_i)=(\angl{q^a_{i+50},p_a},\,\angl{q^b_i,p_b}) & \textrm{for } i\in[30], 
\\[2mm]
\psi(q_z)=(z(\angl{q^a_{50-|z|},p_a}),\,b) & \textrm{for } z\in\{\sigma,\tau\}^*  \textrm{ with } |z|<50, \textrm{and} 
\\[2mm]
\psi(q_z)=(z(a),\,b) & \textrm{for } z\in\{\sigma,\tau\}^* \textrm{ with } |z|=50.
\end{array}
\]
Thus, by Example~\ref{ex:diftup}, the mapping $\psi: Q_N\to\diftup(M)$ is indeed a bijection. 
\qed
\end{example}

From Lemma~\ref{lm:Qdiff} we obtain the following corollary, which, by Example~\ref{ex:Mcounter},
does not hold in the other direction. 

\begin{corollary}\label{co:difffin}
Let $M$ be a total dtla. 
If $M$ is equivalent to a dtop, 
then $\diff(M)$ is finite. 
\end{corollary}
\begin{proof}
If $M$ is a dtop, then $\diff(M)=\emptyset$. Now let $M$ be a dtpla.
By Lemma~\ref{lm:make_uniform} we may assume that $M$ is la-uniform. 
Let $\can(M)$ be a canonical dtpla equivalent to~$M$, 
which exists by Theorem~\ref{th:uniform_earliest}.
If $M$, and hence $\can(M)$, is equivalent to a (total) dtop, 
then $\can(M)$ is equivalent to a canonical dtop, 
by Theorem~\ref{th:uniform_earliest}. 
By Lemmas~\ref{lm:aheadness-mapping} and~\ref{lm:Qdiff}, $\diftup(\can(M))$ is finite,
and so $\diff(\can(M))$ is finite by Lemma~\ref{lm:twodiff}.
Hence $\diff(M)$ is finite by Theorem~\ref{th:uniform_earliest},
because $\maxdiff(M)\leq\maxdiff(\can(M))+\sumfix(M)$.
\qed
\end{proof}

Next we show, for a canonical dtpla $M$, how to compute the axiom $A_{\td(M)}$ of $\td(M)$, 
representing the states of $\td(M)$ by difference tuples. 
For that we need some terminology.

For an la-uniform dtla $M$ and a tree $t\in\T_\Delta(Q(X))$ 
we define $t\Omega\in\T_\Delta(Q\times P)$ by 
$t\Omega=t[q(x_i)\leftarrow\angl{q,\rho(q)}\mid q\in Q,i\in\Nat]$. 
In particular, for a total dtop $N$, 
$t\Omega=t[q(x_i)\leftarrow\angl{q,\bot}\mid q\in Q_N,i\in\Nat]$.
Recall also the definition of the pattern $t\Phi\in\mathcal{P}_\Delta$ 
for $t\in\T_\Delta(Q_N(X))$ or $t\in\T_\Delta(Q_N\times \{\bot\})$ 
after Example~\ref{ex:diftup}: 
each subtree in $Q_N(X)$ or $Q_N\times \{\bot\}$ is replaced by $\bot$.

Intuitively, if $M$, $N$, $\varphi$, and $\psi$ are as in Lemma~\ref{lm:Qdiff}, 
then we expect the pattern $A_N\Phi$ to be the least common prefix 
of all axioms $A_M(p)$ of $M$, or, equivalently of all trees $A_M(p)\Omega$ (with $p\in P_M$). 
Moreover, for every node $v\in V_\bot(A_N\Phi)$ with $A_N/v=q(x_0)$, we expect 
the state $q$ of $N$ to correspond (via $\psi$) to the difference tuple of these trees at the node $v$, 
i.e., the difference tuple $(A_M(\hat{p}_1)\Omega/v,\dots,A_M(\hat{p}_n)\Omega/v)$.  
This intuition is proved in the next lemma. 

\begin{lemma}\label{lm:dtladtopprop2a}
Let $M,N$ be equivalent canonical dtlas such that $M$ is a dtpla and $N$ is a dtop. 
Let $\varphi$ be the aheadness mapping from $N$ to $M$.
Then 
\begin{enumerate}
\item[(1)] $A_N\Phi=\sqcap\{A_M(p)\Omega\mid p\in P_M\}$, and
\item[(2)] for every $v\in V_\bot(A_N\Phi)$ and $q\in Q_N$, \\
if $A_N/v=q(x_0)$, then $\varphi(q,p)=A_M(p)\Omega/v$ for all $p\in P_M$.
\end{enumerate}
\end{lemma}
\begin{proof}
By the semantics of $N^\circ$, $N(\bot)=A_N\Omega$ and 
by the semantics of $M^\circ$, $M(p)=A_M(p)\Omega$ for every $p\in P_M$.
Hence by Lemma~\ref{lm:dtladtopprop2}(1), with $C=\bot$, 
$A_N\Phi=A_N\Omega\Phi=N(\bot)\Phi=\pref(M,\bot)=\sqcap\{M(p)\mid p\in P_M\}=
\sqcap\{A_M(p)\Omega\mid p\in P_M\}$.
If $A_N/v=q(x_0)$ then $N(\bot)/v=A_N\Omega/v=\angl{q,\bot}$ and so 
by Lemma~\ref{lm:dtladtopprop2}(2), with $C=\bot$, 
$\varphi(q,p)=M(p)/v=A_M(p)\Omega/v$ for every $p\in P_M$. 
\qed
\end{proof}

\begin{example}
For $M$ and $N$ of Example~\ref{ex:Mthree}, we have 
$A_M(p_a)=q^a_{80}(x_0)$, $A_M(p_b)=q^b_{30}(x_0)$, and $A_N=q_{30}(x_0)$.
Thus $A_N\Phi=\bot$ and 
$A_M(p_a)\Omega\sqcap A_M(p_b)\Omega=\angl{q^a_{80},p_a}\sqcap\angl{q^b_{30},p_b}=\bot$. 
Moreover, for $v=\varepsilon$, we have
$\varphi(q_{30},p_a)=\angl{q^a_{80},p_a}$ and $\varphi(q_{30},p_b)=\angl{q^b_{30},p_b}$,
cf. Example~\ref{ex:psi}. 

Suppose now that we change the axioms of $M$ into 
$A_M(p_a)=\sigma(q^a_{80}(x_0))$ and $A_M(p_b)=\sigma(q^b_{30}(x_0))$. 
Obviously, an equivalent dtop is obtained by changing the axiom of $N$ into $A_N=\sigma(q_{30}(x_0))$.
It should be clear that the aheadness mapping does not change. 
Now $A_N\Phi=\sigma(\bot)=A_M(p_a)\Omega\sqcap A_M(p_b)\Omega$, 
and for $v=1$ we have the same statement as above. 

Next, suppose that we change the axioms of $M$ into 
$A_M(p_a)=\tau(\sigma(q^a_{80}(x_0)))$ and $A_M(p_b)=\tau(\tau(q^b_{30}(x_0)))$. 
For this $M$, an equivalent canonical dtop $N$ exists, of which we will not give the details. 
It should be clear that its axiom is $A_N=\tau(q_0(x_0))$ for some state $q_0$,
and that for a context $C=w(\bot)$ with $|w|\geq 80$ it computes $N(C)=\angl{q_{z_1,z_2},\bot}$, 
where $z_1$ and $z_2$ are the first 80 and 30 symbols of $w$, respectively; 
for the states $q_{z_1,z_2}$, it has rules 
$q_{z_1,z_2}(a)\to \sigma z_1(a)$ and $q_{z_1,z_2}(b)\to \tau z_2(b)$. 
Now $A_N\Phi=\tau(\bot)=A_M(p_a)\Omega\sqcap A_M(p_b)\Omega$ and, 
after computing the new aheadness mapping $\varphi$ from $N$ to $M$, it can be seen that for $v=1$, 
$\varphi(q_0,p_a)=\sigma(\angl{q^a_{80},p_a})$ and $\varphi(q_0,p_b)=\tau(\angl{q^b_{30},p_b})$.
\qed
\end{example}

Finally we show how to compute the rules of $\td(M)$. 
Again, we need some terminology. In view of the next subsection, 
we consider an arbitrary set $S$ of ``states''
instead of the set of states $Q_N$ of an equivalent canonical dtop $N$ 
(which, of course, motivates the terminology).

Let $M$ be a canonical dtla, $S$ be a set and 
$\varphi: S\times P_M\to \T_\Delta(Q_M\times P_M)$ be a mapping   
such that $\varphi(q,p)\in\T_\Delta(\{\angl{\bar{q},p}\mid\bar{q}\in Q_M,\rho_M(\bar{q})=p\})$ 
for every $q\in S$ and $p\in P_M$. 
Then we define for every $q\in S$, $a\in\Sigma^{(k)}$, and $p_1,\dots,p_k\in P_M$, the tree
$$\rhs_{M,\varphi}(q,a,p_1,\dots,p_k) = \varphi(q,p)[\angl{\bar{q},p}\leftarrow\rhs_M(\bar{q},a,p_1,\dots,p_k)\mid\bar{q}\in Q_M]$$
where $p=\delta_M(a,p_1,\dots,p_k)$. 
Note that by the above condition on $\varphi(q,p)$, 
the right-hand side of the equation is always defined and 
is a tree in $\T_\Delta(Q_M(X_k))$; 
moreover, if it has a subtree $q'(x_i)$, then $\rho_M(q')=p_i$. 
Thus, $\rhs_{M,\varphi}(q,a,p_1,\dots,p_k)\in\T_\Delta(Q_M(X_k))$ and
$\rhs_{M,\varphi}(q,a,p_1,\dots,p_k)\Omega\in\T_\Delta(\{\angl{\bar{q},p}\mid\bar{q}\in Q_M,\rho_M(\bar{q})=p\})$.

Now let $M$, $N$, $\varphi$ (with $S=Q_N$), and $\psi$ be as in Lemma~\ref{lm:Qdiff}, and 
consider $q\in Q_N$ and $a\in \Sigma^{(k)}$. 
Intuitively, in view of Lemma~\ref{lm:dtladtopprop1}, we expect the pattern $\rhs_N(q,a)\Phi$ to be  
the least common prefix of all ``right-hand sides'' 
$\rhs_{M,\varphi}(q,a,p_1,\dots,p_k)\Omega$ with $p_1,\dots,p_k\in P_M$. 
Moreover, in view of Lemma~\ref{lm:dtladtopprop2},
we expect for every node $v\in V_\bot(\rhs_N(q,a)\Phi)$ with $\rhs_N(q,a)/v=q'(x_i)$,
that $\rhs_{M,\varphi}(q,a,p_1,\dots,p_k)/v$ is a tree in $\T_\Delta(Q_M(\{x_i\}))$
that does not depend on $p_j$ for $j\neq i$, and we expect 
the state $q'$ of $N$ to correspond (via $\psi$) to the difference tuple 
of the trees $\rhs_{M,\varphi}(q,a,p_1,\dots,p_k)\Omega$ at the node $v$, for varying $p_i$. 
More precisely, for every $p\in P_M$, 
the set of all trees $\rhs_{M,\varphi}(q,a,p_1,\dots,p_k)/v$ with $p_i=p$ is a singleton $\{t_p\}$ with 
$t_p\in\T_\Delta(Q_M(\{x_i\}))$, and $q'$ corresponds to the difference tuple 
of the trees $t_p\Omega$ at $v$, i.e., 
the difference tuple $(t_{\hat{p}_1}\Omega/v,\dots,t_{\hat{p}_n}\Omega/v)$. 
These intuitions are proved in the next lemma. Examples are given after the proof. 

\begin{lemma}\label{lm:dtladtopprop2b}
Let $M,N$ be equivalent canonical dtlas such that $M$ is a dtpla and $N$ is a dtop. 
Let $\varphi$ be the aheadness mapping from $N$ to $M$.
Let $q\in Q_N$ and $a\in \Sigma^{(k)}$. Then
\begin{enumerate}
\item[(1)] 
$\rhs_N(q,a)\Phi=\sqcap\{\rhs_{M,\varphi}(q,a,p_1,\dots,p_k)\Omega\mid p_1,\dots,p_k\in P_M\}$, and 
\item[(2)] for every $v\in V_\bot(\rhs_N(q,a)\Phi)$, $q'\in Q_N$, and $i\in[k]$, \\
if $\rhs_N(q,a)/v=q'(x_i)$, then
\begin{enumerate}
\item[$(a)$] $\rhs_{M,\varphi}(q,a,p_1,\dots,p_k)/v\in \T_\Delta(Q_M(\{x_i\}))$ 
for all $p_1,\dots,p_k\in P_M$, and
\item[$(b)$] $\varphi(q',p_i)=\rhs_{M,\varphi}(q,a,p_1,\dots,p_k)\Omega/v$
for all $p_1,\dots,p_k\in P_M$. 
\end{enumerate}
\end{enumerate}
\end{lemma}

\noindent
\begin{remark}\label{rem:unique}
Before proving this lemma, we observe that items (a) and (b) determine the number $i$ uniquely 
(and hence also the state $q'$).
Indeed, suppose that (a) and (b) also hold for some $q''$ and some $i'\neq i$. 
Then it follows from (a) that every tree $\rhs_{M,\varphi}(q,a,p_1,\dots,p_k)/v$ is in $\T_\Delta$
(and hence equals $\rhs_{M,\varphi}(q,a,p_1,\dots,p_k)\Omega/v$). 
Moreover, it follows from (a) and (b) together that $\rhs_{M,\varphi}(q,a,p_1,\dots,p_k)/v$ 
does not depend on $p_j$ for $j\neq i$ \emph{and} does not depend on $p_j$ for $j\neq i'$.
In other words, all $\rhs_{M,\varphi}(q,a,p_1,\dots,p_k)/v$ are the same, which contradicts item $(1)$
and the fact that $v\in V_\bot(\rhs_N(q,a)\Phi)$ (by Lemma~\ref{lm:pat}). 
\end{remark}

\begin{proof}
(1) We first show that $\rhs_N(q,a)\Phi\sqsubseteq\rhs_{M,\varphi}(q,a,p_1,\dots,p_k)$ 
for all $p_1,\dots,p_k$ $\in P_M$. 
Consider arbitrary trees $s_1,\dots,s_k\in\T_\Sigma$ 
with $s_i\in\sem{p_i}_M$ for $i\in[k]$, 
and let $s=a(s_1,\dots,s_k)$. 
By Lemma~\ref{lm:elem}(4), 
$$q_N(s)=\rhs_N(q,a)\Psi_N$$ where 
$\Psi_N=[\bar{q}(x_i)\leftarrow {\bar{q}}_N(s_i)\mid \bar{q}\in Q_N, i\in[k]]$. 
On the other hand, by Lemma~\ref{lm:dtladtopprop1} and Lemma~\ref{lm:elem}(4),
$$q_N(s)=\varphi(q,p)[\angl{\bar{q},p}\leftarrow \rhs_M(\bar{q},a,p_1,\dots,p_k)\mid \bar{q}\in Q_M]\Psi_M$$
where
$p=\delta_M(a,p_1,\dots,p_k)$ and 
$\Psi_M= [\bar{q}(x_i)\leftarrow {\bar{q}}_M(s_i)\mid \bar{q}\in Q_M, i\in[k]]$.
Thus 
\begin{equation}\label{eq:rhss}
\rhs_N(q,a)\Psi_N = \rhs_{M,\varphi}(q,a,p_1,\dots,p_k)\Psi_M.
\end{equation}
We have to prove that every node of $\rhs_N(q,a)$ with label $d\in\Delta$ 
is also a node of $\rhs_{M,\varphi}(q,a,p_1,\dots,p_k)$, with the same label. 
We prove this by induction on the length of the nodes of $\rhs_N(q,a)$, similar to the proof of 
Claim~1 of Lemma~\ref{lm:aheadness-mapping}.
Let $v$ be a node of $\rhs_N(q,a)$ with label $d\in\Delta$.  
By induction, $v$ is a node of $\rhs_{M,\varphi}(q,a,p_1,\dots,p_k)$. 
Now suppose that $v$ does not have label $d$ in $\rhs_{M,\varphi}(q,a,p_1,\dots,p_k)$. 
Then Equation~(\ref{eq:rhss}) implies that 
$\rhs_{M,\varphi}(q,a,p_1,\dots,p_k)/v=\bar{q}(x_i)$ and that ${\bar{q}}_M(s_i)$ has root label $d$. 
Since that holds for arbitrary $s_i\in \sem{p_i}_M$, 
it contradicts the fact that $M$ is earliest.
Note that since $M$ is la-uniform, 
$\sem{p_i}_M$ is the domain of $\sem{\bar{q}}_M$.

Now let $\Pi=\{\rhs_{M,\varphi}(q,a,p_1,\dots,p_k)\Omega\mid p_1,\dots,p_k\in P_M\}$.
By the above, $\rhs_N(q,a)\Phi\sqsubseteq\sqcap\Pi$.
It remains to prove for every node $v$ of $\rhs_N(q,a)\Phi$ with label $\bot$
that $v$ has label $\bot$ in $\sqcap\Pi$ too. Let $\rhs_N(q,a)/v=\bar{q}(x_i)$. 
Assume now that $v$ has label $d\in\Delta$ in $\sqcap\Pi$. 
Then $v$ has label $d$ in every tree $\rhs_{M,\varphi}(q,a,p_1,\dots,p_k)$.
Equation~(\ref{eq:rhss})
then implies that $\bar{q}_N(s_i)$ has root label $d$ for every $s_i\in\sem{p_i}_M$. 
Since this holds for every $p_i\in P_M$, $\bar{q}_N(s)$ has root label $d$ for every $s\in\T_\Sigma$, 
which contradicts the fact that $N$ is earliest.

(2) Let $v\in V_\bot(\rhs_N(q,a)\Phi)$ with $\rhs_N(q,a)/v=q'(x_i)$,
and let $p_1,\dots,p_k\in P_M$. 
Before proving (a) and (b) for $p_1,\dots,p_k$, we prove the following claim.

\begin{claim}
Let $s_j\in\sem{p_j}_M$ for every $j\in[k]$, $j\neq i$. Then
$$\varphi(q',p_i)=\rhs_{M,\varphi}(q,a,p_1,\dots,p_k)\Psi_{iM}/v$$
where $\Psi_{iM}$ abbreviates
$[\bar{q}(x_j)\leftarrow\bar{q}_M(s_j)\mid\bar{q}\in Q_M,j\in[k],j\neq i]\Omega$. 
\end{claim}

\noindent
\emph{Proof of Claim.}
We first observe that the effect of 
$\Psi_{iM}$ is to replace every $\bar{q}(x_j)$ with $j\neq i$ 
by a tree in $\T_\Delta$ (viz. the tree $\bar{q}_M(s_j)$) and, through $\Omega$, 
every $\bar{q}(x_i)$ by $\angl{\bar{q},\rho_M(\bar{q})}$. 

Let $\Psi_{iN}$ be defined similar to $\Psi_{iM}$, i.e., $\Psi_{iN}$ 
abbreviates $[\bar{q}(x_j)\leftarrow\bar{q}_N(s_j)\mid\bar{q}\in Q_N,j\in[k],j\neq i]\Omega$.
Let $C\in\C_\Sigma$ and $u\in\Nat^*_+$ be such that $N(C)/u=\angl{q,\bot}$, by Lemma~\ref{lm:reach}, 
and consider the context $C_i=C[a(s_1,\dots,s_{i-1},\bot,s_{i+1},\dots,s_k)]$. 
By Lemma~\ref{lm:semantics} and Lemma~\ref{lm:elem}(4), 
$N(C_i)=N(C)[\angl{\bar{q},\bot}\leftarrow\rhs_N(\bar{q},a)\mid\bar{q}\in Q_N]\Psi_{iN}$. 
Hence 
\begin{equation}\label{eq:rhsN}
N(C_i)/u=\rhs_N(q,a)\Psi_{iN}.
\end{equation} 
Similarly,  
$M(C_i[p_i])=M(C[p])[\angl{\bar{q},p}\leftarrow
\rhs_M(\bar{q},a,p_1,\dots,p_k)\mid\bar{q}\in Q_M]\Psi_{iM}$, 
where $p=\delta_M(a,p_1,\dots,p_k)$.  
Hence, since $M(C[p])/u=\varphi(q,p)$ by Lemma~\ref{lm:dtladtopprop2}(2), 
we obtain that
\begin{equation}\label{eq:etaM}
M(C_i[p_i])/u=\rhs_{M,\varphi}(q,a,p_1,\dots,p_k)\Psi_{iM}.
\end{equation}
Since $\rhs_N(q,a)/v=q'(x_i)$, 
Equation~(\ref{eq:rhsN}) implies that $N(C_i)/uv=\angl{q',\bot}$, and so 
$\varphi(q',p_i)=M(C_i[p_i])/uv$ by Lemma~\ref{lm:dtladtopprop2}(2). 
Then Equation~(\ref{eq:etaM}) gives the required result.
This proves the Claim.

\medskip
(a) We choose trees $s_j\in\sem{p_j}_M$ for $j\in[k]$, $j\neq i$. 
By the Claim, we have that
$\varphi(q',p_i)=\rhs_{M,\varphi}(q,a,p_1,\dots,p_k)\Psi_{iM}/v$. 
Suppose that there exists $w\in\Nat_+^*$ such that 
$\rhs_{M,\varphi}(q,a,p_1,\dots,p_k)/vw=\tilde{q}(x_j)$ with $j\neq i$. 
Let $s_j'\in\sem{p_j}_M$ be such that $\tilde{q}_M(s_j')\neq\tilde{q}_M(s_j)$;
it exists because $M$ is earliest (and so $\tilde{q}_M(s_j')$ and 
$\tilde{q}_M(s_j)$ can even have different root symbols). 
Then, again by the Claim, we have that 
$\varphi(q',p_i)=\rhs_{M,\varphi}(q,a,p_1,\dots,p_k)\Psi_{iM}'/v$ where 
$\Psi_{iM}'$ is obtained from $\Psi_{iM}$ by changing $s_j$ into $s_j'$. 
But the trees $\rhs_{M,\varphi}(q,a,p_1,\dots,p_k)\Psi_{iM}/v$ and
$\rhs_{M,\varphi}(q,a,p_1,\dots,p_k)\Psi_{iM}'/v$ are different: 
their subtrees at node $w$ are $\tilde{q}_M(s_j)$ and $\tilde{q}_M(s_j')$, respectively.
That is a contradiction.

(b) follows immediately from~(a) and the Claim.
\qed
\end{proof}

We observe that for $k\in\{0,1\}$, the statement of Lemma~\ref{lm:dtladtopprop2b} can be simplified.  
For $k=0$, it just states that $\rhs_N(q,a)=\rhs_{M,\varphi}(q,a)$.
For $k=1$, item~(2) of the lemma just states that 
for every $v\in V_\bot(\rhs_N(q,a)\Phi)$ and $q'\in Q_N$, 
if $\rhs_N(q,a)/v=q'(x_1)$, then $\varphi(q',p)=\rhs_{M,\varphi}(q,a,p)\Omega/v$ for all $p\in P_M$.

\begin{example}
We consider again $M$ and $N$ of Example~\ref{ex:Mthree}. 
Since the rank of every symbol in $\Sigma$ is $0$ or $1$, the above observation applies. 

Let us first check Lemma~\ref{lm:dtladtopprop2b} for $q_1\in Q_N$ and for $\gamma\in\{\sigma,\tau\}$. 
To do this we have to compute $\rhs_{M,\varphi}(q_1,\gamma,p)$ for $p\in\{p_a,p_b\}$. 
Since $\varphi(q_1,p_a)=\angl{q^a_{51},p_a}$, we obtain that 
$\rhs_{M,\varphi}(q_1,\gamma,p_a)=\rhs_M(q^a_{51},\gamma,p_a)=\gamma(q^a_{50}(x_1))$.
Since $\varphi(q_1,p_b)=\angl{q^b_1,p_b}$, we obtain that 
$\rhs_{M,\varphi}(q_1,\gamma,p_b)=\rhs_M(q^b_1,\gamma,p_b)=\gamma(b)$. 
Consequently $\rhs_{M,\varphi}(q_1,\gamma,p_a)\Omega\sqcap \rhs_{M,\varphi}(q_1,\gamma,p_b)\Omega=\gamma(\bot)$.
Since $\rhs_N(q_1,\gamma)=\gamma(q_\varepsilon(x_1))$, 
we have that $\rhs_N(q_1,\gamma)\Phi=\gamma(\bot)$,
which shows item~(1) of Lemma~\ref{lm:dtladtopprop2b}. 
To show item~(2), we have to check for $v=1$ 
that $\varphi(q_\varepsilon,p)=\rhs_{M,\varphi}(q_1,\gamma,p)\Omega/v$ for $p\in\{p_a,p_b\}$. 
This is true because $\varphi(q_\varepsilon,p_a)=\angl{q^a_{50},p_a}$ and $\varphi(q_\varepsilon,p_b)=b$. 

Next we check Lemma~\ref{lm:dtladtopprop2b} for $q_z\in Q_N$ with $|z|<50$ 
and for $\gamma\in\{\sigma,\tau\}$, in the same way.  
We first compute $\rhs_{M,\varphi}(q_z,\gamma,p)$ for $p\in\{p_a,p_b\}$. 
Since $\varphi(q_z,p_a)=z(\angl{q^a_{50-|z|},p_a})$, we obtain that 
$\rhs_{M,\varphi}(q_z,\gamma,p_a)=z(\rhs_M(q^a_{50-|z|},\gamma,p_a))$, 
which equals $z\gamma(q^a_{49-|z|}(x_1))$ for $|z|<49$ and $z\gamma(a)$ for $|z|=49$. 
And since $\varphi(q_z,p_b)=b$, we get that 
$\rhs_{M,\varphi}(q_z,\gamma,p_b)=b$. 
Thus $\rhs_{M,\varphi}(q_z,\gamma,p_a)\Omega\sqcap \rhs_{M,\varphi}(q_z,\gamma,p_b)\Omega=\bot$.
Since $\rhs_N(q_z,\gamma)=q_{z\gamma}(x_1)$, we have that $\rhs_N(q_z,\gamma)\Phi=\bot$,
which shows item~(1). 
To show item~(2), we have to check that $\varphi(q_{z\gamma},p)=\rhs_{M,\varphi}(q_z,\gamma,p)\Omega$ 
for $p\in\{p_a,p_b\}$. 
For $p=p_a$ that follows from the fact that $\varphi(q_{z\gamma},p_a)$ equals $z\gamma(\angl{q^a_{50-|z\gamma|},p_a})$
for $|z\gamma|<50$ and $z\gamma(a)$ for $|z\gamma|=50$.
For $p=p_b$, from $\varphi(q_{z\gamma},p_b)=b$.

We now check Lemma~\ref{lm:dtladtopprop2b} for $q_z\in Q_N$ with $|z|\leq 50$ 
and for $a\in\Sigma^{(0)}$. 
For $|z|<50$, we have $\rhs_{M,\varphi}(q_z,a)=z(\rhs_M(q^a_{50-|z|},a))=z(a)=\rhs_N(q_z,a)$.
For $|z|=50$, since $\varphi(q_z,p_a)=z(a)$, we have $\rhs_{M,\varphi}(q_z,a)=z(a)=\rhs_N(q_z,a)$.

To illustrate Lemma~\ref{lm:dtladtopprop2b} also for the case where $k\geq 2$, we extend $M$ with the 
binary input symbols $\ell$ and $r$ (standing for `left' and `right'). We extend $\delta_M$ such that 
$\delta_M(\ell,p,p')=p$ and $\delta_M(r,p',p)=p$ 
for all $p,p'\in\{p_a,p_b\}$, and we extend $R_M$ with the rules 
$q^y_i(\ell(x_1:p_y,x_2:p'))\to q^y_i(x_1)$ and 
$q^y_i(r(x_1:p',x_2:p_y))\to q^y_i(x_2)$ 
for all $(y,i)\in(\{a\}\times[80])\cup(\{b\}\times[30])$ and $p'\in\{p_a,p_b\}$. 
Intuitively, the new symbols $\ell$ and $r$ are signposts that determine 
a unique path through the input tree, and $M$ processes the unary symbols on that path as before. 
To obtain a dtop equivalent to $M$, we extend $N$ with the rules 
$q(\ell(x_1,x_2))\to q(x_1)$ and $q(r(x_1,x_2))\to q(x_2)$ for every $q\in Q_N$. 
It is easy to see that this extension 
does not change the aheadness mapping $\varphi$ from $N$ to~$M$.

Let us now check Lemma~\ref{lm:dtladtopprop2b} for $q_z\in Q_N$ with $|z|<50$ and for the new symbol~$r$. 
For $p'\in\{p_a,p_b\}$, we obtain 
$\rhs_{M,\varphi}(q_z,r,p',p_a)=z(\rhs_M(q^a_{50-|z|},r,p',p_a))=z(q^a_{50-|z|}(x_2))$
and $\rhs_{M,\varphi}(q_z,r,p',p_b)=b$. 
Thus their largest common prefix is $\bot$. 
Since $\rhs_N(q_z,r)=q_z(x_2)$, this shows item~(1). 
For $q'=q_z$ and $i=2$, item~(2)(a) clearly holds 
because the trees $z(q^a_{50-|z|}(x_2))$ and $b$ do not contain the variable $x_1$. 
Item~(2)(b) also holds because 
$\varphi(q_z,p_a)=z(\angl{q^a_{50-|z|},p_a})=\rhs_{M,\varphi}(q_z,r,p',p_a)\Omega$ 
and $\varphi(q_z,p_b)=b=\rhs_{M,\varphi}(q_z,r,p',p_b)\Omega$ 
for every $p'\in\{p_a,p_b\}$. 

Finally, we consider $q_z$ and $r$ with $|z|=50$. 
We now have $\rhs_{M,\varphi}(q_z,r,p',p_a)=z(a)$ and $\rhs_{M,\varphi}(q_z,r,p',p_b)=b$
for every $p'\in\{p_a,p_b\}$, and we have $\varphi(q_z,p_a)=z(a)$ and $\varphi(q_z,p_b)=b$.
From this it is easy to see that items~(1) and~(2) hold, as in the previous paragraph. 
With respect to item~(2)(a) we note that in this case the trees $\rhs_{M,\varphi}(q_z,\ell,p_a,p')$ 
and $\rhs_{M,\varphi}(q_z,\ell,p_b,p')$ do not contain the variable $x_2$ either. 
\qed
\end{example}

The three Lemmas~\ref{lm:Qdiff}, \ref{lm:dtladtopprop2a}, and~\ref{lm:dtladtopprop2b} show that every 
canonical dtpla $M$ that is equivalent to a dtop, 
is equivalent to a unique canonical dtop $\td(M)$, modulo a renaming of states. 
That will be stated  formally in Corollary~\ref{co:uniquecandtop} in the next subsection.

\subsection{The Algorithm}
\label{sec:diffalg}

Let $M=(Q,\Sigma,\Delta,R,A,P,\delta)$ be a canonical dtpla and let $P=\{\hat{p}_1,\dots,\hat{p}_n\}$.
Our aim in this subsection is to construct the canonical dtop $N=\td(M)$ equivalent to $M$, if it exists,
based on the characterizations in Lemmas~\ref{lm:Qdiff}, \ref{lm:dtladtopprop2a}, and~\ref{lm:dtladtopprop2b}. 
By Lemma~\ref{lm:Qdiff}, we may assume that the states of $N$ are difference tuples of $M$, 
and that the aheadness mapping from $N$ to $M$ projects these tuples on their components. 
This leads to the following definitions. 

We define the set $S_M$ of \emph{potential states} to consist of 
all $n$-tuples $(t_1,\dots,t_n)\in \T_\Delta(Q\times P)^n$ such that 
$t_i\in\T_\Delta(\{\angl{q,\hat{p}_i}\mid q\in Q,\rho(q)=\hat{p}_i\})$ 
for every $i\in[n]$. Note that $\diftup(M)\subseteq S_M$. 
Moreover, we define the \emph{projection mapping} $\varphi_M: S_M\times P\to \T_\Delta(Q\times P)$
such that, for every $(t_1,\dots,t_n)\in S_M$ and $i\in[n]$, 
$\varphi((t_1,\dots,t_n),\hat{p}_i)=t_i$. 
In what follows we will denote $S_M$ and $\varphi_M$ just by $S$ and $\varphi$, 
to simplify the notation; it will always be clear to which dtpla they belong. 

The next two definitions are directly based on Lemmas~\ref{lm:dtladtopprop2a} and~\ref{lm:dtladtopprop2b},
respectively. 
For $t\in\T_\Delta(S(X))$ we define the pattern $t\Phi=t[q(x_i)\leftarrow\bot\mid q\in S,i\in\Nat]$.

\begin{definition}\label{def:A}
The \emph{potential axiom} $A_{M,\varphi}$ 
is the unique tree in $\T_\Delta(S(\{x_0\}))$ such that 
\begin{enumerate}
\item[(1)] $A_{M,\varphi}\Phi=\sqcap\{A(p)\Omega\mid p\in P\}$, and
\item[(2)] for every $v\in V_\bot(A_{M,\varphi}\Phi)$ and $q\in S$, \\
if $A_{M,\varphi}/v=q(x_0)$, then $\varphi(q,p)=A(p)\Omega/v$ for all $p\in P$.
\end{enumerate}
\end{definition}

Note that $A_{M,\varphi}$ exists because 
$A(p)\Omega\in\T_\Delta(\{\angl{\bar{q},p}\mid \bar{q}\in Q,\rho(\bar{q})=p\})$
for every $p\in P$.
Obviously, $A_{M,\varphi}$ can be computed from $M$. 

For the next definition, we 
recall that $\rhs_{M,\varphi}(q,a,p_1,\dots,p_k)$ 
was defined before Lemma~\ref{lm:dtladtopprop2b}. It is a tree in $\T_\Delta(Q(X_k))$,
more precisely, in $\T_\Delta(\{\bar{q}(x_i)\mid \bar{q}\in Q,\rho(\bar{q})=p_i\})$.

\begin{definition}\label{def:rhs}
Let $q\in S$ and $a\in \Sigma^{(k)}$. The \emph{potential right-hand side} $\rhs_{M,\varphi}(q,a)$
is the tree in $\T_\Delta(S(X_k))$ such that 
\begin{enumerate}
\item[(1)] 
$\rhs_{M,\varphi}(q,a)\Phi=\sqcap\{\rhs_{M,\varphi}(q,a,p_1,\dots,p_k)\Omega\mid p_1,\dots,p_k\in P\}$, and 
\item[(2)] for every $v\in V_\bot(\rhs_{M,\varphi}(q,a)\Phi)$, $q'\in S$, and $i\in[k]$, \\
if $\rhs_{M,\varphi}(q,a)/v=q'(x_i)$, then
\begin{enumerate}
\item[(a)] $\rhs_{M,\varphi}(q,a,p_1,\dots,p_k)/v\in \T_\Delta(Q(\{x_i\}))$ 
for all $p_1,\dots,p_k\in P$, and
\item[(b)] $\varphi(q',p_i)=\rhs_{M,\varphi}(q,a,p_1,\dots,p_k)\Omega/v$
for all $p_1,\dots,p_k\in P$. 
\end{enumerate}
\end{enumerate}
\end{definition}

For $k=0$ this definition (re)defines $\rhs_{M,\varphi}(q,a)$ consistently to be equal
to the tree $\rhs_{M,\varphi}(q,a)$ as defined before Lemma~\ref{lm:dtladtopprop2b}.

We also observe here that for $k\geq 1$ (more precisely, for $k\geq 2$), 
$\rhs_{M,\varphi}(q,a)$ may not exist. 
However, if it exists then it is unique,
which can be shown as follows (cf. Remark~\ref{rem:unique}). 
First, by item~(1), $\rhs_{M,\varphi}(q,a)\Phi$ always exists and is unique. 
Second, let $v\in V_\bot(\rhs_{M,\varphi}(q,a)\Phi)$ and assume that items (2)(a) and (2)(b) 
hold for $q'$ and $i$, and also for $q''$ and $j$. 
Suppose that $i\neq j$.
Then (a) implies that every $\rhs_{M,\varphi}(q,a,p_1,\dots,p_k)/v$ is in $\T_\Delta$
(and hence equals $\rhs_{M,\varphi}(q,a,p_1,\dots,p_k)\Omega/v$). 
It now follows from (b) that they are all the same tree in $\T_\Delta$. 
Indeed, for all $p_1,\dots,p_k,p'_1,\dots,p'_k\in P$ we have $\rhs_{M,\varphi}(q,a,p_1,\dots,p_k)/v=\rhs_{M,\varphi}(q,a,p'_1,\dots,p'_k)/v$ 
because both $\varphi(q',p_i)$ and $\varphi(q'',p'_j)$ are equal to 
$\rhs_{M,\varphi}(q,a,p_1,\dots,p_{j-1},p'_j,p_{j+1},\dots,p_k)/v$.
This contradicts item (1) and the fact that $v\in V_\bot(\rhs_{M,\varphi}(q,a)\Phi)$, 
by Lemma~\ref{lm:pat}. 
Hence $i=j$. It now follows from item (b) that $\varphi(q',p)=\varphi(q'',p)$ for every $p\in P$, 
and so $q'=q''$. 

In the rest of this subsection we will denote the tree 
$\sqcap\{\rhs_{M,\varphi}(q,a,p_1,\dots,p_k)\Omega\mid p_1,\dots,p_k\in P\}$ 
by $\rhs_{M,\varphi}(q,a)\Phi$, even if $\rhs_{M,\varphi}(q,a)$ does not exist. 

\begin{lemma}\label{le:rhsexist}
For $q\in S$ and $a\in \Sigma^{(k)}$, $\rhs_{M,\varphi}(q,a)$ exists if and only if
for every $v\in V_\bot(\rhs_{M,\varphi}(q,a)\Phi)$ there exists $i\in[k]$ such that
\begin{enumerate}
\item[$(a)$] $\rhs_{M,\varphi}(q,a,p_1,\dots,p_k)/v\in \T_\Delta(Q(\{x_i\}))$ 
for all $p_1,\dots,p_k\in P$, and
\item[$(b)$] for all $p_1,\dots,p_k,p'_1,\dots,p'_k\in P$, if $p_i=p'_i$ then 
$\rhs_{M,\varphi}(q,a,p_1,\dots,p_k)\Omega/v=\rhs_{M,\varphi}(q,a,p'_1,\dots,p'_k)\Omega/v$. 
\end{enumerate}
\end{lemma}

\begin{proof}
The only-if direction is obvious. For the if-direction, let $v\in V_\bot(\rhs_{M,\varphi}(q,a)\Phi)$ and
let $i\in[k]$ satisfy items (a) and (b). 
Then we define $\rhs_{M,\varphi}(q,a)/v=q'(x_i)$
where, for every $p\in P$, $\varphi(q',p)=\rhs_{M,\varphi}(q,a,p,\dots,p)\Omega/v$.
Note that $q'\in S$ because $\rhs_{M,\varphi}(q,a,p,\dots,p)\Omega$ is in 
$\T_\Delta(\{\angl{\bar{q},p}\mid \bar{q}\in Q,\rho(\bar{q})=p\})$. 
This defines $\rhs_{M,\varphi}(q,a)$ such that it satisfies the requirements of Definition~\ref{def:rhs}.
\qed
\end{proof}

This lemma shows that $\rhs_{M,\varphi}(q,a)$ always exists for $k=1$
(and thus, it always exists for $k\in\{0,1\}$).

We now show that $\rhs_{M,\varphi}(q,a)$ can be computed from $M$, $q$, and $a$, as follows.  
First one computes $\rhs_{M,\varphi}(q,a)\Phi$ by Definition~\ref{def:rhs}(1). 
Then one determines whether or not
$\rhs_{M,\varphi}(q,a)$ exists by Lemma~\ref{le:rhsexist}. Thus, one computes for every node
$v\in V_\bot(\rhs_{M,\varphi}(q,a)\Phi)$ a number $i=i_v\in[k]$ that satisfies 
$(a)$ and $(b)$ of Lemma~\ref{le:rhsexist} (provided such numbers exist). If this is successful, then,
according to the proof of Lemma~\ref{le:rhsexist}, one defines, for every such~$v$, 
$\rhs_{M,\varphi}(q,a)/v=q'(x_{i_v})$ where 
$\varphi(q',p)=\rhs_{M,\varphi}(q,a,p,\dots,p)\Omega/v$ for all $p\in P$.

\begin{definition}\label{def:assoc}
Let $M$ be a canonical dtpla. 
A dtop $N$ is \emph{associated with} $M$ if   
\begin{itemize}
\item $N$ is complete, 
\item $N$ has input and output alphabets $\Sigma$ and $\Delta$,
\item $Q_N\subseteq S$, 
\item $A_N=A_{M,\varphi}$, and 
\item $\rhs_N(q,a)=\rhs_{M,\varphi}(q,a)$ for every $q\in Q_N$ and $a\in\Sigma^{(k)}$. 
\end{itemize}
For $h\in\Nat$, $N$ is \emph{associated with $M$ and $h$} if, moreover, 
$\height(q)\leq h$ for every $q\in Q_N$.
\end{definition}

The results of Section~\ref{sec:difftup} can now be summarized as follows. 
Recall from Section~\ref{sec:difftree} that a number $h(M)\in\Nat$ is a 
\emph{difference bound} for $M$ if the following holds:
if $\diff(M)$ is finite, then $\maxdiff(M)\leq h(M)$. 

\begin{corollary}\label{co:assoc}
Let $M$ be a canonical dtpla, let $N$ be a canonical dtop equivalent to~$M$, 
and let $h(M)$ be a difference bound for $M$. 
Then, modulo a renaming of states, $Q_N=\diftup(M)$ and 
$N$ is associated with $M$ and $h(M)$. 
\end{corollary}

\begin{proof}
Let $\bar{N}$ be a canonical dtop equivalent to~$M$. 
As observed after Lemma~\ref{lm:Qdiff}, by renaming the states of $\bar{N}$ 
to difference tuples of $M$, according to the mapping $\psi$ of Lemma~\ref{lm:Qdiff},
we obtain a canonical dtop $N$ such that $Q_N=\diftup(M)\subseteq S$, 
with the projection mapping $\varphi$ as aheadness mapping
(or more precisely, the restriction of $\varphi$ to $Q_N$). 
Since $N$ is la-uniform, it is complete. 
The 4th and 5th condition of Definition~\ref{def:assoc} hold 
by Lemmas~\ref{lm:dtladtopprop2a} and~\ref{lm:dtladtopprop2b}, respectively.
Since $h(M)$ is a difference bound for $M$, 
the height of every $q\in Q_N$ is $\leq h(M)$ by Lemma~\ref{lm:twodiff}. 
\qed
\end{proof}

This corollary has two corollaries. The first shows that, as already observed in Section~\ref{sec:difftup}, 
if $M$ is equivalent to a dtop, then it is equivalent to a unique (reduced) canonical dtop $\td(M)$.  
The second shows that, to find a dtop equivalent to $M$, it suffices to consider all dtops
associated with $M$ and $h(M)$, where $h(M)$ is a difference bound for~$M$. 

\begin{corollary}\label{co:uniquecandtop}
A total dtpla $M$ is equivalent to at most one canonical dtop, modulo a renaming of states. 
\end{corollary}

\begin{proof}
By Lemma~\ref{lm:make_uniform} and Theorem~\ref{th:uniform_earliest} we may assume that $M$ is canonical. 
If $N$ and $N'$ are canonical dtops equivalent to $M$, renamed according to Lemma~\ref{lm:Qdiff}, 
then $Q_N=Q_{N'}$ and $N,N'$ are associated with $M$ and $h(M)$, by Corollary~\ref{co:assoc}.
The 4th and 5th condition of Definition~\ref{def:assoc} then imply that $N=N'$. 
\qed
\end{proof}

\begin{corollary}\label{co:assoc2}
Let $M$ be a canonical dtpla and let $h(M)$ be a difference bound for $M$. 
Then $M$ is equivalent to a dtop if and only if there exists a dtop $N$, associated with $M$ and $h(M)$, 
such that $M$ is equivalent to $N$. 
\end{corollary}

\begin{proof}
If $M$ is equivalent to a dtop, then it is equivalent to 
a canonical dtop by Theorem~\ref{th:uniform_earliest}.
Hence, by Corollary~\ref{co:assoc}, it is equivalent to a dtop that is associated with $M$ and~$h(M)$.
\qed
\end{proof}

The last corollary already allows us to conclude our first main result, for the case that $M$ is canonical. 

\begin{lemma}\label{le:alg}
It is decidable for a given canonical dtpla $M$ and a given difference bound $h(M)$ for $M$
whether there exists a dtop $N$ equivalent to $M$, and if so, such a dtop $N$ can be constructed.
\end{lemma}
\begin{proof}
By Corollary~\ref{co:assoc2}, it suffices to check for every dtop that is associated with $M$ and $h(M)$,
whether it is equivalent to $M$ (and equivalence is decidable by~\cite{DBLP:journals/actaC/Esik81}; 
see also~\cite[Corollary~19]{DBLP:journals/jcss/EngelfrietMS09}). 
As already observed, $A_{M,\varphi}$ and $\rhs_{M,\varphi}(q,a)$ can be computed from $M$.
Thus, for every subset $Q'$ of $\{q\in S\mid \height(q)\leq h(M)\}$ (of which there are only finitely many),
we can compute the unique dtop $N$ associated with $M$ (and $h(M)$) such that $Q_N=Q'$ (if it exists),
and decide its equivalence to $M$. 
Note that such an $N$ may not exist, either because some $\rhs_{M,\varphi}(q,a)$ 
does not exist (which can be determined by Lemma~\ref{le:rhsexist}) or because some potential state (in $S$) 
that occurs in $A_{M,\varphi}$ or some $\rhs_{M,\varphi}(q,a)$, does not belong to $Q'$. 
\qed
\end{proof}

However, it should be clear that the algorithm in the above proof is inelegant and 
not even feasible for small examples. 
Unfortunately, we do not know whether it is decidable if $\diftup(M)$ is finite, 
and whether it can be computed when it is finite. 
In that case it would suffice to consider $Q'=\diftup(M)$, 
and we would not even need the difference bound $h(M)$. 
Nevertheless, we will show in the remainder of this subsection   
that, with a given difference bound $h(M)$, 
there is a straighforward algorithm to compute the (unique) dtop associated with $M$ and $h(M)$,
if it exists. Thus, there is no need to search the space of subsets of 
$\{q\in S\mid \height(q)\leq h(M)\}$. Moreover, equivalence tests are not needed, which we now prove.  

\begin{lemma}\label{lm:assocequi}
Let $M$ be a canonical dtpla and 
let $N$ be a dtop that is associated with~$M$.
Then $N$ is equivalent to $M$.
\end{lemma}
\begin{proof}
We first show, by structural induction on $s\in\T_\Sigma$, that if $\delta_M(s)=p$, then
$q_N(s)=\varphi(q,p)[\angl{\tilde{q},p}\leftarrow {\tilde{q}}_M(s)\mid \tilde{q}\in Q_M]$
for all $q\in Q_N$, cf. Lemma~\ref{lm:dtladtopprop1}.
Let $s=a(s_1,\dots,s_k)$ with $\delta_M(s_i)=p_i$ for $i\in[k]$ and $\delta_M(a,p_1,\dots,p_k)=p$. 
From Lemma~\ref{lm:elem}(4) we obtain that  
$q_N(s)=\rhs_N(q,a)[\bar{q}(x_i)\leftarrow \bar{q}_N(s_i)\mid\bar{q}\in Q_N, i\in[k]]$. 
By induction, $\bar{q}_N(s_i)=
\varphi(\bar{q},p_i)[\angl{\tilde{q},p_i}\leftarrow {\tilde{q}}_M(s_i)\mid \tilde{q}\in Q_M]$. 
Now consider any node $v$ such that $\rhs_N(q,a)/v=\bar{q}(x_i)$.
By Definition~\ref{def:rhs}(2) and the fact that $\rhs_N(q,a)=\rhs_{M,\varphi}(q,a)$,
we have $\varphi(\bar{q},p_i)=\rhs_{M,\varphi}(q,a,p_1,\dots,p_k)\Omega/v$ 
where $\rhs_{M,\varphi}(q,a,p_1,\dots,p_k)/v$ is in $\T_\Delta(Q_M(\{x_i\}))$. 
Consequently 
\[
\begin{array}{lll}
\bar{q}_N(s_i) & = &
(\rhs_{M,\varphi}(q,a,p_1,\dots,p_k)/v)[\tilde{q}(x_i)\leftarrow \tilde{q}_M(s_i)\mid \tilde{q}\in Q_M] \\
& = &(\rhs_{M,\varphi}(q,a,p_1,\dots,p_k)/v)\Psi_M
\end{array}
\]
where $\Psi_M=[\tilde{q}(x_j)\leftarrow \tilde{q}_M(s_j)\mid \tilde{q}\in Q_M, j\in[k]]$.
Hence $q_N(s)=t\Psi_M$ with
$$t=\rhs_N(q,a)\Phi[\rhs_{M,\varphi}(q,a,p_1,\dots,p_k)/v_1,\dots,\rhs_{M,\varphi}(q,a,p_1,\dots,p_k)/v_r]$$ 
where $v_1,\dots,v_r$ are the occurrences of $\bot$ in $\rhs_N(q,a)\Phi$, from left to right. 
By Definition~\ref{def:rhs}(1), 
$\rhs_N(q,a)\Phi\sqsubseteq\rhs_{M,\varphi}(q,a,p_1,\dots,p_k)$ from which it follows that
$t=\rhs_{M,\varphi}(q,a,p_1,\dots,p_k)$. 
So, $q_N(s)=\rhs_{M,\varphi}(q,a,p_1,\dots,p_k)\Psi_M=
\varphi(q,p)[\angl{\tilde{q},p}\leftarrow\rhs_M(\tilde{q},a,p_1,\dots,p_k)\mid\tilde{q}\in Q_M]\Psi_M=
\varphi(q,p)[\angl{\tilde{q},p}\leftarrow {\tilde{q}}_M(s)\mid \tilde{q}\in Q_M]$
where the last equality is by Lemma~\ref{lm:elem}(4).

We now show with a similar proof that $N(s)=M(s)$ for all $s\in\T_\Sigma$, i.e., that $N$ is equivalent to $M$.
By Lemma~\ref{lm:elem}(3), $N(s)=A_N[q(x_0)\leftarrow q_N(s)\mid q\in Q_N]$ and $M(s)=A_M(p)[\tilde{q}(x_0)\leftarrow \tilde{q}_M(s)\mid \tilde{q}\in Q_M]$ where $p=\delta_M(s)$. 
By the above, $q_N(s)=\varphi(q,p)[\angl{\tilde{q},p}\leftarrow {\tilde{q}}_M(s)\mid \tilde{q}\in Q_M]$.
Now consider any node $v$ such that $A_N/v=q(x_0)$. 
By Definition~\ref{def:A}(2) and the fact that $A_N=A_{M,\varphi}$, we have $\varphi(q,p)=A_M(p)\Omega/v$
and consequently $q_N(s)=(A_M(p)/v)[\tilde{q}(x_0)\leftarrow \tilde{q}_M(s)\mid \tilde{q}\in Q_M]$.
Hence $N(s)=t[\tilde{q}(x_0)\leftarrow \tilde{q}_M(s)\mid \tilde{q}\in Q_M]$ with 
$t=A_N\Phi[A_M(p)/v_1,\dots,A_M(p)/v_r]$
where $v_1,\dots,v_r$ are the occurrences of $\bot$ in $A_N\Phi$, from left to right.
By Definition~\ref{def:A}(1), $A_N\Phi\sqsubseteq A_M(p)$ and so $t=A_M(p)$
and $N(s)=A_M(p)[\tilde{q}(x_0)\leftarrow \tilde{q}_M(s)\mid \tilde{q}\in Q_M]=M(s)$.
\qed
\end{proof}

Thus, by Corollary~\ref{co:assoc2}, $M$ is equivalent to a dtop if and only if 
there exists a dtop that is associated with $M$ and $h(M)$, where $h(M)$ is a difference bound for $M$. 
We now show that such a (reduced) dtop is unique. 

\begin{lemma}\label{le:uniqueassoc}
Let $M$ be a canonical dtpla. 
There exists at most one dtop that is associated with $M$. 
\end{lemma}
 
\begin{proof}
Obviously, it suffices to prove that if $N$ and $N'$ are both associated with $M$,
then $Q_N\subseteq Q_{N'}$. We show that $q\in Q_{N'}$ for every $q\in Q_ N$. 
This trivially follows from the reachability of $q$ in $N$, as follows. 
If $q$ occurs in the axiom $A_N$, 
then it occurs in $A_{N'}$ (because $A_N$ and $A_{N'}$ both equal $A_{M,\varphi}$) 
and hence it is in $Q_{N'}$. 
If $\bar{q}$ occurs in $\rhs_N(q,a)$ and we already know that $q\in Q_{N'}$, then 
$\bar{q}$ occurs in $\rhs_{N'}(q,a)$ (because it equals $\rhs_N(q,a)$) and hence $\bar{q}\in Q_{N'}$.
\qed
\end{proof}

Of course, by Corollary~\ref{co:assoc}, this dtop is the unique canonical dtop $\td(M)$ equivalent to $M$.
The following algorithm finds $\td(M)$ by constructing the unique dtop associated with $M$ 
(and difference bound $h(M)$).

\begin{algorithm}\label{alg:alg}
We now present our main algorithm that, for a given canonical dtpla~$M$ and a given difference bound for $M$, 
decides whether there exists a dtop $N$ equivalent to~$M$, and if so, constructs such a dtop $N$. 
Moreover, the constructed dtop $N$ is canonical.

\noindent
\hrulefill

\medskip
\noindent
Input: 

a canonical dtpla $M=(Q,\Sigma,\Delta,R,A,P,\delta)$ and a difference bound $h(M)$ for~$M$. 

\noindent
Output: 

either the unique canonical dtop $N$ equivalent to $M$, 
or the answer `no' if $M$ is not equivalent to a dtop.

\medskip
Let $Q_N$, $\rhs_N$, $A_N$ be variables 
that determine a dtop $N=(Q_N,\Sigma,\Delta,R_N,A_N)$.
The variable $\rhs_N$ is a $Q_N\times\Sigma$ array with $\rhs_N(q,a)\in \T_\Delta(Q_N(X_k))\cup\{\bot\}$
for every $q\in Q_N$ and $a\in\Sigma^{(k)}$. Then $R_N$ consists of all rules 
$q(a(x_1,\dots,x_k))\to\rhs_N(q,a)$ such that $\rhs_N(q,a)\neq\bot$.

The instruction `abort(no)' causes the algorithm to halt and output `no'. 

\medskip
\noindent
Initialization: 

$A_N:= A_{M,\varphi}$\,; 

$Q_N := \{q\in S\mid q \textrm{ occurs in } A_N\}$\,;

\textbf{if } $\{q\in Q\mid \height(q)>h(M)\}\neq\emptyset$ \textbf{ then } abort(no)\,;

$\rhs_N(q,a):=\bot$ \;for every $(q,a)\in Q_N\times\Sigma$\,;

\medskip
\noindent
Loop:

\textbf{while } $N$ is not complete \textbf{ do}

\quad choose \;$(q,a)\in Q_N\times\Sigma$ \;such that $\rhs_N(q,a)=\bot$\,;

\quad \textbf{if } $\rhs_{M,\varphi}(q,a)$ does not exist \textbf{ then } abort(no)\,;

\quad $\rhs_N(q,a):=\rhs_{M,\varphi}(q,a)$\,;

\quad $Q_N := Q_N\cup\{q\in S\mid q \textrm{ occurs in } \rhs_N(q,a)\}$\,;

\quad \textbf{if } $\{q\in Q\mid \height(q)>h(M)\}\neq\emptyset$ \textbf{ then } abort(no)

\medskip
\noindent
Finalization: 

output $N$.

\noindent
\hrulefill
\end{algorithm}

The next theorem shows the correctness of the algorithm. 
Thus it can be viewed as an improvement of the proof of Lemma~\ref{le:alg}. 

\begin{theorem}\label{th:correct}
Algorithm~\ref{alg:alg} is correct.
\end{theorem}

\begin{proof}
We first observe that the algorithm always halts. If it does not abort, 
then $Q_N$ must remain stable after some time because $Q_N$ is then always a subset of the finite set
$\{q\in S\mid \height(q)\leq h(M)\}$, and then $N$ will be complete after at most $|Q_N\times\Sigma|$
repetitions of the while loop. 

We now show that if the algorithm outputs $N$, then $N$ is associated with $M$, 
and hence $N$ is equivalent to $M$ by Lemma~\ref{lm:assocequi}. 
In fact, it is easy to verify that 
the following statements are invariants of the while loop:
\begin{itemize}
\item $N$ has input and output alphabets $\Sigma$ and $\Delta$,
\item $Q_N\subseteq S$, 
\item $A_N=A_{M,\varphi}$,  
\item $\rhs_N(q,a)=\bot$ or $\rhs_N(q,a)=\rhs_{M,\varphi}(q,a)$ 
       for every $q\in Q_N$ and $a\in\Sigma^{(k)}$, 
\end{itemize}
If $N$ is complete, then these statements show that $N$ is associated with $M$.
This proves that if $M$ is not equivalent to a dtop, then the algorithm aborts and outputs `no'. 

Next we show that if $N_0$ is a dtop that is associated with $M$ and $h(M)$ 
(and hence is \emph{the} dtop associated with $M$ by Lemma~\ref{le:uniqueassoc}),
then the algorithm outputs $N_0$. To see this, we claim that the algorithm does not abort and that 
the statement ``$Q_N\subseteq Q_{N_0}$, $R_N\subseteq R_{N_0}$, and $A_N=A_{N_0}$'' 
is an invariant of the while loop. 
Clearly, after lines~1 and~2 of the~Initialization, $A_N=A_{N_0}$ and hence $Q_N\subseteq Q_{N_0}$,
and so line~3 does not abort because $\height(q)\leq h(M)$ for every $q\in Q_{N_0}$. 
Since $R_N=\emptyset$ after line~4, the invariant holds after Initialization.
A similar argument shows the invariance. If $(q,a)\in Q_N\times\Sigma$, then $q$ is in $Q_{N_0}$
and hence $\rhs_{N_0}(q,a)=\rhs_{M,\varphi}(q,a)$. So, line~2 of the while body does not abort,
line~3 makes $\rhs_N(q,a)$ equal to $\rhs_{N_0}(q,a)$, line~4 updates $Q_N$ with the states in
$\rhs_{N_0}(q,a)$, and hence line~5 does not abort. 
This shows that the algorithm does not abort and hence outputs the dtop $N$. By the previous paragraph,
$N$ is associated with $M$, and hence $N=N_0$ by Lemma~\ref{le:uniqueassoc}. 

If $M$ is equivalent to a dtop, then (as in the proof of Corollary~\ref{co:assoc2}) 
it is equivalent to a canonical dtop by Theorem~\ref{th:uniform_earliest}, and 
hence, by Corollary~\ref{co:assoc}, it is equivalent to a canonical dtop $N_0$ 
that is associated with $M$ and $h(M)$. By the previous paragraph, the algorithm outputs $N_0$, 
which, by Corollary~\ref{co:uniquecandtop}, is the unique canonical dtop equivalent to $M$.
\qed
\end{proof}

One might wonder whether the tests on height in Algorithm~\ref{alg:alg}
are necessary, i.e., whether a difference bound for $M$ is needed at all. 
It is easy to see, by a proof very similar to the one of Theorem~\ref{th:correct},
that, without the two lines that test on height, Algorithm~\ref{alg:alg} outputs the unique 
canonical dtop equivalent to $M$, if $M$ is equivalent to a dtop, 
and otherwise either outputs `no' or does not halt. Thus, although it is not guaranteed to halt, 
it is still a useful algorithm. The existence of such an algorithm is
obvious from the fact that equivalence of dtlas is decidable: one can just enumerate all
dtops $N$ and test the equivalence of $M$ and $N$. 
Obviously, our Algorithm~\ref{alg:alg} (without the tests on height) is more direct.

In the remainder of this subsection we consider some examples of Algorithm~\ref{alg:alg}, 
first three examples in which it returns the answer `no', and then
an example in which it successfully returns a dtop equivalent to the given dtla. 

The first example shows that, without the tests on height, Algorithm~\ref{alg:alg} may indeed not halt. 
In fact, in such a case it can be viewed as computing an \emph{infinite} dtop equivalent to $M$.
Of course, with the tests on height, Algorithm~\ref{alg:alg} outputs `no'. 

\begin{example}\label{ex:alg_non_halt}
Consider the dtla $M$ of Example~\ref{ex:Mex} with $M(\sigma^na)=a$ and 
$M(\sigma^nb)=\sigma^nb$ for every $n\in\Nat$.
It has $P_M=\{p_a,p_b\}$ with $\delta(a)=p_a$, $\delta(b)=p_b$,
$\delta(\sigma,p_a)=p_a$, and $\delta(\sigma,p_b)=p_b$. It has $Q_M=\{q\}$, 
its two axioms are $A_M(p_a)=a$ and $A_M(p_b)=q(x_0)$, 
and its rules are $q(\sigma(x_1\ot p_b))\to \sigma(q(x_1))$ and $q(b)\to b$.
In Example~\ref{ex:diftup} we have seen that $\diftup(M)=\{(a,\sigma^n\angl{q,p_b})\mid n\in\Nat\}$.
It is easy to see that $M$ is canonical. 

We now run Algorithm~\ref{alg:alg} with input $M$, without the tests on height. 
Note that it cannot abort, because all symbols in $\Sigma$ are nullary or unary
(see the remark after Lemma~\ref{le:rhsexist}).  
By Definition~\ref{def:A}, $A_N\Phi=a\sqcap\angl{q,p_b}=\bot$
and so $A_N=q_0(x_0)$ with $\varphi(q_0,p_a)=a$ and $\varphi(q_0,p_b)=\angl{q,p_b}$, i.e., 
$q_0=(a,\angl{q,p_b})$. 
Assume now that the algorithm has constructed the state $q_n$ with 
$\varphi(q_n,p_a)=a$ and $\varphi(q_n,p_b)=\sigma^n\angl{q,p_b}$, i.e., 
$q_n$ is the difference tuple $(a,\sigma^n\angl{q,p_b})$ of $M$.
By item (1) of Definition~\ref{def:rhs}, 
$\rhs_N(q_n,b)=\rhs_{M,\varphi}(q_n,b)=
\varphi(q_n,p_b)[\angl{\bar{q},p_b}\leftarrow\rhs_M(\bar{q},b)\mid\bar{q}\in Q_M]=
\varphi(q_n,p_b)[\angl{q,p_b}\leftarrow b]=\sigma^nb$.
Thus, $N$ has the rule $q_n(b)\to \sigma^nb$. 
Similarly, $\rhs_N(q_n,a)=\rhs_{M,\varphi}(q_n,a)=\varphi(q_n,p_a)=a$ 
and so $N$ has the rule $q_n(a)\to a$. 
Next, we compute $\rhs_N(q_n,\sigma)$. 
To do that we need $\rhs_{M,\varphi}(q_n,\sigma,p)$ for every $p\in P_M$. 
For $p=p_b$ we have 
$\rhs_{M,\varphi}(q_n,\sigma,p_b)=\varphi(q_n,p_b)[\angl{q,p_b}\leftarrow\rhs_M(q,\sigma,p_b)]=
\sigma^n\sigma q(x_1)=\sigma^{n+1}q(x_1)$, and for $p=p_a$ we have 
$\rhs_{M,\varphi}(q_n,\sigma,p_a)=\varphi(q_n,p_a)=a$. Thus, by (1) of Definition~\ref{def:rhs},
$\rhs_N(q_n,\sigma)\Phi=a\sqcap \sigma^{n+1}\angl{q,p_b}=\bot$. 
Hence, $\rhs_N(q_n,\sigma)=q(x_1)$ for some state $q$ 
(because $\sigma\in\Sigma^{(1)}$). 
By (2)(b) of Definition~\ref{def:rhs}, 
$\varphi(q,p_y)= \rhs_{M,\varphi}(q_n,\sigma,p_y)\Omega$ for $y\in\{a,b\}$, and so 
$\varphi(q,p_a)=a$ and $\varphi(q,p_b)=\sigma^{n+1}\angl{q,p_b}$.
In other words, $q=q_{n+1}$ and $N$ has the rule $q_n(\sigma(x_1))\to q_{n+1}(x_1)$. 

This shows that the algorithm does not halt. It can be viewed as constructing the \emph{infinite}
dtop $N$ with $Q_N=\{q_n\mid n\in\Nat\}=\diftup(M)$, $A_N=q_0(x_0)$, and rules
$q_n(a)\to a$, $q_n(b)\to \sigma^nb$, and $q_n(\sigma(x_1))\to q_{n+1}(x_1)$ for every $n\in\Nat$. 
Clearly, this infinite dtop $N$ is equivalent to $M$. 

The dtla $M$ is ultralinear and b-erasing (in fact, linear and nonerasing). 
We will see in the proof of Theorem~\ref{th:ultra} that it has difference bound 
$h(M)=1+4\cdot\maxrhs(M)\cdot(|Q|+2)^2\cdot|P|^2=1+4\cdot 2\cdot 3^2\cdot 2^2=289$. 
So, with this difference bound given, Algorithm~\ref{alg:alg}
(\emph{with} the tests on height) will halt with answer `no' when constructing $q_{290}$. 
\qed
\end{example}

In the next two examples we show that Algorithm~\ref{alg:alg} can abort 
because some potential right-hand side $\rhs_{M,\varphi}(q,a)$ does not exist.
According to Lemma~\ref{le:rhsexist} there can be two reasons for such a failure,
either due to condition~(a) or to condition~(b).
We start with~(a). 

\begin{example}\label{ex:eqtest2}
Let $\Sigma=\{\sigma^{(2)},a^{(0)}\}$ and $\Delta=\{\sigma^{(2)},a^{(1)},e^{(0)}\}$,
and consider the following canonical dtla $M$ that translates every subtree $\sigma(a,s)$ 
and $\sigma(s,a)$ into $a(s)$. 
More precisely, $M(a)=e$, $M(\sigma(a,s))=M(\sigma(s,a))=a(M(s))$ for every $s\in\T_\Sigma$,
and $M(\sigma(s_1,s_2))=\sigma(M(s_1),M(s_2))$ for all $s_i\in\T_\Sigma$ with $\lab(s_i,\varepsilon)=\sigma$. 
Its look-ahead automaton has two states $p_a$ and $p_\sigma$ with $\delta(a)=p_a$ and
$\delta(\sigma,p_y,p_z)=p_\sigma$ for all $y,z\in\{a,\sigma\}$.
It has one state $q$ with $\rho(q)=p_\sigma$, its two axioms are
$A_M(p_a)=e$ and $A_M(p_\sigma)=q(x_0)$, and its four rules are 
\[
\begin{array}{lllllll}
q(\sigma(x_1\ot p_a,x_2\ot p_\sigma)) & \to & a(q(x_2)), & \hspace{0.5cm} &
q(\sigma(x_1\ot p_\sigma,x_2\ot p_\sigma)) & \to & \sigma(q(x_1),q(x_2)), \\
q(\sigma(x_1\ot p_a,x_2\ot p_a)) & \to & a(e), & &
q(\sigma(x_1\ot p_\sigma,x_2\ot p_a)) & \to & a(q(x_1)).
\end{array}
\]
We try to construct a dtop $N$ by Algorithm~\ref{alg:alg} (assuming a large difference bound). 
By Definition~\ref{def:A}, $A_N=q_0(x_0)$ with $q_0=(e,\angl{q,p_\sigma})$ 
where $P_M=\{p_a,p_\sigma\}$.   
By (1) of Definition~\ref{def:rhs}, $\rhs_N(q_0,a)=\rhs_{M,\varphi}(q_0,a)=e$ and so
$N$ has the rule $q_0(a)\to e$. 
To compute $\rhs_N(q_0,\sigma)$ by Definition~\ref{def:rhs}, 
we observe that for $y,z\in\{a,\sigma\}$, we have 
\begin{equation}\label{eq:dropphi}
\rhs_{M,\varphi}(q_0,\sigma,p_y,p_z)=\rhs_M(q,\sigma,p_y,p_z).
\end{equation}
Hence $\rhs_N(q_0,\sigma)\Phi=\rhs_{M,\varphi}(q_0,\sigma)\Phi=\bot$ 
by (1) of Definition~\ref{def:rhs}, 
and so $N$ may have a rule of the form 
$q_0(\sigma(x_1,x_2))\to q'(x_i)$.
However, since both $x_1$ and $x_2$ occur in the right-hand sides of the rules of $M$,
condition~(a) of Lemma~\ref{le:rhsexist} is not satisfied for $v=\varepsilon$, for any $i\in\{1,2\}$.
Thus, $\rhs_N(q_0,\sigma)=\rhs_{M,\varphi}(q_0,\sigma)$ does not exist and 
the algorithm aborts with answer `no'. 
\qed
\end{example}

The next example illustrates the failure of condition~(b) of Lemma~\ref{le:rhsexist}.

\begin{example}\label{ex:eqtest}
Let $\Sigma$ and $\Delta$ be as in the previous example, 
except that $\sigma$ has rank~$1$ in $\Delta$ (instead of rank~$2$). 
We now consider a canonical dtla $M$ that translates every tree 
$\sigma(s_1,\sigma(s_2,\dots,\sigma(s_n,a)\cdots))$ into the tree $r_1r_2\cdots r_ne$, 
where $r_i\in\{a,\sigma\}$ is the root symbol of $s_i$ for $i\in[n]$. 
It is the same as the dtla of the previous example, except that it has the following four rules
\[
\begin{array}{lllllll}
q(\sigma(x_1\ot p_a,x_2\ot p_\sigma)) & \to & a(q(x_2)), & \hspace{0.5cm} &
q(\sigma(x_1\ot p_\sigma,x_2\ot p_\sigma)) & \to & \sigma(q(x_2)), \\
q(\sigma(x_1\ot p_a,x_2\ot p_a)) & \to & a(e), & &
q(\sigma(x_1\ot p_\sigma,x_2\ot p_a)) & \to & \sigma(e).
\end{array}
\]
The construction of the dtop $N$ by Algorithm~\ref{alg:alg} is of course the same as in the 
previous example, including Equation~(\ref{eq:dropphi}), and again $N$ may have a rule of the form 
$q_0(\sigma(x_1,x_2))\to q'(x_i)$.
This time only $x_2$ occurs in the right-hand sides of the rules of $M$, and so 
condition~(a) of Lemma~\ref{le:rhsexist} is satisfied for $v=\varepsilon$ and $i=2$
(and not for $i=1$, because $x_2$ actually occurs). 
However, for $i=2$ (and $v=\varepsilon$), condition~(b) of Lemma~\ref{le:rhsexist} is not satisfied
because $\rhs_{M,\varphi}(q_0,\sigma,p_a,p_\sigma)\Omega=a(\angl{q,p_\sigma})\neq
\sigma(\angl{q,p_\sigma})=\rhs_{M,\varphi}(q_0,\sigma,p_\sigma,p_\sigma)\Omega$.
Thus, the algorithm aborts with answer `no'. 

Note that also $\rhs_{M,\varphi}(q_0,\sigma,p_a,p_a)\Omega=a(e)\neq
\sigma(e)=\rhs_{M,\varphi}(q_0,\sigma,p_\sigma,p_a)\Omega$. 
Thus, the algorithm would also abort if, in the rules, 
$a(q(x_2))$ would be changed into $\sigma(q(x_2))$
(or if $a(e)$ would be changed into $\sigma(e)$). 
\qed
\end{example}

Finally, we give an example in which Algorithm~\ref{alg:alg} successfully
returns a dtop $N$ that is equivalent to $M$. 

\begin{example}
Consider the la-uniform version of the dtla $M$ of Example~\ref{ex:Mleaves}, 
as presented in Example~\ref{ex:la-uniform}. As observed after Example~\ref{ex:la-uniform}, 
it is easy to see that $M$ is canonical. 
Recall that its translation is such that
$M(aa)=aa$, $M(ab)=ab$, $M(ba)=ba$, $M(bb)=bb$, and 
$M(\sigma(s_1,s_2))=\sigma(M(s_1),M(s_2),\#(y,z))$ where 
$y\in\{a,b\}$ is the first letter of the label of the left-most leaf of $\sigma(s_1,s_2)$
and $z\in\{a,b\}$ is the second letter of the label of its right-most leaf.
It has $P_M=\{p_{aa},p_{ab},p_{ba},p_{bb}\}$ with  
$\delta(yz)=p_{yz}$ and 
$\delta(\sigma,p_{wx},p_{yz})=p_{wz}$ for all $w,x,y,z\in\{a,b\}$. 
It has $Q_M=\{q_{yz}\mid y,z\in\{a,b\}\}$, its axioms are $A_M(p_{yz})=q_{yz}(x_0)$,
and its rules are $q_{yz}(yz)\to yz$ and 
$$q_{wz}(\sigma(x_1\ot p_{wx},x_2\ot p_{yz}))\to \sigma(q_{wx}(x_1),q_{yz}(x_2),\#(w,z))$$
for all $w,x,y,z\in\{a,b\}$.
In Example~\ref{ex:diftup} we have seen that 
$\diftup(M)$ consists of the three 4-tuples $(a,a,b,b)$, $\;(a,b,a,b)$, and
$(\angl{q_{aa},p_{aa}},\angl{q_{ab},p_{ab}},\angl{q_{ba},p_{ba}},\angl{q_{bb},p_{bb}})$.

We now run Algorithm~\ref{alg:alg} on $M$, constructing a dtop $N$. 
Since $\maxdiff(M)=0$, the construction is the same for every difference bound $h(M)$. 
By Definition~\ref{def:A}, $A_N=q_0(x_0)$ with $\varphi(q_0,p_{yz})=\angl{q_{yz},p_{yz}}$
for $y,z\in\{a,b\}$. 
So, $q_0$ is the 4-tuple $(\angl{q_{aa},p_{aa}},\angl{q_{ab},p_{ab}},\angl{q_{ba},p_{ba}},$ $\angl{q_{bb},p_{bb}})$.
From (1) of Definition~\ref{def:rhs} we conclude that 
$\rhs_N(q_0,yz)\Phi=\rhs_{M,\varphi}(q_0,yz)=
\varphi(q_0,p_{yz})[\angl{q_{yz},p_{yz}}\leftarrow\rhs_M(q_{yz},yz)]=\rhs_M(q_{yz},yz)=yz$,
and hence $N$ has the rules $q_0(yz)\to yz$ for all $y,z\in\{a,b\}$.
To compute $\rhs_N(q_0,\sigma)$, we first observe that for every $w,x,y,z\in\{a,b\}$, 
\begin{eqnarray}\label{eq:rhsMfi}
\lefteqn{\rhs_{M,\varphi}(q_0,\sigma,p_{wx},p_{yz})} \nonumber \\
& \hspace{0.5cm} = & 
   \varphi(q_0,p_{wz})[\angl{q_{wz},p_{wz}}\leftarrow\rhs_M(q_{wz},\sigma,p_{wx},p_{yz})] \nonumber \\
& \hspace{0.5cm} = & \rhs_M(q_{wz},\sigma,p_{wx},p_{yz}) \nonumber \\
& \hspace{0.5cm} = & \sigma(q_{wx}(x_1),q_{yz}(x_2),\#(w,z)).   
\end{eqnarray}
So, by (1) of Definition~\ref{def:rhs},
$\rhs_N(q_0,\sigma)\Phi=\sigma(\bot,\bot,\#(\bot,\bot))$.
Thus, $N$ may have a rule of the form 
$$q_0(\sigma(x_1,x_2))\to \sigma(q_3(x_{i_3}),q_4(x_{i_4}),\#(q_1(x_{i_1}),q_2(x_{i_2}))).$$
We first (try to) determine $\rhs_N(q_0,\sigma)/v=q_1(x_{i_1})$ with $v=(3,1)$.
By Equation~(\ref{eq:rhsMfi}), 
$\rhs_{M,\varphi}(q_0,\sigma,p_{wx},p_{yz})/v=w$ for all look-ahead states. 
So, (2)(a) of Definition~\ref{def:rhs} is satisfied for both $i_1=1$ and $i_1=2$
(and hence does not determine $i_1$). 
Clearly, since $\rhs_{M,\varphi}(q_0,\sigma,p_{wx},p_{yz})\Omega/v$ 
does not depend on $p_{yz}$, condition~(b) of Lemma~\ref{le:rhsexist} is satisfied for $i_1=1$ 
(and not for $i_1=2$). 
So $i_1=1$, and $\varphi(q_1,p_{wx})=w$ for all $w,x\in\{a,b\}$ 
by (2)(b) of Definition~\ref{def:rhs}. Thus, $q_1=(a,a,b,b)$. 
Similarly we obtain for $v=(3,2)$ that $i_2=2$ and $\varphi(q_2,p_{yz})=z$,
and hence $q_2=(a,b,a,b)$.

Next we (try to) determine $\rhs_N(q_0,\sigma)/v=q_3(x_{i_3})$ with $v=1$.
Since Equation~(\ref{eq:rhsMfi}) shows that $\rhs_{M,\varphi}(q_0,\sigma,p_{wx},p_{yz})/v=q_{wx}(x_1)$,
(2)(a) of Definition~\ref{def:rhs} is satisfied for $i_3=1$, but not for $i_3=2$.
Hence $i_3=1$.
Since $\rhs_{M,\varphi}(q_0,\sigma,p_{wx},p_{yz})\Omega/v=\angl{q_{wx},p_{wx}}$
does not depend on $p_{yz}$, condition~(b) of Lemma~\ref{le:rhsexist} is satisfied for $i_3=1$.
Hence $\varphi(q_1,p_{wx})=\angl{q_{wx},p_{wx}}$ by (2)(b) of Definition~\ref{def:rhs}, 
and so $q_3=q_0$. Similarly we obtain for $v=2$ that $i_4=2$ and $q_4=q_0$. 
Consequently, $N$ has the rule 
$$q_0(\sigma(x_1,x_2))\to \sigma(q_0(x_1),q_0(x_2),\#(q_1(x_1),q_2(x_2))).$$
We now (try to) determine the rules for $q_2=(a,b,a,b)$. Clearly, both $\rhs_{M,\varphi}(q_2,yz)$ and 
$\rhs_{M,\varphi}(q_2,\sigma,p_{wx},p_{yz})$ equal $z$. Thus,
$N$ has the rules $q_2(yz)\to z$ and it may have a rule of the form 
$q_2(\sigma(x_1,x_2))\to q(x_i)$. Let $v=\varepsilon$. 
Clearly, (2)(a) of Definition~\ref{def:rhs} is satisfied for $i\in[2]$ because $z\in\T_\Delta$.
Also, condition~(b) of Lemma~\ref{le:rhsexist} is satisfied for $i=2$ 
because $\rhs_{M,\varphi}(q_2,\sigma,p_{wx},p_{yz})$ does not depend on $p_{wx}$.
So $i=2$, and $\varphi(q,p_{yz})=z$ by (2)(b) of Definition~\ref{def:rhs},
which means that $q=q_2$.
Hence, $N$ has the rule $q_2(\sigma(x_1,x_2))\to q_2(x_2)$. 
Similarly it has the rules $q_1(yz)\to y$ and $q_1(\sigma(x_1,x_2))\to q_1(x_1)$. 

Thus, since $N$ is now complete, Algorithm~\ref{alg:alg} halts and outputs 
the dtop $N$ that has axiom $A_N=q_0(x_0)$ and the rules 
\[
\begin{array}{lll}
q_0(yz)\to yz, & \hspace{0.5cm} & 
   q_0(\sigma(x_1,x_2))\to \sigma(q_0(x_1),q_0(x_2),\#(q_1(x_1),q_2(x_2))), \\
q_1(yz)\to y, & & q_1(\sigma(x_1,x_2))\to q_1(x_1), \\
q_2(yz)\to z, & & q_2(\sigma(x_1,x_2))\to q_2(x_2),
\end{array}
\]
for $y,z\in\{a,b\}$. 
It should be clear that $N$ is indeed equivalent to $M$ (cf. the end of Example~\ref{ex:Mleaves}). 

We finally note that the original dtla $M$ of Example~\ref{ex:Mleaves} is 
ultralinear and b-erasing (even linear and nonerasing). The difference bound in the proof of 
Theorem~\ref{th:ultra} for this $M$ is $1+4\cdot 2\cdot 3^2\cdot 4^2=1153$. 
By Lemma~\ref{lm:make_uniform}, the la-uniform version of $M$ 
has the same difference bound. Obviously, this is a rather large bound, 
in view of the fact that $\maxdiff(M)=0$.
\qed
\end{example}

It is left to the reader to show that for the dtla $M$ of Example~\ref{ex:Mthree}, 
Algorithm~\ref{alg:alg} computes the dtop $N$ of Example~\ref{ex:Mthree}, 
which was used as a running example in Section~\ref{sec:difftup}.

\subsection{The First Main Result}
\label{sec:diffdec}

The next theorem is our first main result, for arbitrary total dtlas. 

\begin{theorem}\label{th:alg}
It is decidable for a given total dtla $M$ and a given difference bound for $M$
whether there exists a dtop $N$ such that $\sem{M}=\sem{N}$, and if so,
such a dtop $N$ can be constructed.
\end{theorem}
\begin{proof}
Let $M$ be a total dtla and let $h(M)$ be a difference bound for $M$. 
We may, of course, assume that $M$ is a dtpla.
By Lemma~\ref{lm:make_uniform}, we may additionally assume that $M$ is la-uniform. 
By Theorem~\ref{th:uniform_earliest}, an equivalent canonical dtpla $\can(M)$ can be constructed
such that, as shown after the theorem, $h(M)+\sumfix(M)$ is a (computable) difference bound for $\can(M)$.
We now apply Lemma~\ref{le:alg} to $\can(M)$ and its difference bound $h(\can(M))=h(M)+\sumfix(M)$. 
\qed
\end{proof}

As an algorithm for this problem, we can use the constructions in the proofs of Lemma~\ref{lm:make_uniform}
and Theorem~\ref{th:uniform_earliest}, followed by Algorithm~\ref{alg:alg}. 
Thus, in case of success, the constructed dtop $N$ is the unique canonical dtop equivalent to $M$,
which we previously denoted by $\td(M)$. 

The algorithm is useful for the designer of a dtla, 
cf. Examples~\ref{ex:Mex}--\ref{ex:Mcounter} in Section~\ref{sec:difftree}.
If you have designed a dtla $M$ to satisfy a specification of its translation $\sem{M}$,
then you usually also know how to specify the output $M(C[p])$, 
for every context $C$ and every look-ahead state $p$.
From this it is probably straightforward for you to obtain a specification of $\diff(M)$.
If $\diff(M)$ is infinite (which you probably also can see), 
then $M$ is not equivalent to a dtop by Corollary~\ref{co:difffin}. 
If it is finite, then you can determine $\maxdiff(M)$ (or an upper bound for it), and hence
you have determined a difference bound for $M$. So now you can use the algorithm of Theorem~\ref{th:alg} 
to find out whether $M$ is equivalent to a dtop, and if so, to construct such a dtop. 
On the other hand, if you are \emph{not} able to determine a difference bound for $M$, 
then, as discussed after Theorem~\ref{th:correct}, you can still use the algorithm of Theorem~\ref{th:alg}, 
without the tests on height in Algorithm~\ref{alg:alg}.
If $M$ is equivalent to a dtop, then the algorithm will construct such a dtop; 
otherwise, the algorithm outputs `no' or may not halt 
(as was shown in Example~\ref{ex:alg_non_halt}).

From Theorem~\ref{th:alg} we immediately obtain the following result.

\begin{corollary}\label{co:goal}
Let $\U$ be a class of total dtlas 
with the following property.
\begin{enumerate}
\item [(H)] There is a computable mapping $h: \U\to\Nat$ such that,  
for every $M\in\U$, \\
$h(M)$ is a difference bound for $M$.
\end{enumerate}
Then it is decidable for a given dtla $M\in\U$ whether 
there exists a dtop $N$ such that $\sem{M}=\sem{N}$, and if so,
such a dtop $N$ can be constructed.
\end{corollary}
\begin{proof}
For a dtla $M$ in $\U$, compute the difference bound $h(M)$ and run the 
algorithm of Theorem~\ref{th:alg}.
\qed
\end{proof}

Let $\U$ be the class of total ultralinear b-erasing dtlas. 
Our goal in Sections~\ref{sec:origins}--\ref{sec:graphs} is to prove that $\U$ has Property~(H),
i.e., to compute a difference bound for the dtlas in the class $\U$.

\section{Links and Origins}
\label{sec:origins}

In this section we define two basic concepts for dtlas, 
and discuss some of their properties.
Thus, this section can be viewed as a sequel to Section~\ref{sec:dtop}.

\medskip
{\bf Convention.} In this section and the next two sections
we assume that $M$ is a dtla that 
is {\bf initialized} with initial states $q_{0,p}$ for $p\in P$
and {\bf la-uniform} with la-map $\rho: Q\to P$,
cf. Lemmas~\ref{lm:make_initialized} and~\ref{lm:make_uniform}. 
Since $M$ is la-uniform, all its initial states $q_{0,p}$ are distinct.

\medskip
If $M(s)=t$, then every node $v$ of the output tree 
$t\in\T_\Delta$ is produced (together with its label)
at a certain node $u$ of the input tree $s\in\T_\Sigma$ by the application
of a rule $q(a(x_1\ot p_1,\dots,x_k\ot p_k))\to \zeta$ at $u$.
The node $v$ is an instantiation of a node $z$ of $\zeta$ with
label in $\Delta$. The triple $(q,u,z)$ will be called 
the ``origin'' of $v$, cf.~\cite{deursen}.

In the special case that $z=\varepsilon$, there exist 
pairs $(q',u')$ for which there is a reachable sentential form $\xi$ for $s$
such that $\xi/v=q'(u')$. In such a case $v$ is already a node of $\xi$, 
but does not yet have a label in $\Delta$. In fact, $(q,u)$ is such a pair,
but there can be more due to the presence of erasing rules: if $(q',u')$ is
such a pair and $\rhs(q',s,u')=q''(x_i)$, then $(q'',u'i)$ is also such a pair.
A pair $(q',u')$ will be called a ``link'' to $v$  (cf.~\cite{DBLP:conf/mol/Maletti11}). 
For convenience it will be denoted as a triple $(q',u',\#)$, and the origin $(q,u,z)$ of $v$ will 
also be called a link to $v$. 

We now formally define links. 
For $s\in\T_\Sigma$ and all $v\in\Nat_+^*$, the sets $\lk_s(v)$ of triples
$(q,u,z)$ with $q\in Q$, $u\in V(s)$, and $z\in\Nat_+^*\cup\{\#\}$, 
are defined recursively to be the smallest sets such that:
\begin{enumerate}
\item[(1)] If $\delta(s)=p$, then $(q_{0,p},\varepsilon,\#)\in\lk_s(\varepsilon)$.
\item[(2)] If $(q,u,\#)\in\lk_s(v)$ and $\rhs(q,s,u)=\zeta$, then 
\begin{enumerate}
\item $(q,u,z)\in\lk_s(vz)$ for every $z\in V_\Delta(\zeta)$, and
\item $(\bar{q},ui,\#)\in\lk_s(vz)$ for every $z\in V(\zeta)$ with $\zeta/z=\bar{q}(x_i)$.
\end{enumerate}
\end{enumerate}
A triple $(q,u,z)$ in $\lk_s(v)$ is called a \emph{link} from $u$ to $v$; 
if $z\in\Nat_+^*$, then it is also called an \emph{origin} of $v$. 
The above intuition about links and origins is proved in the next two lemmas. 
It will follow that $\lk_s(v)$ contains exactly one origin. 

\begin{lemma}\label{lm:linkdef}
Let $s\in\T_\Sigma$, $q\in Q$, $u\in V(s)$, and $z,v\in\Nat_+^*$. Then
\begin{enumerate}
\item[(1)] $(q,u,\#)\in\lk_s(v)$ if and only if there is a sentential form 
$\xi$ of $M$ such that $q_{0,\delta(s)}(\varepsilon)\Rightarrow_s^*\xi$ and $\xi/v=q(u)$, 
\item[(2)] $(q,u,z)\in\lk_s(v)$ if and only if there are $\hat{v}\in\Nat_+^*$ 
and a sentential form $\xi$ of $M$ such that 
$v=\hat{v}z$, $\,q_{0,\delta(s)}(\varepsilon)\Rightarrow_s^*\xi$,  
$\,\xi/\hat{v}=q(u)$, and $z\in V_\Delta(\rhs(q,s,u))$, 
\item[(3)] $\lk_s(v)\neq\emptyset$ if and only if $v\in V(M(s))$, and
\item[(4)] if $(q,u,\#)\in\lk_s(v)$ or $(q,u,z)\in\lk_s(v)$, then $\rho(q)=\delta(s/u)$.
\end{enumerate}
\end{lemma}
\begin{proof}
Since the definition of $\lk_s$ closely follows the semantics of $M$, (1) and~(2) are easy to prove
by recursion induction on the definition of $\lk_s$ in one direction, and by induction on the length 
of the computation $q_{0,\delta(s)}(\varepsilon)\Rightarrow_s^*\xi$ in the other direction. 

Clearly, the right-hand side of~(1) implies that $v\in V(M(s))$, 
by Lemma~\ref{lm:elem}(2) applied to $\xi\Rightarrow_s^* M(s)$,
and the right-hand side of~(2) also implies that $v\in V(M(s))$ because there is a 
computation step $\xi \Rightarrow_s \xi'=\xi[\hat{v}\leftarrow \rhs(q,s,u)[x_i\leftarrow ui\mid i\in\Nat_+]]$
with $v\in V(\xi')$. Thus, if $\lk_s(v)\neq\emptyset$ then $v\in V(M(s))$. 
Now let $v\in V(M(s))$ and consider,  
in a computation of $M$ that translates $s$ into $M(s)$, 
the first sentential form $\xi$ such that $v\in V_{\Delta\cup Q}(\xi)$. 
If $v\in V_Q(\xi)$, then $\lk_s(v)$ contains a triple $(q,u,\#)$ by~(1). 
If $v\in V_\Delta(\xi)$, then $v$ is produced in the computation step $\xi' \Rightarrow_s \xi$ 
where $\xi'$ is the previous sentential form, and then $\lk_s(v)$ contains a triple $(q,u,z)$ 
by~(2) applied to $\xi'$. Thus, $\lk_s(v)\neq\emptyset$.

(4) follows from~(1) and~(2) by Lemma~\ref{lm:prop_la-uniform}(4). 
\qed
\end{proof}

\begin{lemma}\label{lm:linkform}
For every $s\in\T_\Sigma$ and $v\in V(M(s))$, 
either $\lk_s(v)=\{(q,u,z)\}$ with $z\in\Nat_+^*$ and $z\neq\varepsilon$, 
or $\lk_s(v)=\{(q_1,u_1,\#),\dots,(q_n,u_n,\#),(q_n,u_n,\varepsilon)\}$ with $n\geq 1$ and 
for every $j\in[n-1]$ there exists $i_j\in\Nat_+$ such that 
$\rhs(q_j,s,u_j)=q_{j+1}(x_{i_j})$ and $u_{j+1}=u_ji_j$.
\end{lemma}
\begin{proof}
Let us say that a mapping $\lambda$ from $\Nat_+^*$ to the finite subsets of 
$Q\times V(s)\times (\Nat^*\cup\{\#\})$
is an \emph{approximation} of $\lk_s$ if it is obtained by 
a finite number of applications of requirements~(1) and~(2) of the definition of $\lk_s$,
starting with $\lambda(v)=\emptyset$ for all $v\in\Nat_+^*$ 
(and so $\lambda(v)\subseteq\lk_s(v)$ for all $v$).
It is straightforward to show by induction on the number of such applications that 
for every $v\in\Nat_+^*$, either $\lambda(v)=\emptyset$, 
or $\lambda(v)=\{(q,u,z)\}$ with $z\in\Nat_+^*-\{\varepsilon\}$, 
or $\lambda(v)=\{(q_1,u_1,\#),\dots,(q_n,u_n,\#),(q_n,u_n,\varepsilon)\}$ 
with the condition stated in the lemma,
or $\lambda(v)=\{(q_1,u_1,\#),\dots,(q_n,u_n,\#)\}$ with that same condition. 
In the last case, Lemma~\ref{lm:linkdef}(4) implies that $\rhs(q_n,s,u_n)$ is defined, and hence 
$\lambda(v)$ is properly included in $\lk_s(v)$ by the definition of $\lk_s$. 
The first case does not occur when $v\in V(M(s))$, by Lemma~\ref{lm:linkdef}(3).
Since $\lk_s$ is itself an approximation of $\lk_s$, this proves the lemma. 
\qed
\end{proof}

Lemma~\ref{lm:linkform} shows that every node $v$ of $M(s)$ has exactly one origin. 
Thus, for $s\in\T_\Sigma$ and $v\in V(M(s))$, 
we define $\orig_s(v)\in Q\times V(s)\times\Nat_+^*$, 
called \emph{the} origin of $v$, 
by $\orig_s(v)=(q,u,z)$ if $(q,u,z)\in\lk_s(v)$.
We denote the node $u$ of $s$ also by $\orignode_s(v)$ and call it the \emph{origin node} of $v$.

In the remainder of this section we state some elementary properties of links and origins. 

\begin{lemma}\label{lm:orig_prop1}
Let $s\in\T_\Sigma$, $v\in V(M(s))$, $q\in Q$, $u\in V(s)$, and $z\in\Nat_+^*$.
\begin{enumerate}
\item[(1)] If $(q,u,\#)\in\lk_s(v)$, then $M(s)/v=q_M(s/u)$.
\item[(2)] If $\orig_s(v)=(q,u,z)$ and $\zeta=\rhs(q,s,u)$, then $z\in V_\Delta(\zeta)$ and 
\[
M(s)/v=q_M(s/u)/z=\zeta/z[\bar{q}(x_i)\leftarrow \bar{q}_M(s/ui)\mid\bar{q}\in Q,\,i\in\Nat_+].
\]
\end{enumerate}
\end{lemma}
\begin{proof}
If $(q,u,\#)\in\lk_s(v)$ then, by Lemma~\ref{lm:linkdef}(1), 
there is a reachable sentential form $\xi$ for $s$ such that $\xi/v=q(u)$. 
Since $\xi \Rightarrow_s^*M(s)$, we obtain that $\xi/v \Rightarrow_s^*M(s)/v$
and so $M(s)/v=q_M(s/u)$ by Lemma~\ref{lm:elem}(2). 

If $\orig_s(v)=(q,u,z)$ then, by Lemma~\ref{lm:linkdef}(2), there are $\hat{v}\in\Nat^*$ and 
a reachable sentential form $\xi$ for $s$ 
such that $v=\hat{v}z$, $\,\xi/\hat{v}=q(u)$, and $z\in V_\Delta(\zeta)$.
Hence $(q,u,\#)\in\lk_s(\hat{v})$ by Lemma~\ref{lm:linkdef}(1), 
and so $M(s)/\hat{v}=q_M(s/u)$ by the previous paragraph. 
Thus $M(s)/v=q_M(s/u)/z$.
Since $q_M(s/u)=\zeta[\bar{q}(x_i)\leftarrow \bar{q}_M(s/ui)\mid\bar{q}\in Q,\,i\in\Nat_+]$
by Lemma~\ref{lm:elem}(4), this shows the result. 
\qed
\end{proof}

The following four lemmas state relationships between the $\lk$ sets and the ancestor relation. 

Recall that (since $M$ is initialized) 
$\maxrhs(M)$ is the maximal height of the right-hand sides of rules of $M$.

\begin{lemma}\label{lm:link_anc}
Let $s\in\T_\Sigma$ and let $v_i\in V(M(s))$, $q_i\in Q$, $u_i\in V(s)$, and 
$z_i\in\Nat_+^*\cup\{\#\}$ for $i\in[2]$.
If $(q_1,u_1,z_1)\in\lk_s(v_1)$ and $(q_2,u_2,z_2)\in\lk_s(v_2)$, and  
$v_1$ is a proper ancestor of $v_2$, then 
\begin{enumerate} 
\item[(1)] $u_1$ is an ancestor of $u_2$, and 
\item[(2)] if $u_1=u_2$ and $z_1,z_2\in\Nat_+^*$, then $|v_2|-|v_1|\leq\maxrhs(M)$. 
\end{enumerate}
\end{lemma}
\begin{proof}
It is immediate from the definition of $\lk_s$ that 
if $v_1$ is a proper ancestor of $v_2$, then $u_1$ is an ancestor of $u_2$. 
Moreover, if $u_1=u_2=u$ and $z_1,z_2\in\Nat_+^*$, then $q_1=q_2=q$ and 
$z_1,z_2\in V_\Delta(\rhs(q,s,u))$, and 
there exists $v\in\Nat_+^*$ such that $v_1=vz_1$ and $v_2=vz_2$; 
hence $|v_2|-|v_1|=|z_2|-|z_1|\leq\height(\rhs(q,s,u))\leq\maxrhs(M)$. 
\qed
\end{proof}

The next lemma is an immediate corollary of Lemma~\ref{lm:link_anc}.

\begin{lemma}\label{lm:orig_anc}
Let $s\in\T_\Sigma$ and $v,\hat{v}\in V(M(s))$. 
\begin{enumerate}
\item[(1)] If $\hat{v}$ is an ancestor of $v$, 
then $\orignode_s(\hat{v})$ is an ancestor of $\orignode_s(v)$.
\item[(2)] If $\hat{v}$ is an ancestor of $v$ and 
$\orignode_s(\hat{v})=\orignode_s(v)$, 
then $|v|-|\hat{v}|\leq\maxrhs(M)$.
\end{enumerate}
\end{lemma}

\begin{lemma}\label{lm:orig_prop2}
Let $s\in\T_\Sigma$ and $v\in V(M(s))$.
If $\orig_s(v)=(q,u,z)$, then there exists $\hat{v}\in\Nat_+^*$ 
such that $v=\hat{v}z$ and $(q,u,\#)\in\lk_s(\hat{v})$.
\end{lemma}
\begin{proof} Immediate by (2) and (1) of Lemma~\ref{lm:linkdef}. 
\qed
\end{proof}

\begin{lemma}\label{lm:anc_preserve}
Let $s\in\T_\Sigma$ and $v\in V(M(s))$.
If $(q,u,z)\in\lk_s(v)$ and $\hat{u}$ is an ancestor of $u$ in $s$,
then there exist an ancestor $\hat{v}$ of $v$ and a state $q'\in Q$ 
such that $(q',\hat{u},\#)\in\lk_s(\hat{v})$.
\end{lemma}
\begin{proof}
Straightforward by recursion induction on the definition of $\lk_s$, as follows. 
For $(q_{0,\delta(s)},\varepsilon,\#)\in\lk_s(\varepsilon)$ we have $\hat{u}=\varepsilon$ and
we take $\hat{v}=\varepsilon$ and $q'=q_{0,\delta(s)}$. 
Assume now that the statement holds for $(q,u,\#)\in\lk_s(v)$.
For $(q,u,z)\in\lk_s(vz)$ and an ancestor $\hat{u}$ of $u$ we obtain by induction
that $(q',\hat{u},\#)\in\lk_s(\hat{v})$ for an ancestor $\hat{v}$ of $v$, 
which is also an ancestor of $vz$. 
For $(\bar{q},ui,\#)\in\lk_s(vz)$ and an ancestor $\hat{u}$ of $u$ (and hence of $ui$) 
the previous argument also holds; 
for the ancestor $\hat{u}=ui$ of $ui$ we take $\hat{v}=vz$ and $q'=\bar{q}$. 
\qed
\end{proof}

Intuitively, a fact such as $(q,u,\#)\in\lk_s(v)$ does not depend on the whole of $s$, 
but only on the proper ancestors of $u$ and their children. This is proved in the next lemma.

Let $s,s'\in\T_\Sigma$ and $u\in V(s)\cap V(s')$. 
We will say that $u$ is \emph{similar in $s$ and $s'$} if
\begin{enumerate}
\item[$(1)$] $\lab(s,u)=\lab(s',u)$, and 
\item[$(2)$] $\delta(s/ui)=\delta(s'/ui)$ 
for every child $ui$ of $u$.
\end{enumerate}
This implies that $\rhs(q,s,u)=\rhs(q,s',u)$ for every $q\in Q$.

\begin{lemma}\label{lm:link_preserve}
Let $s,s'\in\T_\Sigma$ be such that $\delta(s)=\delta(s')$, and let 
$u\in V(s)\cap V(s')$ be such that every proper ancestor of $u$ is similar in $s$ and $s'$. 
Let $q\in Q$ and $v,z\in\Nat_+^*$.
\begin{enumerate}
\item[(1)] If $(q,u,\#)\in\lk_s(v)$ then $(q,u,\#)\in\lk_{s'}(v)$.
\item[(2)] Let, moreover, $u$ be similar in $s$ and $s'$. \\
If $(q,u,z)\in\lk_s(v)$ then $(q,u,z)\in\lk_{s'}(v)$.
\end{enumerate}
\end{lemma}
\begin{proof}
We prove~(1) by induction on $|u|$. 
By the definition of $\lk_s$, if $(q,\varepsilon,\#)\in\lk_s(v)$, 
then $q=q_{0,\delta(s)}$ and $v=\varepsilon$, 
and so $(q,\varepsilon,\#)\in\lk_{s'}(v)$ by the definition of $\lk_{s'}$.
For the induction step we consider, by the definition of $\lk_s$,
that $(\bar{q},ui,\#)\in\lk_s(vz)$ where 
$(q,u,\#)\in\lk_s(v)$, $\rhs(q,s,u)=\zeta$, and $\zeta/z=\bar{q}(x_i)$.
By induction, $(q,u,\#)\in\lk_{s'}(v)$. 
Since $u$ is similar in $s$ and $s'$, $\rhs(q,s',u)=\zeta$. 
Hence $(\bar{q},ui,\#)\in\lk_{s'}(vz)$ by the definition of $\lk_{s'}$.

To prove~(2) we consider, by the definition of $\lk_s$, 
that $(q,u,z)\in\lk_s(vz)$ where 
$(q,u,\#)\in\lk_s(v)$, $\rhs(q,s,u)=\zeta$, and $z\in V_\Delta(\zeta)$.
By~(1), $(q,u,\#)\in\lk_{s'}(v)$, 
and since $u$ is similar in $s$ and $s'$ by assumption, $\rhs(q,s',u)=\zeta$. 
Hence $(q,u,z)\in\lk_{s'}(vz)$ by the definition of $\lk_{s'}$.
\qed
\end{proof}

Note that in this lemma, by reasons of symmetry, the implications are actually equivalences. 

In the next section we wish to change (in fact, pump) parts of the input tree $s$ 
in such a way that a given node $v$ of $M(s)$ is preserved, together with the labels 
of all its ancestors (cf. Lemma~\ref{lm:diff_char}). Intuitively, this can be done as long as 
we do not change the following in $s$: the origin node $u$ of $v$, 
the labels of all ancestors of $u$, and for each ancestor of $u$ 
the look-ahead states at its children. This is proved in the next lemma.

\begin{lemma}\label{lm:orig_preserve}
Let $s\in\T_\Sigma$ and $v\in V(M(s))$, 
and let $\orignode_s(v)=u\in V(s)$.
Moreover, let $s'\in\T_\Sigma$ be such that $u\in V(s')$
and every ancestor of $u$ (including $u$ itself) is similar in $s$ and $s'$. 
Then $v\in V(M(s'))$ and 
$\lab(M(s'),\hat{v})=\lab(M(s),\hat{v})$ for every ancestor $\hat{v}$ of $v$ 
(including $v$ itself). 
\end{lemma}
\begin{proof}
Since, by Lemma~\ref{lm:orig_anc}(1), $\orignode_s(\hat{v})$ is an ancestor of $u$,  
it suffices to prove the lemma for $\hat{v}=v$. Let $\orig_s(v)=(q,u,z)$.
Since the root $\varepsilon$ is similar in $s$ and $s'$, $\delta(s)=\delta(s')$.
It now follows from Lemma~\ref{lm:link_preserve}(2) that $v\in V(M(s'))$ and
$\orig_{s'}(v)=(q,u,z)=\orig_s(v)$. 
Using this, and the fact that $\rhs(q,s',u)=\rhs(q,s,u)$ (because $u$ is similar in $s$ and $s'$),
we obtain from Lemma~\ref{lm:orig_prop1}(2) that $\lab(M(s'),v)=\lab(M(s),v)$. 
\qed
\end{proof}


\section{Auxiliary Bounds}
\label{sec:upper}

If $M$ is an initialized la-uniform dtla, then so is $M^\circ$, 
cf. Lemma~\ref{lm:prop_la-uniform}(3). 
In this section and the next, 
the lemmas of the previous section will be applied to $M^\circ$ instead of $M$.
 
We recall that $\maxrhs(M)$ is the maximal height of the right-hand sides of rules of~$M$.

\medskip
In view of Theorem~\ref{th:alg} and Corollary~\ref{co:goal}, 
we wish to compute a difference bound for $M$, 
i.e., an upper bound for $\maxdiff(M)$ 
when $\diff(M)$ is finite. Thus, we are looking for an upper bound
on the height of all difference trees of $M$, i.e., 
all trees $M(C[p])/v$ where $C$ is a $\Sigma$-context, $p\in P$, and 
$v$ is a difference node of $M(C[p])$ and $M(C[p'])$ where $p'\in P$.
Let $u$ and $u'$ be the respective origin nodes of $v$, 
i.e., $u=\orignode_{C[p]}(v)$ and $u'=\orignode_{C[p']}(v)$. 

We first compute an upper bound for the case where $u$ is \emph{not} 
a proper ancestor of $u'$. The idea is that if the height of $M(C[p])/v$ is larger than 
that upper bound, then we can pump the subtree $C[p]/uj$ at one of $u$'s children $uj$
(without changing the look-ahead state at $uj$ in both $C[p]$ and $C[p']$), 
thus turning $C$ into $\bar{C}$, 
in such a way that $M(\bar{C}[p])/v$ becomes
arbitrarily large. Since the pumping does not change the labels of the ancestors of $u$ and $u'$, 
nor the look-ahead states at the children of those ancestors,  
$v$ is still a difference node of $M(\bar{C}[p])$ and $M(\bar{C}[p'])$
by Lemmas~\ref{lm:diff_char} and~\ref{lm:orig_preserve}.

To express an upper bound for the height of $M(C[p])/v$, 
we use an auxiliary bound defined as follows.

\begin{definition}\label{df:output_bound}
A number $h_{\rm o}(M)\in \Nat$ is an \emph{output bound} for $M$ 
if it has the following two properties, for every $q\in Q$ and $p,p',p_1,p'_1\in P$:
\begin{enumerate}
\item[(1)] if the set $\{q_M(s)\mid s\in\T_\Sigma,\,\delta(s)=p_1\}$ is finite, 
then $\height(t)\leq h_{\rm o}(M)$ for every tree~$t$ in this set, 
and
\item[(2)] 
if the set $\{q_M(C[p])\mid C\in\C_\Sigma,\,\delta(C[p])=p_1,\,\delta(C[p'])=p'_1\}$ is finite, 
then $\height(t)\leq h_{\rm o}(M)$ for every tree $t$ in this set. 
\end{enumerate}
\end{definition}
Note that since $M$ is la-uniform, it suffices to consider the case where $p_1=\rho(q)$.
Condition~(1) of this definition is needed for the case where the subtree $C[p]/uj$ discussed above
is in $\T_\Sigma$ (i.e., does not contain $p$), whereas Condition~(2) is needed 
when $C[p]/uj$ is of the form $D[p]$ for a context $D\in\C_\Sigma$ (i.e., contains $p$). 
The restrictions on $\delta$ in the sets of Conditions~(1) and~(2) are needed to ensure that 
the pumping of $C[p]/uj$ preserves the look-ahead state of $uj$ in $C[p]$ and $C[p']$. 

The minimal output bound $h_{\rm o}(M)$ can be computed from $M$, because 
every set mentioned in Definition~\ref{df:output_bound} is the image of a regular tree language
by a dtla translation (as will be shown in the proof of Theorem~\ref{th:output_bound}), 
and because it is decidable whether or not such an image is finite, and if so,
the elements of that image can be computed (see~\cite[Theorem~4.5]{DBLP:journals/iandc/DrewesE98} and
note that every dtla translation can be realized by a macro tree transducer).
Also, for completeness sake, we will prove by a pumping argument 
that $\maxrhs(M)\cdot|Q|\cdot(|P|+2)$ is an output bound for $M$, 
in Lemma~\ref{lm:range_bound} and Theorem~\ref{th:output_bound}. 

We now show that the upper bound discussed above is $\maxrhs(M)+h_{\rm o}(M)$.
For later use (in the proof of Lemma~\ref{lm:new_prop_ancestor}) 
we prove a slightly more general result; the case discussed above is obtained
by taking $\bar{v}=v$.

\begin{lemma}\label{lm:new_not_ancestor}
Let $h_{\rm o}(M)$ be an output bound for $M$. 
Let $C\in\C_\Sigma$ and $p,p'\in P$, let $v$ be a difference node of $M(C[p])$ and $M(C[p'])$, 
and let $\bar{v}$ be a descendant of $v$ in $M(C[p])$ such that
$\orignode_{C[p]}(\bar{v})$ is not a proper ancestor of $\orignode_{C[p']}(v)$.
If $\diff(M)$ is finite, then 
$$\height(M(C[p])/\bar{v})\leq \maxrhs(M)+h_{\rm o}(M).$$
\end{lemma}
\begin{proof}
Assume that $\height(M(C[p])/\bar{v})>\maxrhs(M)+h_{\rm o}(M)$.
Let $u=\orignode_{C[p]}(v)$ and $u'=\orignode_{C[p']}(v)$, 
and let $(q,\bar{u},z)=\orig_{C[p]}(\bar{v})$.
By Lemma~\ref{lm:orig_anc}(1), $u$ is an ancestor of $\bar{u}$. 
By Lemma~\ref{lm:orig_prop1}(2), 
$M(C[p])/\bar{v}=\zeta/z[\bar{q}(x_i)\leftarrow \bar{q}_M(C[p]/\bar{u}i)\mid\bar{q}\in Q,\,i\in\Nat_+]$
where $\zeta=\rhs(q,C[p],\bar{u})$.  
Since $\height(M(C[p])/\bar{v})>\maxrhs(M)+h_{\rm o}(M)$, 
there exist $y\in\Nat_+^*$, $q'\in Q$, and $j\in\Nat$ such that 
$\zeta/zy=q'(x_j)$ and $\height(q'_M(C[p]/\bar{u}j))> h_{\rm o}(M)$.
Note that $M(C[p])/\bar{v}y=q'_M(C[p]/\bar{u}j)$.
Note also that, by the definition of $\lk$, $(q',\bar{u}j,\#)\in\lk_{C[p]}(\bar{v}y)$.
The situation is shown in Figure~\ref{fi:notanc} 
(but note that it is also possible that $u$ is an ancestor of $u'$, 
as in Figure~\ref{fi:anc}).
\begin{figure}
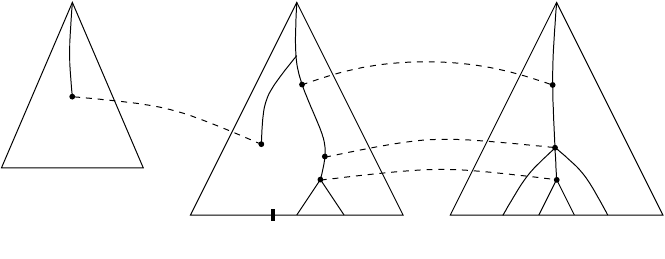
\caption{\quad $\bar{u}=\orignode_{C[p]}(\bar{v})$ is not a proper ancestor of $u'=\orignode_{C[p']}(v)$
\label{fi:notanc}}
\end{figure}%

\noindent
There are now two cases: $p$ does or does not occur in $C[p]/\bar{u}j$.
Let $p_j=\delta(C[p]/\bar{u}j)$.

\emph{Case 1}: $p$ does not occur in $C[p]/\bar{u}j$, 
i.e., $C[p]/\bar{u}j=C/\bar{u}j\in\T_\Sigma$. 
In Figure~\ref{fi:notanc}, the vertical thick line represents the node of $C$ with label $\bot$; 
so, it shows Case~1.
By Definition~\ref{df:output_bound}(1) of $h_{\rm o}(M)$, the set 
$\{q'_M(s)\mid s\in\T_\Sigma,\,\delta(s)=p_j\}$ is infinite.  
Hence there exist trees $s_n\in\T_\Sigma$ such that $\delta(s_n)=p_j$ and 
$\height(q'_M(s_n))>n$ for every $n\in\Nat$.
Let $C_n=C[\bar{u}j\leftarrow s_n]$, the context $C$ 
in which the subtree at $\bar{u}j$ is replaced by $s_n$.
Since $\delta(s_n)=\delta(C[p]/\bar{u}j)=\delta(C[p']/\bar{u}j)$ and
$\bar{u}$ is not a proper ancestor of $u$ or $u'$, 
it follows from Lemma~\ref{lm:orig_preserve} that 
$\lab(M(C_n[p]),\hat{v})=\lab(M(C[p]),\hat{v})$ for every ancestor $\hat{v}$ of $v$, 
and similarly for $p'$. 
Hence, by Lemma~\ref{lm:diff_char}, 
$v$ is a difference node of $M(C_n[p])$ and $M(C_n[p'])$,
and so $M(C_n[p])/v$ is in $\diff(M)$.
Also, since $(q',\bar{u}j,\#)\in\lk_{C[p]}(\bar{v}y)$, 
we obtain from Lemma~\ref{lm:link_preserve}(1) that $(q',\bar{u}j,\#)\in\lk_{C_n[p]}(\bar{v}y)$.
Consequently, by Lemma~\ref{lm:orig_prop1}(1), 
$M(C_n[p])/\bar{v}y=q'_M(C_n[p]/\bar{u}j)=q'_M(s_n)$, which implies that 
$\height(M(C_n[p])/v)\geq\height(M(C_n[p])/\bar{v}y) >n$. Hence $\diff(M)$ is infinite.

\emph{Case 2}: $p$ occurs in $C[p]/\bar{u}j$, 
i.e., $C[p]/\bar{u}j=D[p]$ where $D=C/\bar{u}j\in\C_\Sigma$. 
Let $p_j'=\delta(C[p']/\bar{u}j)$.
By Definition~\ref{df:output_bound}(2) of $h_{\rm o}(M)$, the set 
$\{q'_M(D[p])\mid D\in\C_\Sigma, \,\delta(D[p])=p_j, \,\delta(D[p'])=p_j'\}$ is infinite.
Hence there exist contexts $D_n$ such that 
$\delta(D_n[p])=p_j$, $\delta(D_n[p'])=p_j'$, and $\height(q'_M(D_n[p]))>n$.
Let $C_n=C[\bar{u}j\leftarrow D_n]$.
It can now be shown in the same way as in Case~1 that 
$M(C_n[p])/v$ is in $\diff(M)$ and $\height(M(C_n[p])/v)>n$,
and hence $\diff(M)$ is infinite.
Note that the condition $\delta(D_n[p'])=p_j'$ is needed to ensure that
Lemma~\ref{lm:orig_preserve} is applicable to $u'$. 
\qed
\end{proof}

We now consider the case where $u$ \emph{is} a proper ancestor of $u'$, 
and again wish to obtain an upper bound for the height of $M(C[p])/v$.
It follows from Lemma~\ref{lm:new_not_ancestor} that 
$\height(M(C[p])/\bar{v})\leq \maxrhs(M)+h_{\rm o}(M)$ 
for every descendant $\bar{v}$ of $v$ 
of which the origin node $\bar{u}$ is not a proper ancestor of $u'$. 
But what about a descendant $vw$ of $v$ of which the origin node $x$ 
\emph{is} a proper ancestor of $u'$? (see Figure~\ref{fi:anc} with $w=w_1$).
Then, by the previous observation, it suffices to obtain an upper bound on $|w|$.
Now we observe that, roughly speaking, when $M$ arrives at node $x$ of $C[p]$ 
it generates node $vw$ of $M(C[p])$, which is a descendant of $v$, 
but when $M$ arrives at node $x$ of $C[p']$ it generates 
a node $\hat{v}$ of $M(C[p'])$ that is an ancestor of $v$. 
Hence, it suffices to have an upper bound on $|vw|-|\hat{v}|$, 
which measures how much the translation of $M(C[p])$ is ahead of 
the translation of $M(C[p'])$ when $M$ arrives at $x$.
Note that by Lemma~\ref{lm:diff_char}, 
$\lab(M(C[p]),\tilde{v})=\lab(M(C[p']),\tilde{v})$ 
for every proper ancestor $\tilde{v}$ of $\hat{v}$.

Thus, to express an upper bound for the height of $M(C[p])/v$, 
we use an auxiliary bound defined as follows. 

\begin{definition}\label{df:ancestral_bound}
A number $h_{\rm a}(M)\in\Nat$ is an \emph{ancestral bound} for $M$ if
$\diff(M)$ is infinite or the following holds 
for every $C\in\C_\Sigma$, $p,p'\in P$,
$x\in V(C)$, $y\in V(M(C[p]))$, $y'\in V(M(C[p']))$, and $q,q'\in Q$:
 
\medskip
\noindent
if 
\begin{enumerate}
\item[(1)] $(q,x,\#)\in\lk_{C[p]}(y)$ and $\;(q',x,\#)\in\lk_{C[p']}(y')$, 
\item[(2)] $y'$ is a prefix of $y$, hence an ancestor of $y$ in $M(C[p])$, and
\item[(3)] $\lab(M(C[p]),\hat{y}')=\lab(M(C[p']),\hat{y}')$ for every proper ancestor $\hat{y}'$ of $y'$,
\end{enumerate}
then $|y|-|y'|\leq h_{\rm a}(M)$. 
\end{definition}

We will show in the next section (Lemma~\ref{lm:no_split}) that every dtla $M$ 
has an ancestral bound $h_{\rm a}(M)$.
Unfortunately, we do not know whether an ancestral bound can be computed from $M$. 
We will also show in the next section (Theorem~\ref{th:ancestral_bound}) that it can be computed 
in the restricted case where $M$ is ultralinear and b-erasing. 

\begin{lemma}\label{lm:new_prop_ancestor}
Let $h_{\rm o}(M)$ be an output bound and $h_{\rm a}(M)$ an ancestral bound for $M$. 
Let $C\in\C_\Sigma$ and $p,p'\in P$, and let $v$ be a difference node of $M(C[p])$ and $M(C[p'])$ 
such that $\orignode_{C[p]}(v)$ is a proper ancestor of $\orignode_{C[p']}(v)$.
If $\diff(M)$ is finite, then 
$$\height(M(C[p])/v)\leq 2\cdot\maxrhs(M)+h_{\rm o}(M)+h_{\rm a}(M)+1.$$
\end{lemma}
\begin{proof}
Let $\orignode_{C[p]}(v)=u$ and $\orignode_{C[p']}(v)=u'$.
Consider an arbitrary leaf $w$ of $M(C[p])/v$.
We have to show that $|w|\leq 2\cdot\maxrhs(M)+h_{\rm o}(M)+h_{\rm a}(M)+1$.
The proof is illustrated in Figure~\ref{fi:anc}.
\begin{figure}
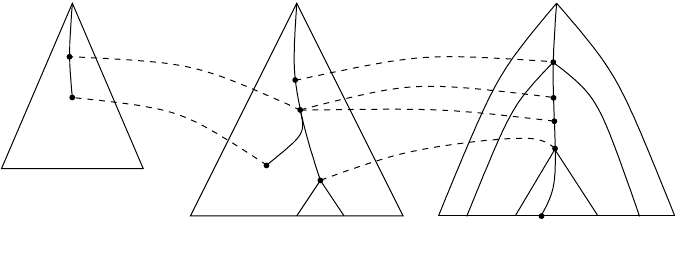
\caption{\quad $u=\orignode_{C[p]}(v)$ is a proper ancestor of $u'=\orignode_{C[p']}(v)$
\label{fi:anc}}
\end{figure}%

Let $w=w_1w_2$ where $w_1$ is the longest ancestor of $w$ such that 
$\orignode_{C[p]}(vw_1)$ is a proper ancestor of $u'$.
If $w_2=jw_2'$ with $j\in\Nat_+$, then $\orignode_{C[p]}(vw_1j)$ is not a proper ancestor of $u'$, 
and so $|w_2|=|w_2'|+1\leq \maxrhs(M)+h_{\rm o}(M)+1$ 
by Lemma~\ref{lm:new_not_ancestor} (with $\bar{v}=vw_1j$).
It now suffices to prove that $|w_1|\leq\maxrhs(M)+h_{\rm a}(M)$.

Let $\orig_{C[p]}(vw_1)=(q,x,z)$, and 
note that $x$ is a descendant of $u$ by Lemma~\ref{lm:orig_anc}(1)
(and a proper ancestor of $u'$ by definition of $w_1$).
If $x=u$, then $|w_1|\leq\maxrhs(M)$ by Lemma~\ref{lm:orig_anc}(2).
Otherwise, by Lemma~\ref{lm:orig_prop2} (applied to $vw_1$) and Lemma~\ref{lm:link_anc}(1), 
$w_1=w_1'z$ such that $(q,x,\#)\in\lk_{C[p]}(vw_1')$.
Since $|z|\leq\maxrhs(M)$, 
it now remains to show that $|w_1'|\leq h_{\rm a}(M)$.

By Lemma~\ref{lm:anc_preserve}, since $x$ is an ancestor of $u'$, 
there is an ancestor $\hat{v}$ of $v$
such that $(q',x,\#)\in\lk_{C[p']}(\hat{v})$ for some $q'\in Q$.
We now apply Definition~\ref{df:ancestral_bound} of $h_{\rm a}(M)$ 
to the nodes $y=vw_1'$ of $M(C[p])$ and $y'=\hat{v}$ of $M(C[p'])$, and obtain that 
$|w_1'|\leq |vw_1'|-|\hat{v}|\leq h_{\rm a}(M)$. 
Note that condition~(3) in Definition~\ref{df:ancestral_bound} is satisfied 
by Lemma~\ref{lm:diff_char}, because $\hat{v}$ is an ancestor of the 
difference node $v$ of $M(C[p])$ and $M(C[p'])$. 
\qed
\end{proof}

\begin{theorem}\label{th:upper_bound}
If $h_{\rm o}(M)$ is an output bound and $h_{\rm a}(M)$ an ancestral bound for $M$,
then $h(M)=2\cdot\maxrhs(M)+h_{\rm o}(M)+h_{\rm a}(M)+1$ is a difference bound for $M$.
\end{theorem}
\begin{proof}
Immediate from Lemmas~\ref{lm:new_not_ancestor} (for $\bar{v}=v$) 
and~\ref{lm:new_prop_ancestor}.
\qed
\end{proof}

We end this section by proving that every dtla $M$ has a computable output bound 
(Theorem~\ref{th:output_bound}). 
The reader who believes that this follows from \cite[Theorem~4.5]{DBLP:journals/iandc/DrewesE98}
can skip the rest of this section. However, the proof of the next lemma can also serve 
as an introduction to the pumping technique used in Section~\ref{sec:graphs} for computing an
ancestral bound for $M$. 

\begin{lemma}\label{lm:range_bound}
Let $p\in P$ such that the set $\{M(s)\mid s\in\sem{p}\}$ is finite.
Then $\height(M(s))\leq \maxrhs(M)\cdot|Q|$ for every $s\in\sem{p}$.
\end{lemma}

\begin{proof}
The (obvious) idea for the proof is that we consider an output path from the root to a leaf $v$ in $M(s)$ 
that is longer than $\maxrhs(M)\cdot|Q|$. Then we find two ancestors $\hat{u}$ and $\bar{u}$
of the origin node $u$ of $v$ where the computation of $M$ on $s$ is in a cycle 
(i.e., arrives in the same state at $\hat{u}$ and $\bar{u}$) and 
produces a nonempty part of the output path.  
And then we pump the part of $s$ between $\hat{u}$ and $\bar{u}$, 
thus pumping the path to $v$ in $M(s)$ and obtaining arbitrarily long output paths. 
For a formal proof it is convenient to build a graph $G^{\rm o}_M$ of which each path
corresponds to a pair $(u,v)$ and a state $q$ such that there is a sentential form $\xi$ with 
$q_{0,p}\Rightarrow^*_s \xi$ and $\xi/v=q(u)$, i.e., such that $(q,u,\#)\in\lk_s(v)$, see
Lemma~\ref{lm:linkdef}(1).
The pumping of $s$ then corresponds to the repetition of a cycle in $G^{\rm o}_M$. 
For the notion of a directed edge-labeled graph see Section~\ref{sec:prelim}.

We construct a directed edge-labeled \emph{dependency graph} $G^{\rm o}_M$ 
with node set $Q$ and 
an edge with label $(j,z)$ from $q$ to $q'$ if there is a rule
$q(a(x_1\ot p_{i,1},\ldots,x_k\ot p_{i,k}))\to \zeta$ in $R$ such that 
$\;z\in V(\zeta)$ and $\zeta/z=q'(x_j)$.
Note that $|z|\leq \maxrhs(M)$.
Each path $e_1\cdots e_n$ in $G^{\rm o}_M$ (where each $e_i$ is an edge) has 
a label in $\Nat_+^*\times\Nat_+^*$, obtained by the 
component-wise concatenation of the labels of $e_1,\dots,e_n$. 

\smallskip
{\bf Claim.} There is a path $\pi$ from $q_{0,p}$ to $q$ with label $(u,v)$ in $G^{\rm o}_M$ 
if and only if
there is a tree $s\in\sem{p}$ such that $(q,u,\#)\in\lk_s(v)$.

\smallskip
Proof of Claim. ($\Rightarrow$) We proceed by induction on the length of $\pi$. 
If $\pi$ is empty, then $q=q_{0,p}$ and $(u,v)=(\varepsilon,\varepsilon)$, 
and so $(q,u,\#)\in\lk_s(v)$ for every $s\in\sem{p}$ (by the definition of $\lk$).
Now let $\pi'=\pi e$ where $\pi$ is a path from $q_{0,p}$ to $q$ with label $(u,v)$,
and $e$ is an edge from $q$ to $q'$ with label $(j,z)$. Thus, $\pi'$ is a path from 
$q_{0,p}$ to $q'$ with label $(uj,vz)$. 
By the definition of $G^{\rm o}_M$, the edge $e$ is obtained from a rule 
$q(a(x_1\ot p_1,\ldots,x_k\ot p_k))\to \zeta$ such that 
$z\in V(\zeta)$ and $\zeta/z=q'(x_j)$.
By induction, $(q,u,\#)\in\lk_s(v)$. Choose $s_m\in\sem{p_m}$ for $m\in[k]$, and 
let $s'=s[u\leftarrow a(s_1,\dots,s_k)]$. 
By Lemma~\ref{lm:linkdef}(4), we have $\delta(s/u)=\rho(q)$ and so 
$\delta(s/u)=\delta(a(s_1,\dots,s_k))$ because $M$ is la-uniform. 
Hence $\delta(s)=\delta(s')$ and every proper ancestor of $u$ is similar in $s$ and $s'$.
Consequently $(q,u,\#)\in\lk_{s'}(v)$ by Lemma~\ref{lm:link_preserve}(1).
Since $\rhs(q,s',u)=\zeta$, we obtain that $(q',uj,\#)\in\lk_{s'}(vz)$ from the definition of $\lk$.

($\Leftarrow$) We proceed by induction on the length of $u$.
If $u=\varepsilon$, then $q=q_{0,p}$ and $v=\varepsilon$ by the definition of $\lk$,
and so the empty path satisfies the conditions. 
Now let $u'=uj$ with $u\in\Nat_+^*$ and $j\in\Nat_+$, 
and let $s\in\sem{p}$ such that $(q',u',\#)\in\lk_s(v')$.
By the definition of $\lk$ there is a rule 
$q(a(x_1\ot p_1,\ldots,x_k\ot p_k))\to \zeta$ such that 
$a=\lab(s,u)$ and $p_m=\delta(s,um)$ for every $m\in[k]$; moreover, 
there exist $v\in\Nat_+^*$ and $z\in V(\zeta)$ such that $v'=vz$, 
$(q,u,\#)\in\lk_s(v)$, and $\zeta/z=q'(x_j)$.
By induction, there is a path $\pi$ from $q_{0,p}$ to $q$ with label $(u,v)$.
By the definition of $G^{\rm o}_M$ there is an edge $e$ from $q$ to $q'$ with label $(j,z)$. 
Hence $\pi e$ is a path from $q_{0,p}$ to $q'$ with label $(uj,vz)=(u',v')$.
This ends the proof of the Claim.

\smallskip
Suppose that $\pi_0$ is a path from $q_{0,p}$ to $q$ with label $(u_0,v_0)$, and that 
$\pi$ is a cycle from $q$ to itself with label $(u,v)$ such that $v\neq \varepsilon$.
Consider the ``pumped'' path $\pi_0\pi^k$ from $q_{0,p}$ to $q$, for any $k\geq 1$. 
By the Claim, there exists a tree $s_k\in\sem{p}$ such that
$(q,u_0u^k,\#)\in\lk_{s_k}(v_0v^k)$. 
Since $v_0v^k\in V(M(s_k))$ and $|v_0v^k|\geq k$, we obtain that $\height(M(s_k))\geq k$. 
That contradicts the fact that $\{M(s)\mid s\in\sem{p}\}$ is finite.

Now consider a tree $s\in\sem{p}$ and a leaf $v$ of $M(s)$. 
Let $\orig_s(v)=(q,u,z)$. 
Then, by Lemma~\ref{lm:orig_prop2},
there exists an ancestor $\hat{v}$ of $v$ such that 
$(q,u,\#)\in\lk_s(\hat{v})$ and $v=\hat{v}z$.
By the Claim, there is a path $\pi$ from $q_{0,p}$ to $q$ with label $(u,\hat{v})$.
As shown above, if $\pi$ contains a cycle, then its output label is empty. Hence
there is a path $\pi_1$ without cycles from $q_{0,p}$ to $q$ with label $(u_1,\hat{v})$.
Then $\pi_1$ has at most $|Q|-1$ edges. Since the second component of the label of
every edge has length at most $\maxrhs(M)$, we obtain that 
$|\hat{v}|\leq \maxrhs(M)\cdot(|Q|-1)$. 
Hence $|v|=|\hat{v}|+|z|\leq |\hat{v}|+\maxrhs(M)\leq \maxrhs(M)\cdot|Q|$.
\qed
\end{proof}

\begin{theorem}\label{th:output_bound}
The number $\maxrhs(M)\cdot|Q|\cdot(|P|+2)$ is an output bound for $M$.
\end{theorem}

\begin{proof}
For a set $\{q_M(s)\mid s\in\T_\Sigma,\,\delta(s)=p_1\}$ 
as in Definition~\ref{df:output_bound}(1), with $\rho(q)=p_1$, 
we construct the dtla $M_1$ from $M$ by changing its $p_1$-axiom into $q(x_0)$. 
Application of Lemma~\ref{lm:range_bound} to $M_1$ and $p_1$ gives the upper bound 
$\maxrhs(M)\cdot|Q|$ on the height of the trees in the given set. 

For a set $\{q_M(C[p])\mid C\in\C_\Sigma,\,\delta(C[p])=p_1,\,\delta(C[p'])=p'_1\}$ 
as in Definition~\ref{df:output_bound}(2), with $\rho(q)=p_1$,
we first change $M$ into $M_1$ as before. It remains to consider the set
$\{M_1(C[p])\mid C\in\C_\Sigma,\,\delta(C[p'])=p'_1\}$.
In order to apply Lemma~\ref{lm:range_bound}, we modify the dtla $M_1^\circ$.
We use a simple product construction to obtain from $M_1^\circ$ an equivalent dtla $M_2$ that 
``recognizes'' input trees of the form $C[p]$ and computes $\delta(C[p'])$ by additional look-ahead.
The dtla $M_2$ has the same input alphabet $\Sigma\cup P$ and 
output alphabet $\Delta\cup(Q\times P)$ as $M^\circ$ and $M_1^\circ$. 
It has the same set $Q$ of states as $M$ and $M_1$, 
and the set of look-ahead states $P\times(P\cup\{1,0\})$.
Its transition function $\bar{\delta}$ is constructed in such a way that 
$\bar{\delta}(s)=(\bar{p},\bar{p}')$ for $s\in\T_\Sigma(P)$, where $\bar{p}$ and $\bar{p}'$
are given as follows:
$\bar{p}=\delta(s)$, 
$\bar{p}'=1$ if $s\in\T_\Sigma$,
$\bar{p}'=0$ if $s$ contains more than one occurrence of $p$ 
or at least one occurrence of an element of $P-\{p\}$, and 
$\bar{p}'=\delta(C[p'])$ if $s=C[p]$ for some $C\in\C_\Sigma$.
We leave the easy construction of $\bar{\delta}$ to the reader.
The rules of $M_2$ are defined by: 
$\rhs_{M_2}(\bar{q},a,(\bar{p}_1,\bar{p}_1'),\dots,(\bar{p}_k,\bar{p}_k'))=
\rhs_{M_1^\circ}(\bar{q},a,\bar{p}_1,\dots,\bar{p}_k)$, 
and the $(\bar{p},\bar{p}')$-axiom of $M_2$ is the $\bar{p}$-axiom of $M_1$. 
Thus, $M_2$ ignores the additional look-ahead and hence is equivalent to $M_1^\circ$. 
Note that $M_2$ is initialized, but need not be la-uniform. 
We finally turn $M_2$ into an equivalent initialized la-uniform dtla $M_3$ 
by an obvious variant of the construction in the proof of Lemma~\ref{lm:make_uniform}, 
such that $M_3$ has state set $Q_3=Q\times(P\cup\{1,0\})$ 
and la-map $\rho_3$ with $\rho_3(\angl{\bar{q},\bar{p}'})=\rho(\bar{q})$. 
We also leave this construction to the reader.
Note that $\maxrhs(M_3)=\maxrhs(M)$. 
Application of Lemma~\ref{lm:range_bound} to $M_3$ and $(p_1,p_1')$ gives the upper bound 
$\maxrhs(M_3)\cdot|Q_3|=\maxrhs(M)\cdot|Q|\cdot(|P|+2)$ on the height of the trees in the given set. 
\qed
\end{proof}

\section{A Dependency Graph for Output Branches}
\label{sec:graphs}

\noindent
{\bf Convention.} In this section
we assume additionally that the initalized la-uniform dtla $M$ is 
{\bf ultralinear} with mapping $\mu: Q\to \Nat$ 
and {\bf b-erasing} with graph $E_M$, see Section~\ref{sec:dtop}. 
These properties will be used (only) in the proof of Lemma~\ref{lm:cycle}.

For this $M$ we will compute an ancestral bound $h_{\rm a}(M)$,
as defined in Definition~\ref{df:ancestral_bound}.
In view of this definition (and Lemma~\ref{lm:diff_char}), 
it is technically convenient to combine a node $v$ of an output tree
with the sequence of labels of the proper ancestors of $v$, as follows. 

\medskip
For the output alphabet $\Delta$, we define the \emph{branch alphabet} $\Delta_\B$
by:
\[
\Delta_\B = 
\{(d,j)\mid d\in\Delta,\,\rk(d)\geq 1,\,j\in[\rk(d)]\}.
\]
A string in $\Delta_\B^*$ is called a \emph{branch}. 
For a branch $v\in\Delta_\B^*$ we define $\nod(v)\in\Nat_+^*$ to be 
the sequence of numbers obtained from $v$ by changing every $(d,j)$ into $j$.
For a tree $t=d(t_1,\ldots,t_k)\in \T_\Delta(Q\times P)$
with $k\in\Nat$, $d\in(\Delta\cup (Q\times P))^{(k)}$, and $t_1,\ldots,t_k\in \T_\Delta(Q\times P)$, 
we define the set $B(t)\subseteq \Delta_\B^*$ of \emph{branches of $t$} inductively as follows:
\[
B(d(t_1,\ldots,t_k))=\{\varepsilon\}\cup\{(d,j)\,v\mid j\in[k], v\in B(t_j)\}.
\]
The mapping $\nod$, restricted to $B(t)$, is a bijection from $B(t)$ to $V(t)$. 
Intuitively, a branch $v$ contains the node $\nod(v)$ and 
the labels of its proper ancestors (from the root to the node).
For example, if $v=(a,2)(b,1)(b,3)$ is a branch of $t$, 
then it corresponds to the node $\nod(v)=(2,1,3)$ of $t$, and moreover, the root of $t$ has label $a$,
node $2$ has label $b$, and node $(2,1)$ has label $b$ too.
For a branch $v\in B(t)$, we define $t/v=t/\nod(v)$ and $\lab(t,v)=\lab(t,\nod(v))$. 

We will need the following lemma on branches. 
Roughly, it says that if $M(C[p])$ has a branch that is longer than any branch of $M(C[p'])$,
then a prefix of that branch corresponds to a difference node of $M(C[p])$ and $M(C[p'])$.
In this section, we rename $p,p'$ into $p_1,p_2$. 

\begin{lemma}\label{lm:height-diff}
Let $C\in\C_\Sigma$ and $p_1,p_2\in P$. 
Let $v$ be a branch of both $M(C[p_1])$ and $M(C[p_2])$, 
and let $w\in\Delta_\B^*$ be such that 
\begin{enumerate}
\item[(1)] $vw$ is a branch of $M(C[p_1])$, and
\item[(2)] $|w|>\height(M(C[p_2])/v)$.
\end{enumerate}
Then there is a prefix $w'$ of $w$ such that 
$\nod(vw')$ is a difference node of $M(C[p_1])$ and $M(C[p_2])$.
\end{lemma}
\begin{proof}
Since $|w|>\height(M(C[p_2])/v)$, $vw$ is not a branch of $M(C[p_2])$.
Let $w'$ be the longest prefix of $w$ such that $vw'$ is a branch of $M(C[p_2])$.
Since $vw'$ is also a branch of $M(C[p_1])$, all proper ancestors of $\nod(vw')$
have the same label in $M(C[p_1])$ and $M(C[p_2])$. 
By Lemma~\ref{lm:diff_char}, it remains to show that $\nod(vw')$ 
has different labels in $M(C[p_1])$ and $M(C[p_2])$.
Since $w'\neq w$, there exist $k\geq 1$, $d\in\Delta^{(k)}$, and $j\in[k]$ 
such that $w'(d,j)$ is a prefix of $w$. Then $d=\lab(M(C[p_1]),vw')$.
Suppose that also $d=\lab(M(C[p_2]),vw')$. 
Then $vw'(d,j)$ is a branch of $M(C[p_2])$, contradicting the choice of $w'$. 
\qed
\end{proof}

Again in view of Definition~\ref{df:ancestral_bound}, 
we extend the definition of $\lk$ to branches, and call it $\blk$; 
we are, however, only interested in triples $(q,u,\#)$. 
For $s\in\T_\Sigma(P)$ and all $v\in\Delta_\B^*$, the sets $\blk_s(v)$ of triples
$(q,u,\#)$ with $q\in Q$ and $u\in V(s)$, 
are defined inductively as follows:
\begin{enumerate}
\item[(1)] If $\delta(s)=p$, then $(q_{0,p},\varepsilon,\#)\in\blk_s(\varepsilon)$.
\item[(2)] If $(q,u,\#)\in\blk_s(v)$ and $\rhs(q,s,u)=\zeta$, \\
then $(\bar{q},uj,\#)\in\blk_s(vz)$ for every $z\in B(\zeta)$ with $\zeta/z=\bar{q}(x_j)$.
\end{enumerate}

\begin{lemma}\label{lm:blink}
For every $s\in\T_\Sigma(P)$, $v\in\Delta_\B^*$, $q\in Q$, and $u\in V(s)$,
the following three statements are equivalent:
\begin{enumerate}
\item[(1)] $(q,u,\#)\in\blk_s(v)$; 
\item[(2)] $(q,u,\#)\in\lk_s(\nod(v))$ and $v\in B(M(s))$;
\item[(3)] there is a sentential form $\xi$ of $M^\circ$ such that 
$q_{0,\delta(s)}(\varepsilon)\Rightarrow_s^*\xi$, $\;v\in B(\xi)$, and $\xi/v=q(u)$. 
\end{enumerate}
\end{lemma}
\begin{proof}
Since the definition of $\blk_s$ closely follows the semantics of $M^\circ$, 
it is straightforward to show the equivalence of~(1) and~(3). 
This proves the equivalence of~(1) and~(2), 
by Lemmas~\ref{lm:linkdef}(1) and~\ref{lm:elem}(2). 
\qed
\end{proof}

With these definitions, we can reformulate Definition~\ref{df:ancestral_bound} as follows.

\begin{lemma}\label{lm: ancbound_redef1}
A number $h_{\rm a}(M)\in\Nat$ is an ancestral bound for $M$ if and only if 
$\diff(M)$ is infinite or the following holds 
for every $C\in\C_\Sigma$, $p_1,p_2\in P$,
$u\in V(C)$, $v_1\in B(M(C[p_1]))$, $v_2\in B(M(C[p_2]))$, and $q_1,q_2\in Q$: \\
if $(q_1,u,\#)\in\blk_{C[p_1]}(v_1)$, $\;(q_2,u,\#)\in\blk_{C[p_2]}(v_2)$, and
$v_2$ is a prefix of $v_1$, \\
then $|v_1|-|v_2|\leq h_{\rm a}(M)$. 
\end{lemma}

Thus, to determine an ancestral bound for $M$, we are interested in a node $u$
of a $\Sigma$-context $C$ and in two computations of $M$ 
on the path from the root of $C$ to $u$, 
one with input $C[p_1]$ and the other with input $C[p_2]$.
The idea is, roughly, 
to find two ancestors $\hat{u}$ and $\bar{u}$ of $u$ in $C$
where each of these computations 
is in a cycle (i.e., arrives in the same state at $\hat{u}$ and $\bar{u}$), 
and then to pump the part of $C$ 
between $\hat{u}$ and $\bar{u}$, leading to an infinite $\diff(M)$. 
For this it is technically helpful to build a graph $G_M$
of which each path captures two such computations: 
those corresponding to $(q_1,u,\#)\in\blk_{C[p_1]}(v_1)$ and 
$(q_2,u,\#)\in\blk_{C[p_2]}(v_2)$ according to Lemma~\ref{lm:blink}(3).
The pumping of $C$ then corresponds to the repetition of a cycle in the graph $G_M$. 
In this graph, we have to distinguish between the case that $u$ is on or off 
the spine of $C$, where the spine of a context $C$ is the path from the root 
to the unique occurrence of $\bot$ (as in the proof of Lemma~\ref{lm:new_not_ancestor}).
Formally, for $C\in\C_\Sigma$, the \emph{spine} of $C$ is the set 
$\spi(C)=\{u\in V(C)\mid C/u\in\C_\Sigma\}$. 

For an introduction to the pumping technique using a graph, 
the reader is advised to study the proof of Lemma~\ref{lm:range_bound}, 
where each path of the graph captures just one computation of $M$, corresponding to $(q,u,\#)\in\lk_s(v)$.
For the notion of a directed edge-labeled graph see Section~\ref{sec:prelim}.

\medskip
For the dtla $M$, we define a finite
directed edge-labeled \emph{dependency graph} $G_M$ 
whose nodes are all 3-tuples $(q_1,q_2,b)$ with
$q_1,q_2\in Q$ and $b\in\{0,1\}$,
and whose edge labels are in 
$R\times R\times \Nat_+\times\Delta_\B^*\times\Delta_\B^*$,
more precisely, in the finite set of all 5-tuples $(r_1,r_2,j,z_1,z_2)$ such that 
$r_1,r_2\in R$, $j\in[\rk(a)]$ for some $a\in\Sigma$,  
and $z_i\in \Delta_\B^*$ is a branch of the right-hand side of $r_i$ for $i=1,2$.
The boolean $b$ is called the \emph{type} of the node;
intuitively, $b=1$ stands for ``on the spine'', and $b=0$ for ``off the spine''.
The edges of the dependency graph $G_M$ are defined as follows, together with its ``entry nodes''.

\begin{enumerate}
\item[(Ge)] 
For every $\bar{p}_1,\bar{p}_2\in P$, $(q_{0,\bar{p}_1},q_{0,\bar{p}_2},1)$ 
is an \emph{entry node} of $G_M$.
\end{enumerate}
\begin{enumerate}
\item[(G1)] Let $k\geq 1$, $a\in\Sigma^{(k)}$, and $\ell,j\in[k]$.
Let $(q_1,q_2,1)$ be a node of $G_M$ and, for $i=1,2$,
let $r_i=q_i(a(x_1\ot \bar{p}_{i,1},\ldots,x_k\ot \bar{p}_{i,k}))\to \zeta_i$
be a rule of $M$, let $z_i\in\Delta_\B^*$, and let $q_i'\in Q$, such that 
\begin{enumerate}
\item $\bar{p}_{1,m}=\bar{p}_{2,m}$ for all $m\in[k]-\{\ell\}$, 
\item $z_i\in B(\zeta_i)$ and $\zeta_i/z_i=q'_i(x_j)$, for $i=1,2$.
\end{enumerate}
Then there is an edge labeled $(r_1,r_2,j,z_1,z_2)$ 
from $(q_1,q_2,1)$ to $(q_1',q_2',b')$,
where $b'=1$ if and only if $\ell=j$.
\end{enumerate}
\begin{enumerate}
\item[(G0)] Let $k\geq 1$, $a\in\Sigma^{(k)}$, and $j\in[k]$.
Let $(q_1,q_2,0)$ be a node of $G_M$ and, for $i=1,2$,
let $r_i=q_i(a(x_1\ot \bar{p}_{i,1},\ldots,x_k\ot \bar{p}_{i,k}))\to \zeta_i$
be a rule of $M$, let $z_i\in\Delta_\B^*$, and let $q_i'\in Q$, such that 
\begin{enumerate}
\item $\bar{p}_{1,m}=\bar{p}_{2,m}$ for all $m\in[k]$,
\item $z_i\in B(\zeta_i)$ and $\zeta_i/z_i=q'_i(x_j)$, for $i=1,2$.
\end{enumerate}
Then there is an edge labeled $(r_1,r_2,j,z_1,z_2)$ 
from $(q_1,q_2,0)$ to $(q_1',q_2',0)$.
\end{enumerate}

\noindent
Note that there are no edges from a node of type $0$ to a node of type $1$.
Note also that if $(q_1,q_2,b)$ is a node of $G_M$, then so is $(q_2,q_1,b)$;
moreover, if there is an edge from $(q_1,q_2,b)$ to $(q'_1,q'_2,b')$ 
with label $(r_1,r_2,j,z_1,z_2)$, then there is an edge from 
$(q_2,q_1,b)$ to $(q'_2,q'_1,b')$ with label $(r_2,r_1,j,z_2,z_1)$.

A path $\pi=e_1\cdots e_n$ in $G_M$ (with $n\geq 0$ and $e_i\in E$ for every $i\in[n]$) 
has a \emph{label} $(u,v_1,v_2)\in\Nat_+^*\times\Delta_\B^*\times\Delta_\B^*$ 
obtained by the component-wise concatenation of 
the last three components of the labels of $e_1,\dots,e_n$.
The \emph{output label} of $\pi$ is the pair of branches $(v_1,v_2)$,  
denoted by $\out(\pi)$.
We say that $\pi$ is an \emph{entry path} if it starts at an entry node, and
that it is a \emph{$(q_1,q_2,b)$-path} if it is an entry path that  
ends at the node $(q_1,q_2,b)$. We will only be interested in the entry paths of $G_M$.

Let us now consider the intuition behind items (Ge), (G1), and (G2),  
continuing the discussion before the definition of $G_M$.  
An entry path of $G_M$, more precisely a $(q_1,q_2,b)$-path with label $(u,v_1,v_2)$, 
is meant to capture two computations of $M$, 
corresponding to $(q_1,u,\#)\in\blk_{C[p_1]}(v_1)$ and 
$(q_2,u,\#)\in\blk_{C[p_2]}(v_2)$ according to Lemma~\ref{lm:blink}(3),
and moreover, the boolean $b$ indicates whether or not $u$ is on the spine of $C$. 
This is formally stated in Lemma~\ref{lm:path_in_graph}.
Since these two computations start at the root of $C$ in the initial states 
$q_{0,\bar{p}_1}$ and $q_{0,\bar{p}_2}$ with $\bar{p}_i=\delta(C[p_i])$, 
and since the root of $C$ is on its spine, this explains item (Ge) for the empty path
at that entry node and for $u=v_1=v_2=\varepsilon$,
see also condition~(1) of the definition of $\blk$. 
Items (G1) and (G0) extend a $(q_1,q_2,b)$-path with label $(u,v_1,v_2)$ with one more edge,
assuming that $u$ is on/off the spine, respectively. 
The computations of $M$ have arrived at node $u$ of $C$, 
and now move to its child $uj$ according to condition~(2) of the definition of $\blk$. 
To do this, $M$ applies the rules $r_1$ with right-hand side $\zeta_1=\rhs(q_1,C[p_1],u)$ 
and $r_2$ with right-hand side $\zeta_2=\rhs(q_2,C[p_2],u)$
in $C[p_1]$ and $C[p_2]$, respectively, 
and it picks branches $z_i$ in $\zeta_i$ that lead to $q_i'(x_j)$.
If $u$ is on the spine, 
let $u\ell$ be the (unique) child of $u$ that is on the spine. 
Then the look-ahead state at each child $um$ of $u$ with $m\neq \ell$ 
is the same in $C[p_1]$ and $C[p_2]$ 
(and if $u$ is not on the spine, then this holds for all its children). 
The path is now extended by one edge to a $(q_1',q_2',b')$-path with label $(uj,v_1z_1,v_2z_2)$,
where $b'=1$ if $b=1$ and $j=\ell$ (i.e., $uj$ is the unique child of $u$ on the spine),
and $b'=0$ otherwise. 
The edge is labeled additionally with the rules $r_1$ and $r_2$ for technical reasons.

We now give two examples of a dependency graph. 

\begin{example}\label{ex:depgraphsimple}
We first consider a very simple initialized la-uniform dtla $M$ 
that changes the label of every node $u$ of the input tree $s$ 
into the label of the left-most leaf of $s/u$, and changes the label of every leaf into $e$. 
Let $\Sigma=\{\sigma^{(2)},a^{(0)},b^{(0)}\}$ and $\Delta=\{a^{(2)},b^{(2)},e^{(0)}\}$.
The look-ahead states of $M$ are $p_a$ and $p_b$, 
which compute the label of the left-most leaf:
$\delta(y)=p_y$ and $\delta(\sigma,p_y,p_z)=p_y$ for all $y,z\in\{a,b\}$.
The states of $M$ are $q_a$ and $q_b$.
For $y,w\in\{a,b\}$, the axioms of $M$ are $A(p_y)=q_y(x_0)$ and its rules are
$q_y(y)\to e$ and $r_{yw} = q_y(\sigma(x_1\ot p_y,x_2\ot p_w)) \to y(q_y(x_1),q_w(x_2))$. 

The dependency graph $G_M$ has eight nodes $(q_y,q_z,\overline{b})$, 
for $y,z\in\{a,b\}$ and $\overline{b}\in\{0,1\}$. 
By (Ge), the four nodes $(q_y,q_z,1)$ are the entry nodes of $G_M$. 
Applying (G1) to these entry nodes the following 32 edges are obtained, for all $y,z,w,w'\in\{a,b\}$:
\begin{enumerate}
\item[(i)] $j=1$, $\ell=1$: from $(q_y,q_z,1)$ to itself labeled $(r_{yw},r_{zw},1,(y,1),(z,1))$
\item[(ii)] $j=1$, $\ell=2$: from $(q_y,q_y,1)$ to $(q_y,q_y,0)$ labeled $(r_{yw},r_{yw'},1,(y,1),(y,1))$
\item[(iii)] $j=2$, $\ell=2$: from $(q_y,q_y,1)$ to $(q_w,q_{w'},1)$ labeled $(r_{yw},r_{yw'},2,(y,2),(y,2))$
\item[(iv)] $j=2$, $\ell=1$: from $(q_y,q_z,1)$ to $(q_w,q_w,0)$ labeled $(r_{yw},r_{zw},2,(y,2),(z,2))$.
\end{enumerate}
Then, applying (G0) to all nodes $(q_y,q_y,0)$, the following 8 edges are obtained, 
for all $y,w\in\{a,b\}$:
\begin{enumerate}
\item[(v)] $j=1$: from $(q_y,q_y,0)$ to itself labeled $(r_{yw},r_{yw},1,(y,1),(y,1))$
\item[(vi)] $j=2$: from $(q_y,q_y,0)$ to $(q_w,q_w,0)$ labeled $(r_{yw},r_{yw},2,(y,2),(y,2))$.
\end{enumerate}
Note that the nodes $(q_a,q_b,0)$ and $(q_b,q_a,0)$ are not reachable from an entry node 
and hence we do not care about the edges that leave them (in fact, there are none). 

Let us consider the $\Sigma$-context $C=\sigma(b,\sigma(C',b))$ with $C'=\sigma(\bot,\sigma(a,\sigma(b,a)))$. 
Clearly, we have $M(C[p_a])=b(e,a(t_a,e))$ with $t_a=a(\angl{q_a,p_a},a(e,b(e,e)))$, and 
$M(C[p_b])=b(e,b(t_b,e))$ with $t_b=b(\angl{q_b,p_b},a(e,b(e,e)))$.
Now consider the input node $u=(2,1,2,2)$, i.e., the node with $C/u=\sigma(b,a)$, 
and note that $u\notin\spi(C)$. 
From Lemma~\ref{lm:blink}(3), or from the definition of $\blk$, it should be clear that 
$(q_b,u,\#)\in\blk_{C[p_a]}(v_a)$ and $(q_b,u,\#)\in\blk_{C[p_b]}(v_b)$ for the branches 
$v_a=(b,2)(a,1)(a,2)(a,2)$ and $v_b=(b,2)(b,1)(b,2)(a,2)$.
According to the intuition described before, 
this corresponds to the $(q_b,q_b,0)$-path $e_1e_2e_3e_4$ with label $(u,v_a,v_b)$ where,
using (iii), (i), (iv), and (vi) above, respectively:
\begin{itemize}
\item $e_1$ is from entry node $(q_b,q_b,1)$ to $(q_a,q_b,1)$ labeled $(r_{ba},r_{bb},2,(b,2),(b,2))$,
\item $e_2$ is from $(q_a,q_b,1)$ to itself labeled $(r_{ab},r_{bb},1,(a,1),(b,1))$,
\item $e_3$ is from $(q_a,q_b,1)$ to $(q_a,q_a,0)$ labeled $(r_{aa},r_{ba},2,(a,2),(b,2))$, and 
\item $e_4$ is from $(q_a,q_a,0)$ to $(q_b,q_b,0)$ labeled $(r_{ab},r_{ab},2,(a,2),(a,2))$.
\end{itemize}
Note that the prefixes of this path correspond in a similar way to the ancestors of $u$. 
\qed
\end{example}

The dtla $M$ of the above example is very simple.  
If $(q,u,\#)\in\blk_{C[p]}(v)$ then $|u|=|v|$ and correspondingly, 
if $(u,v_1,v_2)$ is the label of an entry path in $G_M$ then $|v_1|=|v_2|$.
The next example is more complicated in that respect, and 
hence can serve as an example later on. 

\begin{example}\label{ex:depgraph}
Let $\Sigma=\{\tau^{(2)},\sigma^{(1)},a^{(0)},b^{(0)}\}$ and let
$\Delta=\{\tau^{(2)},\sigma_a^{(1)},\sigma_b^{(1)},a^{(0)},b^{(0)},$ $e^{(0)}\}$.
We consider an initialized la-uniform dtla $M$ such that,  
for all $m,n\in\Nat$ and $y\in\{a,b\}$,
$\;M(\tau(\sigma^m y,\sigma^n a))=\tau(\sigma_y^my,M_y(\sigma^n a))$ where $M_y(a)=a$, 
$$M_a(\sigma^{n+1} a)=\tau(M_a(\sigma^n a),a) \mbox{\; and \;} 
M_b(\sigma^{n+1} a)=\tau(\tau(a,M_b(\sigma^n a)),a).$$ 
We will not be interested in other input trees, which are translated by $M$ into trees
with at least one occurrence of $e$.  
The look-ahead states of $M$ are $p_a$ and $p_b$, 
which, as in the previous example, compute the label of the left-most leaf:
$\delta(y)=p_y$, $\delta(\sigma,p_y)=p_y$, and $\delta(\tau,p_y,p_z)=p_y$ for all $y,z\in\{a,b\}$.
The states of $M$ are all $q_{iy}$ with $i\in\{0,1,2\}$ and $y\in\{a,b\}$; 
the la-map of $M$ is $\rho(q_{0y})=\rho(q_{1y})=p_y$ and $\rho(q_{2y})=p_a$. 
For $y\in\{a,b\}$, the axioms of $M$ are $A(p_y)=q_{0y}(x_0)$, and it has the following rules, 
where the missing rules all have right-hand side $e$. 
First, it has the rules
\[
\begin{array}{lll}
r_{0y} & = & q_{0y}(\tau(x_1\ot p_y,x_2\ot p_a))  \to  \tau(q_{1y}(x_1),q_{2y}(x_2)), \\
r_{1y} & = & q_{1y}(\sigma(x_1\ot p_y))  \to  \sigma_y(q_{1y}(x_1)), \\
r_{2a} & = & q_{2a}(\sigma(x_1\ot p_a))  \to  \tau(q_{2a}(x_1),a), \mbox{ and} \\
r_{2b} & = & q_{2b}(\sigma(x_1\ot p_a))  \to  \tau(\tau(a,q_{2b}(x_1)),a).
\end{array}
\]
Second, it has the rules $q_{1y}(y)\to y$ and $q_{2y}(a)\to a$.

The dependency graph $G_M$ has the four entry nodes $(q_{0y},q_{0z},1)$ for $y,z\in\{a,b\}$.
We will not construct all edges of $G_M$, but just mention four interesting ones.
First, by~(G1) with $\ell=j=1$, 
there is an edge $e_{01}$ from $(q_{0a},q_{0b},1)$ to $(q_{1a},q_{1b},1)$
with label $(r_{0a},r_{0b},1,(\tau,1),(\tau,1))$.
Second, by (G1) with $\ell=1$ and $j=2$, 
there is an edge $e_{02}$ from $(q_{0a},q_{0b},1)$ to $(q_{2a},q_{2b},0)$
with label $(r_{0a},r_{0b},2,(\tau,2),(\tau,2))$.
Third, by (G1), 
there is an edge $e_1$ from $(q_{1a},q_{1b},1)$ to itself
with label $(r_{1a},r_{1b},1,(\sigma_a,1),(\sigma_b,1))$.
And finally, by~(G0), 
there is an edge $e_2$ from $(q_{2a},q_{2b},0)$ to itself
with the label $(r_{2a},r_{2b},1,(\tau,1),(\tau,1)(\tau,2))$.
\qed
\end{example}

As observed above, we express in Lemma~\ref{lm:path_in_graph} the meaning of the entry paths in $G_M$. 
But we will need a stronger version of the implication $(1)\Rightarrow(2)$ of that lemma,
which we prove now. 
Recall from the paragraph before Theorem~\ref{th:uniform_earliest} that 
$\fix(M)$ is a fixed set of representatives of the equivalence classes $\sem{p}$, $p\in P$. 

\begin{lemma}\label{lm:path_to_context}
Let $\pi=e_1\cdots e_n$, $n\geq 0$, be an entry path in $G_M$, 
where $e_m$ is an edge of $G_M$ for every $m\in[n]$. 
For every $m$, $0\leq m\leq n$, let the prefix $e_1\cdots e_m$ of $\pi$ 
be a $(q_1^{(m)},q_2^{(m)},b^{(m)})$-path with label $(u^{(m)},v_1^{(m)},v_2^{(m)})$. 
Then there are $C\in\C_\Sigma$ and $p_1,p_2\in P$
such that for every $m$, $0\leq m\leq n$, and $i=1,2$, 
\begin{enumerate}
\item[$(a)$] $(q_i^{(m)},u^{(m)},\#)\in\blk_{C[p_i]}(v_i^{(m)})$, 
\item[$(b)$] $b^{(m)}=1$ if and only if $u^{(m)}\in\spi(C)$, 
\item[$(c)$] if $m\leq n-1$ and $e_{m+1}$ has label $(r_1,r_2,j,z_1,z_2)$, then 
\begin{enumerate}
\item[$(1)$] $\rhs(q_i^{(m)},C[p_i],u^{(m)})=\rhs(r_i)$ and
\item[$(2)$] for every $j'\in\Nat_+$ such that $u^{(m)}j'$ is a node of $C$, 
if $j'\neq j$ or $m=n-1$, then $C/u^{(m)}j'\in\fix(M)\cup\{\bot\}$. 
\end{enumerate}
\end{enumerate}
\end{lemma}
\begin{proof}
We proceed by induction on the length $n$ of $\pi$. 
If $n=0$, i.e., $\pi$ is empty, then $m=0$, $u^{(0)}=v_1^{(0)}=v_2^{(0)}=\varepsilon$, and 
$(q_1^{(0)},q_2^{(0)},b^{(0)})=(q_{0,p_1},q_{0,p_2},1)$ for $p_1,p_2\in P$.
Let $C=\bot$. It is easy to check requirements (a) and (b) for these $C$, $p_1$ and $p_2$;
requirement (c) holds trivially.  

Now let $\pi' = \pi e$ where 
$\pi$ is a $(q_1,q_2,b)$-path with label $(u,v_1,v_2)$, and where 
the last edge $e$ has label $(r_1,r_2,j,z_1,z_2)$ and ends at node $(q_1',q_2',b')$. 
Then $\pi'$ is a $(q_1',q_2',b')$-path with label $(u',v_1',v_2')$ where $u'=uj$ and $v_i'=v_iz_i$. 
Let $r_i$ be the rule
$q_i(a(x_1\ot \bar{p}_{i,1},\ldots,x_k\ot \bar{p}_{i,k}))\to \zeta_i$ of $M$,
hence $k\geq 1$, $a\in\Sigma^{(k)}$, $j\in[k]$, and 
$z_i\in B(\zeta_i)$ with $\zeta_i/z_i = q_i'(x_j)$ for $i=1,2$. 
By induction, requirements~(a), (b), and~(c) hold for some $C\in\C_\Sigma$ and $p_1,p_2\in P$,
for every prefix of $\pi$. In particular (for $m=n$), 
$(q_i,u,\#)\in\blk_{C[p_i]}(v_i)$, and
$b=1$ if and only if $u\in\spi(C)$.
We first consider the case where $b=1$, i.e., the last edge $e$ is obtained by (G1). 
Then there exists $\ell\in[k]$ such that  
$\bar{p}_{1,m}=\bar{p}_{2,m}$ for all $m\in[k]-\{\ell\}$,  
and $b'=1$ if and only if $\ell=j$.
Let $p_1'=\bar{p}_{1,\ell}$ and $p_2'=\bar{p}_{2,\ell}$.
Let $s_\ell = \bot$
and for $m\in[k]-\{\ell\}$, 
let $s_m$ be the unique tree in $\fix(M)$ such that $\delta(s_m)=\bar{p}_{1,m}=\bar{p}_{2,m}$.
Define $C'=C[u\leftarrow C_a]$ where
$C_a$ is the $\Sigma$-context $a(s_1,\dots,s_k)$; 
since $u\in\spi(C)$, $C'$ is a $\Sigma$-context. 
We now claim that requirements~(a), (b), and~(c) hold for $C'$ and 
$p_1',p_2'$, for every prefix of $\pi'$.

Obviously, $u'=uj$ is a branch of $C'$, and $C'/u\ell=\bot$.
Hence, $u'\in\spi(C')$ if and only if $\ell=j$ if and only if $b'=1$. 
For every prefix $e_1\cdots e_m$ of the path $\pi$, by the definition of the label of a path, 
$u^{(m)}$ is a prefix of $u$, i.e., an ancestor of $u$. 
Hence, $u^{(m)}\in\spi(C')$ if and only if 
$u^{(m)}\in\spi(C)$ if and only if $b^{(m)}=1$
(and in fact, since $b=1$, they are all true).
This proves requirement~(b). 

Since $(q_i,u,\#)\in\blk_{C[p_i]}(v_i)$, Lemmas~\ref{lm:blink} and~\ref{lm:linkdef}(4) imply that 
$\delta(C[p_i]/u)=\rho(q_i)$ which equals $\delta(C_a[p_i'])$ 
by definition of $C_a$ (and because $M$ is la-uniform). 
Hence $\delta(C[p_i]/u)=\delta(C'[p_i']/u)$, and so $\delta(C[p_i])=\delta(C'[p_i'])$
and every proper ancestor of $u$ is similar in $C[p_i]$ and $C'[p_i']$. 
It now follows from Lemmas~\ref{lm:link_preserve}(1) and~\ref{lm:blink} that 
$(q_i,u,\#)\in\blk_{C'[p_i']}(v_i)$. 
Since $\rhs(q_i,C'[p_i'],u)=\rhs(q_i,C_a[p_i'],\varepsilon)=\zeta_i$, 
the definition of $\blk$ implies that $(q_i',uj,\#)\in\blk_{C'[p_i']}(v_iz_i)$, i.e., 
that $(q_i',u',\#)\in\blk_{C'[p_i']}(v_i')$.
Since $u^{(m)}$ is an ancestor of $u$ for every prefix $e_1\cdots e_m$ of $\pi$,
Lemmas~\ref{lm:link_preserve}(1) and~\ref{lm:blink} also imply that 
$(q_i^{(m)},u^{(m)},\#)\in\blk_{C'[p_i']}(v_i^{(m)})$.
This proves requirement~(a).

It should be clear from the observations in the previous two paragraphs 
that requirement~(c) holds for every proper prefix $e_1\cdots e_m$ of $\pi$.
In fact, $u^{(m)}$ is a proper ancestor of $u$, and $\rhs(q,C[p_i],\hat{u})=\rhs(q,C'[p_i'],\hat{u})$ 
for every proper ancestor $\hat{u}$ of $u$ (and every $q$), 
because $\hat{u}$ is similar in $C[p_i]$ and $C'[p_i']$. 
Moreover, if $u^{(m)}j'\neq u^{(m+1)}$ then $C/u^{(m)}j'=C'/u^{(m)}j'$.
For the proper prefix $\pi$ of $\pi'$ and the edge $e$, 
requirement (c) holds by definition of the context $C'$. 

The proof for the case where $e$ is obtained by (G0) is similar, 
with $b=b'=0$. 
Since $u$ is not on the spine of $C$, we let 
$s_m\in \fix(M)$ with $\delta(s_m)=\bar{p}_{1,m}=\bar{p}_{2,m}$ for every $m\in[k]$, 
and we take $p_1'=p_1$ and $p_2'=p_2$.
\qed
\end{proof}

\begin{example}\label{ex:path_to_context}
For $n\in\Nat_+$, consider the entry path $\pi=e_{01}e_1^{n-1}$ in $G_M$, 
where $M$ is the dtla of Example~\ref{ex:depgraph}. 
It is a path from $(q_{0a},q_{0b},1)$ to $(q_{1a},q_{1b},1)$, and it  
has label $(1^n,(\tau,1)(\sigma_a,1)^{n-1},(\tau,1)(\sigma_b,1)^{n-1})$
where $1^n$ is the sequence $(1,\dots,1)$ of length~$n$. 
Let $\fix(M)=\{a,b\}$. 
Then the requirements of Lemma~\ref{lm:path_to_context} are fulfilled for
$C=\tau(\sigma^{n-1}\bot,a)$ and for $p_1=p_a$ and $p_2=p_b$.

Now consider the entry path $\pi=e_{02}e_2^{n-1}$ 
from $(q_{0a},q_{0b},1)$ to $(q_{2a},q_{2b},0)$
with label $(2\cdot 1^{n-1},(\tau,2)(\tau,1)^{n-1},(\tau,2)((\tau,1)(\tau,2))^{n-1})$. 
The requirements of Lemma~\ref{lm:path_to_context} for this path are fulfilled for
$C=\tau(\bot,\sigma^{n-1}a)$ and for $p_1=p_a$ and $p_2=p_b$.
\qed
\end{example}

In the next lemma we express the meaning of the entry paths in $G_M$.

\begin{lemma}\label{lm:path_in_graph}
Let $q_1,q_2\in Q$, $b\in\{0,1\}$, $u\in\Nat_+^*$, and $v_1,v_2\in\Delta_\B^*$. \\
Then the following two statements are equivalent:
\begin{enumerate}
\item[(1)] there is a $(q_1,q_2,b)$-path in $G_M$ with label $(u,v_1,v_2)$;
\item[(2)] there are $C\in\C_\Sigma$ and $p_1,p_2\in P$ 
such that 
\begin{enumerate}
\item $(q_1,u,\#)\in\blk_{C[p_1]}(v_1)$ and $(q_2,u,\#)\in\blk_{C[p_2]}(v_2)$, and 
\item $b=1$ if and only if $u\in\spi(C)$. 
\end{enumerate}
\end{enumerate}
\end{lemma}
\begin{proof}
$(1)\Rightarrow(2)$.\quad
Immediate from Lemma~\ref{lm:path_to_context} for $m=n$ (and disregarding (c)). 

$(2)\Rightarrow(1)$.\quad
This proof is similar to the proof of Lemma~\ref{lm:path_to_context}, 
but it is easier because in the induction step we can use the same $C$, $p_1$, and $p_2$. 
An illustration of the proof can be found in Example~\ref{ex:depgraphsimple}.
Here we proceed by induction on the length of $u$.
Let $u=\varepsilon$. Then $u\in\spi(C)$ and so $b=1$. 
Since $(q_i,\varepsilon,\#)\in\blk_{C[p_i]}(v_i)$, 
it follows from the definition of $\blk$ that $v_i=\varepsilon$, 
and that $q_i=q_{0,\bar{p}_i}$ where $\bar{p}_i=\delta(C[p_i])$.
Hence $(q_1,q_2,b)=(q_{0,\bar{p}_1},q_{0,\bar{p}_2},1)$ is an entry node. 
This proves statement~(1) for this case, because the empty path from 
$(q_1,q_2,b)$ to itself has label $(\varepsilon,\varepsilon,\varepsilon)$. 
 
Now let $u'=uj$ with $u\in\Nat_+^*$ and $j\in\Nat_+$.
Let $C\in\C_\Sigma$ and $p_1,p_2\in P$ (and consider $q_i'$, $v_i'$, and $b'$) such that 
$(q_i',u',\#)\in\blk_{C[p_i]}(v_i')$ for $i=1,2$ and 
such that $b'=1$ if and only if $u'\in\spi(C)$. 
By the definition of $\blk$, there exists 
a rule $r_i$ of the form $q_i(a(x_1\ot \bar{p}_{i,1},\ldots,x_k\ot \bar{p}_{i,k}))\to \zeta_i$
such that $a=\lab(C,u)$ with $k\geq 1$, $a\in\Sigma^{(k)}$, and $j\in[k]$, and
such that $\bar{p}_{i,m}=\delta(C[p_i]/um)$ for every $m\in[k]$;  
moreover, there exist $v_i\in\Delta_\B^*$ and $z_i\in B(\zeta_i)$ such that $v_i'=v_iz_i$, 
$\;(q_i,u,\#)\in\blk_{C[p_i]}(v_i)$, and $\zeta_i/z_i=q_i'(x_j)$.
We first consider the case where $u\in\spi(C)$. 
Then, by induction, there is a $(q_1,q_2,1)$-path $\pi$ in $G_M$ with label $(u,v_1,v_2)$.
Let $\ell\in[k]$ be such that $u\ell\in\spi(C)$;
so, $b'=1$ if and only if $\ell=j$. 
We observe that $\bar{p}_{1,m}=\bar{p}_{2,m}$ for all $m\in[k]-\{\ell\}$, 
because $C/um\in\T_\Sigma$.
So, by (G1), there is an edge $e$ labeled $(r_1,r_2,j,z_1,z_2)$
from $(q_1,q_2,1)$ to $(q_1',q_2',b')$.
Hence, $\pi e$ is a $(q_1',q_2',b')$-path with label $(uj,v_1z_1,v_2z_2)=(u',v_1',v_2')$.

The proof for the case where $u\notin\spi(C)$ is similar. 
Then, by induction, there is a $(q_1,q_2,0)$-path $\pi$. 
Of course $u'\notin\spi(C)$, and so $b'=0$. 
Since now $\bar{p}_{1,m}=\bar{p}_{2,m}$ for all $m\in[k]$,
there is an edge $e$ from $(q_1,q_2,0)$ to $(q_1',q_2',b')$ by~(G0).
\qed
\end{proof}

For strings $v_1,v_2$, we define $\diff(v_1,v_2)$ to be 
the pair of strings $(w_1,w_2)$ 
such that $v_1=vw_1$ and $v_2=vw_2$ 
where $v$ is the longest common prefix of $v_1$ and $v_2$. 
Note that if $\diff(v_1,v_2)=(w_1,w_2)$, 
then $\diff(v_1z_1,v_2z_2)=\diff(w_1z_1,w_2z_2)$ for all strings $z_1$ and $z_2$.
For a path $\pi$ in $G_M$,
we define $\diff(\pi)=\diff(\out(\pi))$. 
So, if $\pi=\pi_1\pi_2$ then $\diff(\pi)=\diff(\diff(\pi_1)\cdot\out(\pi_2))$,
where $\cdot$ is component-wise concatenation.

We say that a pair of strings $(w_1,w_2)\in\Delta_\B^*\times\Delta_\B^*$ is \emph{ancestral} 
if $w_1=\varepsilon$ or $w_2=\varepsilon$ (or both).
A path $\pi$ in $G_M$ is \emph{ancestral} if $\diff(\pi)$ is ancestral.
Thus, if $\out(\pi)=(v_1,v_2)$, then $\pi$ is ancestral if and only if 
$v_1$ is a prefix of $v_2$ or vice versa. 
Clearly, if $\pi$ is ancestral then every prefix of $\pi$ is ancestral:
if $\pi=\pi_1\pi_2$ and $\diff(\pi_1)$ is not ancestral 
(i.e., both its components are nonempty and their first symbols differ), 
then $\diff(\pi_1)\cdot\out(\pi_2)$ is not ancestral and $\diff(\pi)=\diff(\pi_1)\cdot\out(\pi_2)$. 

By Lemmas~\ref{lm: ancbound_redef1} and~\ref{lm:path_in_graph}, and using the above definitions, 
we can again reformulate the definition of ancestral bound (Definition~\ref{df:ancestral_bound}), 
as follows.

\begin{lemma}\label{lm: ancbound_redef2}
A number $h_{\rm a}(M)\in\Nat$ is an ancestral bound for $M$ if and only if 
$\diff(M)$ is infinite or the following holds 
for every ancestral entry path $\pi$ in $G_M$: \\
if $\diff(\pi)=(w,\varepsilon)$, then $|w|\leq h_{\rm a}(M)$.
\end{lemma}

In the next lemma we show that every dtla $M$ has an ancestral bound. It will be 
convenient to prove a slightly stronger result, where we consider more entry paths in $G_M$
than the ancestral ones. 

We say that a pair of strings $(w_1,w_2)\in\Delta_\B^*\times\Delta_\B^*$ \emph{splits} 
if there exist $d\in\Delta^{(k)}$ and $j_1,j_2\in[k]$ with $j_1\neq j_2$ 
such that the first symbol of $w_i$ is $(d,j_i)$. 
For $(v_1,v_2)\in\Delta_\B^*\times\Delta_\B^*$, the fact that $\diff(v_1,v_2)$ splits 
means that there exists a tree $t\in\T_\Delta$ such that $v_1$ and $v_2$ are branches 
of $t$ that correspond to distinct leaves of $t$. 
A path $\pi$ in $G_M$ is \emph{nonsplitting} if $\diff(\pi)$ does not split,
and \emph{splitting} if $\diff(\pi)$ splits.
Note that an ancestral pair of strings does not split, and hence every ancestral path in $G_M$
is nonsplitting. 

Recall again from the paragraph before Theorem~\ref{th:uniform_earliest} that 
$\fix(M)$ is a fixed set of representatives of the equivalence classes $\sem{p}$, $p\in P$. 
We define 
$$\maxfix(M)=\max\{\height(q_M(s))\mid q\in Q,\,s\in\fix(M),\,\rho(q)=\delta(s)\}.$$
Note that since $\fix(M)$ is finite, $\maxfix(M)\in\Nat$.

The next lemma implies that $\maxfix(M)+\maxdiff(M)$ is an 
ancestral bound for $M$ when $\diff(M)$ is finite, and so every dtla has an ancestral bound. 

\begin{lemma}\label{lm:no_split}
Let $\pi$ be a nonsplitting entry path in $G_M$. 
If $\diff(\pi)=(w_1,w_2)$, then $|w_i|\leq \maxfix(M)+\maxdiff(M)$ for $i=1,2$. 
\end{lemma}
\begin{proof}
Let $\pi$ be a $(q_1,q_2,b)$-path with label $(u,v_1,v_2)$ and let $\diff(v_1,v_2)=(w_1,w_2)$.
For reasons of symmetry, it suffices to show that $|w_1|\leq \maxfix(M)+\maxdiff(M)$.
Thus, we assume that $w_1\neq\varepsilon$ (and hence $\pi\neq\varepsilon$). 
We consider two cases, depending on the value of $b$.

\smallskip
\emph{Case 1}: $b=1$. In this case we show that $|w_1|\leq \maxdiff(M)$.
By Lemma~\ref{lm:path_to_context} 
there are $C\in\C_\Sigma$ and $p_1,p_2\in P$ such that 
$(q_i,u,\#)\in\blk_{C[p_i]}(v_i)$ for $i=1,2$, and $C/u=\bot$.
Hence, by Lemma~\ref{lm:blink}, 
there is a reachable sentential form $\xi$ for $C[p_i]$ such that
$v_i\in B(\xi)$ and $\xi/v_i=q_i(u)$. 
Since $C[p_i]/u=p_i$, Lemma~\ref{lm:elem}(2) implies that $v_i\in B(M(C[p_i]))$ 
and $M(C[p_i])/v_i=\angl{q_i,p_i}$. 
Let $v$ be the longest common prefix of $v_1$ and $v_2$.
So $v_i=vw_i$. Since $(w_1,w_2)$ does not split, $\nod(v)$ is 
a difference node of $M(C[p_1])$ and $M(C[p_2])$
by Lemma~\ref{lm:diff_char}:
each of its proper ancestors has the same label in $M(C[p_1])$ and $M(C[p_2])$,
but the node itself does not have the same label in $M(C[p_1])$ and $M(C[p_2])$.
In fact,
if $w_1$ and $w_2$ are both nonempty, 
then their first symbols are $(d_1,j_1),(d_2,j_2)$ with $d_1\neq d_2$, 
and $\nod(v)$ has label $d_i$ in $M(C[p_i])$. 
If $w_2=\varepsilon$, then $v=v_2$ and  
hence $\nod(v)$ has label $\angl{q_2,p_2}$ in $M(C[p_2])$;
also, $\nod(vw_1)$ has label $\angl{q_1,p_1}$ in $M(C[p_1])$ 
and hence the label of $\nod(v)$ is in $\Delta$ (because $w_1\neq\varepsilon$).
Thus, since $\nod(v)$ is a difference node of $M(C[p_1])$ and $M(C[p_2])$, 
$|w_1|\leq \height(M(C[p_1])/v)\leq \maxdiff(M)$. 

\smallskip
\emph{Case 2}: $b=0$. 
By Lemma~\ref{lm:path_to_context}, 
there exist $C\in\C_\Sigma$ and $p_1,p_2\in P$
such that $(q_i,u,\#)\in\blk_{C[p_i]}(v_i)$ and $C/u\in\fix(M)$.
Note that $v_i$ is a branch of $M(C[p_i])$ by Lemma~\ref{lm:blink}, 
and that $M(C[p_i])/v_i=q_{iM}(C/u)$ by Lemma~\ref{lm:orig_prop1}(1).

First assume that $w_2\neq\varepsilon$. This is similar to Case~1 above.
Let $v$ be the longest common prefix of $v_1$ and $v_2$.
So $v_i=vw_i$. Since $(w_1,w_2)$ does not split and both $w_1$ and $w_2$ are nonempty, 
$\nod(v)$ is a difference node of $M(C[p_1])$ and $M(C[p_2])$.
So $|w_1|\leq \height(M(C[p_1])/v)\leq\maxdiff(M)$.

Now assume that $w_2=\varepsilon$, and so $v_1=v_2w_1$.
If $|w_1|\leq\maxfix(M)$ then we are ready.
If $|w_1|>\maxfix(M)$, then $|w_1|>\height(q_{2M}(C/u))$ because $C/u\in\fix(M)$, 
and so $|w_1|>\height(M(C[p_2])/v_2)$. Thus, the branch
$v_2w_1$ of $M(C[p_1])$ is longer than any branch of $M(C[p_2])$ with prefix $v_2$.
By Lemma~\ref{lm:height-diff} (with $v:=v_2$ and $w:=w_1$) 
there is a difference node $\nod(v_2w_1')$ of $M(C[p_1])$ and $M(C[p_2])$
where $w_1'$ is a prefix of $w_1$. Since $\nod(v_2w_1')$ is a node of $M(C[p_2])$, 
$\;|w_1'|\leq\height(M(C[p_2])/v_2)=\height(q_{2M}(C/u))$. 
Hence, 
$|w_1|\leq |w_1'|+\height(M(C[p_1])/v_2w_1')\leq \height(q_{2M}(C/u))+\maxdiff(M)$.
This shows that $|w_1|\leq\maxfix(M)+\maxdiff(M)$
by the definition of $\maxfix(M)$.
\qed
\end{proof}

\begin{corollary}\label{cor:no_split_fin}
If $\diff(M)$ is finite, then the set  
$\{\diff(\pi)\mid \pi$ is a nonsplitting entry path in $G_M\}$
is finite.
\end{corollary}

We now turn to the pumping of $\Sigma$-contexts, as discussed before. In terms of $G_M$
that corresponds to the repetition of a cycle. In the next two lemmas we show that pumping 
(i.e., repeating) a cycle in an ancestral entry path produces again an ancestral path, 
and in Lemma~\ref{lm:bound_graph} we show that this leads to infinitely many 
$\diff(\pi)$ values and hence to an infinite $\diff(M)$
by Corollary~\ref{cor:no_split_fin}. 
 
A path $\pi$ in $G_M$ is a cycle, and in particular a \emph{$(q_1,q_2,b)$-cycle}, 
if it is nonempty and leads from the node $(q_1,q_2,b)$ to itself; 
note that every node of $\pi$ is of type $b$. 
For every $i\geq 1$, we denote by $\pi^i$ the $i$-fold concatenation $\pi\cdots \pi$.

\begin{lemma}\label{lm:drop_prime}
Let $\diff(M)$ be finite.
Let $\pi_0$ be a $(q_1,q_2,b)$-path in $G_M$, and 
$\pi$ be a $(q_1,q_2,b)$-cycle in $G_M$ with output label $(v_1,v_2)$
such that $v_1v_2\neq\varepsilon$.
If $\pi_0\pi$ is nonsplitting, then it is ancestral. 
\end{lemma}
\begin{proof}
The idea of the proof is that if $\pi_0\pi$ is not ancestral,
then we can pump $\pi$ and thus obtain nonsplitting paths with arbitrarily large $\diff$ values,
contradicting Corollary~\ref{cor:no_split_fin}.

Let $\diff(\pi_0\pi)=(w_1,w_2)$.
Suppose that $\pi_0\pi$ is not ancestral, 
i.e., both $w_1$ and $w_2$ are nonempty.
Since $(w_1,w_2)$ is nonsplitting, $(w_1,w_2)=((d_1,j_1)w_1',(d_2,j_2)w_2')$ 
with $d_1\neq d_2$. 
For every $k\geq 1$, consider the $(q_1,q_2,b)$-path $\pi_0\pi^{k+1}$. 
Now $\diff(\pi_0\pi^{k+1})=\diff(\diff(\pi_0\pi)\cdot\out(\pi^k))=
\diff(w_1v_1^k,w_2v_2^k)=((d_1,j_1)w_1'v_1^k,(d_2,j_2)w_2'v_2^k)$.
Hence $\pi_0\pi^{k+1}$ is nonsplitting.
Also since $v_1v_2\neq\varepsilon$, 
we obtain that $|(d_i,j_i)w_i'v_i^k|>k$ for $i=1$ or $i=2$.
Since this holds for every $k$, 
the set $\{\diff(\pi_0\pi^{k+1})\mid k\geq 1\}$ is infinite, 
contradicting Corollary~\ref{cor:no_split_fin}.
\qed
\end{proof}

\begin{example}\label{ex:drop_prime}
Let $M$ be the dtla of Example~\ref{ex:depgraph} and let $\pi_0=e_{01}$ and $\pi=e_1$.
Then $\pi$ has output label $((\sigma_a,1),(\sigma_b,1))$ and $(\sigma_a,1)(\sigma_b,1)\neq\varepsilon$.
The path $\pi_0\pi$ has output label $((\tau,1)(\sigma_a,1),(\tau,1)(\sigma_b,1))$
and so $\diff(\pi_0\pi)=((\sigma_a,1),(\sigma_b,1))$, which means that 
$\pi_0\pi$ is nonsplitting but not ancestral. 
Indeed, consider the pumped path $\pi_0\pi^{k+1}$.
By Example~\ref{ex:path_to_context}, 
it has label $(1^{k+2},(\tau,1)(\sigma_a,1)^{k+1},(\tau,1)(\sigma_b,1)^{k+1})$
and satisfies the requirements of Lemma~\ref{lm:path_to_context} for
$C=\tau(\sigma^{k+1}\bot,a)$ and for $p_1=p_a$ and $p_2=p_b$.
By the proof of Lemma~\ref{lm:no_split}, that implies that the node $\nod((\tau,1))=1$
is a difference node of $M(C[p_a])$ and $M(C[p_b])$, and that
$M(C[p_a])/1=(\sigma_a,1)^{k+1}\angl{q_{1a},p_a}$ is a difference tree of $M$. 
Hence $\diff(M)$ is infinite. 
\qed
\end{example}

In the proof of the next, technical pumping lemma we use that $M$ is
ultralinear (with mapping $\mu: Q\to \Nat$) 
and b-erasing (with graph $E_M$).
We note here the obvious fact that if $e$ is an edge in $G_M$ 
from $(q_1,q_2,b)$ to $(q_1',q_2',b')$,
then $\mu(q_i)\leq\mu(q_i')$; 
and hence the same holds for paths in $G_M$.
Thus, if $e$ is part of a cycle in $G_M$, then $\mu(q_i)=\mu(q_i')$.
Also, if $e$ has output label $(v_1,v_2)$ and 
$v_i=\varepsilon$, then there is an edge from $q_i$ to $q_i'$ in $E_M$; 
and hence the same holds for paths (of the same length) 
in $G_M$ and $E_M$. 
Thus, if a $(q_1,q_2,b)$-cycle in $G_M$ has output label $(\bar{v}_1,\bar{v}_2)$, 
then both $\bar{v}_1$ and $\bar{v}_2$ are nonempty. 

\begin{lemma}\label{lm:cycle}
Let $\diff(M)$ be finite.
Let $\pi_0$ be a $(q_1,q_2,b)$-path in $G_M$ and
$\pi$ be a $(q_1,q_2,b)$-cycle in $G_M$.
If $\pi_0\pi$ is nonsplitting, 
then $\pi_0\pi^k$ is nonsplitting for every $k\geq 1$.
\end{lemma}
\begin{proof}
Clearly, it suffices to show that $\pi_0\pi^2$ is nonsplitting.
Suppose that there exist $\pi_0$ and $\pi$ such that $\pi_0\pi^2$ is splitting.
Let $\pi=e_1\cdots e_n$ with $n\geq 1$ and $e_m$ an edge of $G_M$ for $m\in[n]$.
Without loss of generality, we may assume that $\pi_0\pi e_1$ is splitting. 
In fact, suppose that $e_j$ is the first edge of $\pi$ that causes splitting,
i.e., $\pi_0\pi e_1\cdots e_{j-1}$ is nonsplitting and $\pi_0\pi e_1\cdots e_j$ is splitting.
Let $e_{j-1}$ lead to node $(\widetilde{q}_1,\widetilde{q}_2,b)$, and 
let $\widetilde{\pi}_0=\pi_0 e_1\cdots e_{j-1}$ and $\widetilde{\pi}=e_j\cdots e_ne_1\cdots e_{j-1}$.
Then $\widetilde{\pi}_0$ is a $(\widetilde{q}_1,\widetilde{q}_2,b)$-path, 
$\widetilde{\pi}$ is a $(\widetilde{q}_1,\widetilde{q}_2,b)$-cycle, 
$\widetilde{\pi}_0\widetilde{\pi}$ is nonsplitting, and 
$\widetilde{\pi}_0\widetilde{\pi}e_j$ is splitting.

So, suppose that $\pi_0\pi e$ is splitting, where $e=e_1$. 
Let $\pi_0\pi$ have label $(u,v_1,v_2)$, 
let $\diff(\pi_0\pi)=\diff(v_1,v_2)=(w_1,w_2)$ and let
$\pi$ have output label $(\bar{v}_1,\bar{v}_2)$,   
where $\bar{v}_i$ is a postfix of $v_i$.
Since $M$ is b-erasing, $\bar{v}_1\neq\varepsilon$ and $\bar{v}_2\neq\varepsilon$
(because if $\bar{v}_i=\varepsilon$, then there is a cycle of length $n$ from $q_i$ to $q_i$ in $E_M$).
Consequently, $\pi_0\pi$ is ancestral by Lemma~\ref{lm:drop_prime}, 
i.e., $w_1=\varepsilon$ or $w_2=\varepsilon$.
Let $e$ end at the node $(q_1',q_2',b)$ and let
$(r_1,r_2,j,z_1,z_2)$ be its label. 
Since $\pi_0\pi e$ is splitting, 
$\diff(\pi_0\pi e)=\diff(v_1z_1,v_2z_2)=\diff(w_1z_1,w_2z_2)=((d,j_1)y_1,(d,j_2)y_2)$ 
with $j_1\neq j_2$.
Assume that $w_1=\varepsilon$, i.e., $v_2=v_1w_2$ 
(the case where $w_2=\varepsilon$ is analogous).
So $\diff(z_1,w_2z_2)=((d,j_1)y_1,(d,j_2)y_2)$.
Let $y$ be the longest common prefix of $z_1$ and $w_2z_2$, i.e., 
$z_1=y(d,j_1)y_1$ and $w_2z_2=y(d,j_2)y_2$.
Then $v_1y$ is the longest common prefix of $v_1z_1$ and $v_2z_2$, because
$v_1z_1=v_1y(d,j_1)y_1$ and $v_2z_2=v_1w_2z_2=v_1y(d,j_2)y_2$.
Now recall that $z_1$ is a branch of the right-hand side $\zeta_1$ of $r_1$ such that 
$\zeta_1/z_1=q'_1(x_j)$.
Since $M$ is ultralinear, we have $\mu(q_1)\leq\mu(q_1')$. 
We also have $\mu(q_1')\leq\mu(q_1)$, 
because $e_2\cdots e_n$ is a path in $G_M$ from $(q_1',q_2',b)$ to $(q_1,q_2,b)$.
Hence $\mu(q_1)=\mu(q_1')$, and so, again by ultralinearity, 
there is no other occurrence of $x_j$ in $\zeta_1$. 
In particular, since $\zeta_1/y(d,j_1)y_1=\zeta_1/z_1=q'_1(x_j)$, 
the subtree $\zeta_1/y(d,j_2)$ does not contain $x_j$.

Let $\pi'=e_2\cdots e_n$, so $\pi=e\pi'$.
Let $k\in\Nat$ be such that $k>\maxdiff(M)+\maxrhs(M)+\maxfix(M)$, and let 
$\pi_k$ be the pumped path $\pi_0\pi^{k+2}=\pi_0\pi e\pi'\pi^k$. 
By Lemma~\ref{lm:path_to_context}, applied to $\pi_k$, 
there exist $C\in\C_\Sigma$ and $p_1,p_2\in P$ such that 
requirements~(a), (b), and~(c) of that lemma hold for the prefix $\pi_0\pi$ of $\pi_k$.  
By~(a), $(q_1,u,\#)\in\blk_{C[p_1]}(v_1)$. 
Hence, since $\rhs(q_1,C[p_1],u)=\rhs(r_1)=\zeta_1$ by (c)(1),   
it follows from Lemma~\ref{lm:blink} and the definition of $\lk$ that 
$(q_1,u,\nod(y))\in\lk_{C[p_1]}(\nod(v_1y))$ and hence 
$\orig_{C[p_1]}(\nod(v_1y))=(q_1,u,\nod(y))$.
So, we obtain from Lemma~\ref{lm:orig_prop1}(2) that  
$M(C[p_1])/v_1y(d,j_2)=
\zeta_1/y(d,j_2)[\bar{q}(x_{j'})\leftarrow \bar{q}_M(C[p_1]/uj')\mid\bar{q}\in Q,\,j'\in\Nat_+]$.
For every $j'\in\Nat_+$ with $j'\neq j$ and $uj'\in V(C)$, 
the node $uj'$ is not on the spine of $C$
because, by~(b), $u$ and $uj$ are both on or both off the spine. Hence, by (c)(2), 
$C[p_1]/uj'=C/uj'\in\fix(M)$ and so $\height(\bar{q}_M(C[p_1]/uj'))\leq\maxfix(M)$. 
Since as observed above, $\zeta_1/y(d,j_2)$ does not contain $x_j$, 
we get that $\height(M(C[p_1])/v_1y(d,j_2))\leq\maxrhs(M)+\maxfix(M)<k$.

Let $\pi'$ have label $(v_1',v_2')$.
Then $\out(\pi_k)=(v_1z_1v_1'\bar{v}_1^k,v_2z_2v_2'\bar{v}_2^k)$
and since (a) also holds for the path $\pi_k$ itself, 
$v_2z_2v_2'\bar{v}_2^k$ is a branch of $M(C[p_2])$ by Lemma~\ref{lm:blink}. 
Recall that $v_2z_2=v_1w_2z_2=v_1y(d,j_2)y_2$. Hence 
$y_2v_2'\bar{v}_2^k$ is a branch of $M(C[p_2])/v_1y(d,j_2)$.
Since $\bar{v}_2\neq\varepsilon$, $|y_2v_2'\bar{v}_2^k|\geq k$
(in the case where $w_2=\varepsilon$, we need here that $\bar{v}_1\neq\varepsilon$).
By Lemma~\ref{lm:height-diff} (with $v:= v_1y(d,j_2)$ and $w:=y_2v_2'\bar{v}_2^k$, 
and with $p_1$ and $p_2$ interchanged), there is a prefix $w'$ of $y_2v_2'\bar{v}_2^k$
such that $\nod(v_1y(d,j_2)w')$ is a difference node of $M(C[p_1])$ and $M(C[p_2])$.
Since $\nod(w')$ is a node of $M(C[p_1])/v_1y(d,j_2)$, we have 
$|w'|\leq \maxrhs(M)+\maxfix(M)$. 
So $\height(M(C[p_2])/v_1y(d,j_2)w')\geq |y_2v_2'\bar{v}_2^k|-|w'|\geq 
k-(\maxrhs(M)+\maxfix(M))> \maxdiff(M)$, a contradiction.
\qed
\end{proof}

\begin{example}\label{ex:cycle}
Let $M$ be the dtla of Example~\ref{ex:depgraph} and let $\pi_0=e_{02}$ and $\pi=e_2$.
The path $\pi_0\pi$ has output label $((\tau,2)(\tau,1),(\tau,2)(\tau,1)(\tau,2))$
and so $\diff(\pi_0\pi)=(\varepsilon,(\tau,2))$, which means that 
$\pi_0\pi$ is ancestral. On the other hand, the path $\pi_0\pi^2$ has output label $((\tau,2)(\tau,1)(\tau,1),(\tau,2)(\tau,1)(\tau,2)(\tau,1)(\tau,2))$ which implies that
$\diff(\pi_0\pi^2)=((\tau,1),(\tau,2)(\tau,1)(\tau,2))$,  
and so $\pi_0\pi^2$ is splitting. 
Indeed, consider the pumped path $\pi_k=\pi_0\pi^{k+2}$, 
which has label 
$(2\cdot 1^{k+2},(\tau,2)(\tau,1)^{k+2},(\tau,2)((\tau,1)(\tau,2))^{k+2})$  
and satisfies the requirements of Lemma~\ref{lm:path_to_context} for
$C=\tau(\bot,\sigma^{k+2}a)$ and for $p_1=p_a$ and $p_2=p_b$, 
by Example~\ref{ex:path_to_context}. 
Now $M(C[p_a])=\tau(\angl{q_{1a},p_a},\tau(\tau(q_{2aM}(\sigma^ka),a),a))$ and 
$M(C[p_b])=\tau(\angl{q_{1a},p_a},\tau(\tau(a,\tau(\tau(a,q_{2bM}(\sigma^ka)),a)),a))$.
Consequently, the node $\nod((\tau,2)(\tau,1)(\tau,2))=(2,1,2)$ is a difference node
of $M(C[p_a])$ and $M(C[p_b])$, with $M(C[p_a])/(2,1,2)=a$
and $M(C[p_b])/(2,1,2)=\tau(\tau(a,q_{2bM}(\sigma^ka)),a)$.  
So, $\tau(\tau(a,q_{2bM}(\sigma^ka)),a)$ is a difference tree of $M$ 
and has a branch $((\tau,1)(\tau,2))^{k+1}$, for every $k$. 
Hence $\diff(M)$ is infinite, 
which, of course, we already knew from Example~\ref{ex:drop_prime}. 
\qed
\end{example}

Finally, we prove that $(2\cdot|Q|^2-1)\cdot\maxrhs(M)$ is an ancestral bound for $M$. 

\begin{lemma}\label{lm:bound_graph}
Let $\diff(M)$ be finite. 
Let $\pi$ be an ancestral entry path in $G_M$. \\
If $\diff(\pi)=(w,\varepsilon)$,
then $|w|\leq (2\cdot|Q|^2-1)\cdot\maxrhs(M)$. 
\end{lemma}
\begin{proof}
Suppose that $|w|> (2\cdot|Q|^2-1)\cdot\maxrhs(M)$.
For a prefix $\pi'$ of $\pi$, we will denote the first component of $\diff(\pi')$ by 
$\diff_1(\pi')$. 
So, $|\diff_1(\varepsilon)|=0$ and $|\diff_1(\pi)|> (2\cdot|Q|^2-1)\cdot\maxrhs(M)$. 
We also observe that, by definition of $G_M$, every edge $e$ of $\pi$ adds 
at most $\maxrhs(M)$ to $|\diff_1(\pi')|$; 
formally, if $\pi'e$ is a prefix of $\pi$, 
then $|\diff_1(\pi'e)| \leq |\diff_1(\pi')|+\maxrhs(M)$, 
because $\diff(\pi'e)=\diff(\diff(\pi')\cdot\out(e))$ and so 
$\diff_1(\pi'e)$ is a postfix 
of $\diff_1(\pi')z_1$ if $\out(e)=(z_1,z_2)$. 

Let $\pi_0=\varepsilon$ and for $1\leq i\leq 2\cdot|Q|^2$, 
let $\pi_i$ be the shortest prefix of $\pi$
such that $|\diff_1(\pi_i)|>|\diff_1(\pi_{i-1})|$. 
Obviously, $\pi_{i-1}$ is a proper prefix of $\pi_i$. 
Moreover, by the above observation, $|\diff_1(\pi_i)| \leq |\diff_1(\pi_{i-1})|+\maxrhs(M)$ and 
hence $|\diff_1(\pi_i)| \leq i\cdot\maxrhs(M)$, 
which shows that $\pi_i$ is well defined for every $i\leq 2\cdot|Q|^2$. 
Note that $\pi_i$ is ancestral
and $\diff(\pi_i)=(w_i,\varepsilon)$ for some branch $w_i$. 
Since we have defined $2\cdot|Q|^2+1$ prefixes of $\pi$, 
there exist $\pi_i$ and $\pi_j$ with $i<j$ that are $(q_1,q_2,b)$-paths 
for the same node $(q_1,q_2,b)$ of $G_M$. 
So, $\pi_i$ is a prefix of $\pi_j$ and $|w_j|>|w_i|$.
Let $\bar{\pi}$ be the $(q_1,q_2,b)$-cycle such that $\pi_j=\pi_i\bar{\pi}$, 
and let $\out(\bar{\pi})=(v_1,v_2)$. 
Then $\diff(\pi_j)=\diff(\diff(\pi_i)\cdot\out(\bar{\pi}))$ and so
$(w_j,\varepsilon)=\diff(w_iv_1,v_2)$, i.e., $w_iv_1=v_2w_j$. 
Since $|w_i|<|w_j|$, we obtain that $|v_1|>|v_2|$. 
We now pump $\bar{\pi}$ and consider the path $\pi_i\bar{\pi}^k$ for every $k\geq 1$. 
Let $(w_{1,k}',w_{2,k}')=\diff(\pi_i\bar{\pi}^k)=\diff(\diff(\pi_i)\cdot\out(\bar{\pi}^k))=
\diff(w_iv_1^k,v_2^k)$. 
By Lemmas~\ref{lm:cycle} and~\ref{lm:drop_prime}, $\pi_i\bar{\pi}^k$ is ancestral. 
Since $|w_iv_1^k|>|v_2^k|$, it follows that $v_2^k$ is a prefix of $w_iv_1^k$, 
and so $w_{2,k}'=\varepsilon$ and $w_iv_1^k=v_2^kw_{1,k}'$. 
Hence $|w_{1,k}'|=|w_iv_1^k|-|v_2^k|=|w_i|+k(|v_1|-|v_2|)\geq k$. 
Since this holds for every $k\geq 1$, it contradicts Corollary~\ref{cor:no_split_fin}. 
\qed
\end{proof}

\begin{theorem}\label{th:ancestral_bound}
The number $(2\cdot|Q|^2-1)\cdot\maxrhs(M)$ is an ancestral bound for $M$.
\end{theorem}
\begin{proof}
Immediate from Lemmas~\ref{lm: ancbound_redef2} and~\ref{lm:bound_graph}.
\qed
\end{proof}

We now present our second main result.

\begin{theorem}\label{th:ultra}
It is decidable for a total ultralinear bounded erasing 
dtla $M$ whether there exists a dtop $N$ such that $\sem{M}=\sem{N}$, and if so,
such a dtop $N$ can be constructed.
\end{theorem}
\begin{proof}
We use Corollary~\ref{co:goal} for the class $\U$ of total ultralinear b-erasing dtlas.
We first consider an initialized and la-uniform $M\in\U$. 
As observed in Section~\ref{sec:upper}, an output bound $h_{\rm o}(M)$ for $M\in\U$ 
can be computed from $M$; in fact, by Theorem~\ref{th:output_bound}, 
$h_{\rm o}(M)=\maxrhs(M)\cdot|Q|\cdot(|P|+2)$ is an output bound for $M$.
By Theorem~\ref{th:ancestral_bound}, 
$h_{\rm a}(M)=(2\cdot|Q|^2-1)\cdot\maxrhs(M)$
is an ancestral bound for $M$ (which, of course, can be computed from $M$).
Finally, Theorem~\ref{th:upper_bound} shows that 
$2\cdot\maxrhs(M)+h_{\rm o}(M)+h_{\rm a}(M)+1=
1+\maxrhs(M)\cdot(|Q|\cdot(|P|+2)+2\cdot|Q|^2+1)$ 
is a difference bound for $M$, and hence so is the larger number 
$1+2\cdot\maxrhs(M)\cdot(|Q|+|P|)^2$.
It now follows from Lemmas~\ref{lm:make_initialized} and~\ref{lm:make_uniform}
that we obtain a difference bound for \emph{every} $M\in\U$  
by modifying this expression as follows: first change
$\maxrhs(M)$ into $2\cdot\maxrhs(M)$ and $|Q|$ into $(|Q|+1)\cdot|P|$, and then 
take the maximum with $\maxrhs(M)$ 
(cf. the paragraphs after Lemmas~\ref{lm:make_initialized} and~\ref{lm:make_uniform}). 
Thus, $\U$ has Property~(H) of Corollary~\ref{co:goal} with the computable mapping
$h:\U\to\Nat$ such that $h(M)=1+4\cdot\maxrhs(M)\cdot(|Q|+2)^2\cdot|P|^2$.
\qed
\end{proof}

The same result holds for a total output-monadic dtla $M$, 
where \emph{output-monadic} means that $\rk(d)\leq 1$ for every $d\in\Delta$.
The reason is that Lemma~\ref{lm:cycle}, 
which is the only result that needs the assumption that 
$M$ is ultralinear and bounded erasing,
is trivial if $M$ is output-monadic, because in that case 
every entry path in $G_M$ is nonsplitting.  
Note that every output-monadic dtla is (ultra)linear, 
but not necessarily bounded erasing.
This result is a slight generalization of the result of \cite{DBLP:journals/tcs/Choffrut77}
for string transducers (more precisely, subsequential functions). 

Theorem~\ref{th:ultra} also holds for a total initialized depth-uniform dtla $M$, 
where \emph{depth-uniform} means that for every $a\in\Sigma^{(k)}$ and every $j\in[k]$ 
there exists a number $d_{a,j}\in\Nat$ such that for every rule 
$q(a(x_1\ot p_1,\dots,x_k\ot p_k))\to\zeta$ and every $z\in V(\zeta)$, 
if $\zeta/z=q'(x_j)$ then $|z|=d_{a,j}$. In other words, every translation of the $j$th child 
of a node with label $a$ is at the same depth of the right-hand side of every rule for $a$.
This property implies that $|z_1|=|z_2|$ for every label $(r_1,r_2,j,z_1,z_2)$ of an edge of $G_M$,
hence $|v_1|=|v_2|$ for every output label $(v_1,v_2)$ of an entry path in $G_M$, and hence 
$\diff(\pi)=(\varepsilon,\varepsilon)$ for every ancestral entry path in $G_M$. 
Thus, $h_{\rm a}(M)=0$ is an ancestral bound for $M$, by Lemma~\ref{lm: ancbound_redef2}. 

\begin{example}
Let $\Sigma=\{\sigma^{(1)},\tau^{(1)},a^{(0)},b^{(0)}\}$ and 
$\Delta=\{\sigma_a^{(2)},\sigma_b^{(2)},a^{(0)},b^{(0)}\}$.
Consider the initialized la-uniform dtla $M$ 
that translates every input tree $s$ with $n$ occurrences of $\sigma$ 
into the full binary tree of height~$n$ over $\{\sigma_y,y\}$, 
where $y$ is the label of the leaf of $s$.
For instance, $M(\sigma\tau\sigma a)= \sigma_a(\sigma_a(a,a),\sigma_a(a,a))$. 
It has look-ahead states $p_a$ and $p_b$, with the usual transitions, and  
states $q_a$ and $q_b$. For $y\in\{a,b\}$, it has axioms $A(p_y)=q_y(x_0)$ and rules 
$q_y(\sigma(x_1\ot p_y))\to \sigma_y(q_y(x_1),q_y(x_1))$,
$\;q_y(\tau(x_1\ot p_y))\to q_y(x_1)$, and
$q_y(y)\to y$. 
Clearly $M$ is depth-uniform, with $d_{\sigma,1}=1$ and $d_{\tau,1}=0$.
Note that $M$ is neither ultralinear nor b-erasing.
\qed
\end{example}

The depth-uniform property can be weakened by defining it as follows: 
there is a number $h'_{\rm a}(M)\in\Nat$ such that for every entry path $\pi$ in $G_M$,
if $\out(\pi)=(v_1,v_2)$ then $|\,|v_1|-|v_2|\,|\leq h'_{\rm a}(M)$, i.e., the distance 
between $|v_1|$ and $|v_2|$ is at most $h'_{\rm a}(M)$. 
Clearly, such a number $h'_{\rm a}(M)$ is an ancestral bound for $M$ by Lemma~\ref{lm: ancbound_redef2}.
This weak depth-uniform property is decidable, and if it holds then a bound $h'_{\rm a}(M)$
can be computed. In fact, it is easy to see that $M$ is weak depth-uniform if and only if 
$|v_1|=|v_2|$ for every output label $(v_1,v_2)$ of a cycle in $G_M$. This can be checked
by considering all simple cycles in $G_M$, 
i.e., cycles in which no node of $G_M$ occurs more than once. 
The bound $h'_{\rm a}(M)$ can then be computed by considering all entry paths in $G_M$
that do not contain a cycle.

\section{Conclusion}

We have tried to answer the following questions: 
is there a method to determine 
for a given top-down tree transducer \emph{with regular look-ahead}, 
whether or not its translation can be realized by a 
top-down tree transducer \emph{without regular look-ahead}?
And if so, can such a transducer be constructed from the given one?

We have obtained some partial answers to this question for total dtlas 
(i.e., total deterministic top-down tree transducers with regular look-ahead).
In Theorem~\ref{th:alg} we have given a general algorithm to decide 
for a given total dtla $M$ with a given difference bound, 
whether $M$ is equivalent to a deterministic top-down tree transducer 
(without regular look-ahead), and if so, to construct such a dtop. 
In Sections~\ref{sec:upper} and~\ref{sec:graphs} we have shown that 
a difference bound can be computed for total dtlas that are 
ultralinear and bounded erasing, or output-monadic, or
initialized and depth-uniform.

Of course, we would like to prove that a difference bound can be computed 
for \emph{all} total dtlas. To do this it suffices, 
as shown in Theorems~\ref{th:upper_bound} and~\ref{th:output_bound},
to prove that an ancestral bound can be computed for all total dtlas.
We would also like to extend our results to the non-total case where a dtla
realizes a partial function, and to the case where the dtla and the dtop are 
restricted to a given regular tree language. 
And more generally, we would like to have an algorithm that for a given dtla 
constructs an equivalent dtla with a minimal number of look-ahead states. 

We did not investigate the precise time complexity of our algorithm. 
Consider a total ultralinear and bounded erasing dtla $M$. 
We have shown in the proof of Theorem~\ref{th:ultra} that 
$M$ has a difference bound $h(M)$ that is polynomial in the size of $M$.
Since $\sumfix(M)$ can be taken exponential in the size of $M$, 
the canonical dtla $\can(M)$ can be built in exponential time
and has an exponential difference bound $h(\can(M))$, 
see Theorem~\ref{th:alg}.
Clearly, the running time of Algorithm~\ref{alg:alg} on input $\can(M)$ 
is at most the number of tuples of trees of height $\leq h(\can(M))$ 
in $\T_\Delta(Q\times P)^n$ (with $n=|P|$), 
which is double exponential in $h(\can(M))$.
Thus, the time complexity of our algorithm is at most 
triple exponential in the size of $M$. 
On the other hand, it is at least exponential in the size of $M$,
because the size of a dtop equivalent to $M$ can be exponential in the size of $M$. 
To see this, consider Example~\ref{ex:Mthree} and, for every $n\in\Nat$, let $M_n$ be the dtla $M$
obtained from $M$ by changing $80$ into $n$, and let $N_n$ be the dtop obtained from $N$ 
by changing $50$ into $n-30$. Clearly, on input $M_n$ our algorithm computes the dtop $N_n$.
The size of $M_n$ is polynomial in $n$, whereas, due to the states $q_z$ of $N_n$ 
with $z\in\{\sigma,\tau\}^*$ and $|z|\leq n-30$, 
the size of $N_n$ is exponential in $n$. Thus, our algoritm takes exponential time on input $M_n$. 
We do not know the precise time complexity of our algorithm. Also, we do not know the precise upper bound 
on the size of a dtop $N$ equivalent to $M$ (if such an $N$ exists). 

We have shown in Corollary~\ref{co:difffin} and Lemma~\ref{lm:twodiff} that, 
for $M$ to be equivalent to a dtop, $\diftup(M)$ must be finite. But we do not know 
whether it is decidable if $\diftup(M)$ is finite.
If so, and if, moreover, we could compute $\diftup(M)$ from $M$, then Algorithm~\ref{alg:alg} 
could be simplified: we could just (try to) compute the unique dtop $N$ with $Q_N=\diftup(M)$ 
that is associated with $M$ (see Lemma~\ref{lm:Qdiff}). 
Thus, we would like to get more grip on the difference tuples of $M$.
We do not even know whether $\diftup(M)$ is a decidable subset of $\T_\Delta(Q\times P)^n$.  

As observed at the end of the Introduction, regular look-ahead can always be removed 
from a macro tree transducer \cite{DBLP:journals/jcss/EngelfrietV85}. 
Hence another more general question is: 
Is it decidable for a given macro tree transducer
whether it is equivalent to a top-down tree transducer?
For total deterministic transducers, this problem can be reduced to our problem 
by first proving the decidability of the following related question: 
Is it decidable for a given macro tree transducer
whether it is equivalent to a top-down tree transducer with regular look-ahead?
This is an open problem of which it was recently shown in~\cite{DBLP:conf/icalp/FiliotMRT15} 
that the answer is `yes' when the transducers have origin-semantics, 
which means that they should associate 
the same origins in the input tree to the nodes of the output tree.

As observed by one of the referees, a problem related to our problem is 
whether it is decidable if a given \emph{nondeterministic}
top-down tree transducer is equivalent to a \emph{deterministic} one (both without regular look-ahead). 
Since it is decidable 
whether a nondeterministic top-down tree transducer is functional (i.e., realizes a function), 
see~\cite{DBLP:journals/actaC/Esik81}, we may assume that the given nondeterministic transducer is functional. 
Hence, it can be transformed into a deterministic top-down tree transducer with regular look-ahead
by the result of~\cite{DBLP:journals/ipl/Engelfriet78}. This transformation clearly preserves the properties 
of ultralinearity and bounded erasing (appropriately defined for nondeterministic transducers). 
Thus, it is decidable for a given total ultralinear and b-erasing nondeterministic top-down tree 
transducer whether it is equivalent to a deterministic top-down tree transducer. 
This also holds for (appropriately defined) output-monadic,
or initialized and depth-uniform nondeterministic top-down tree transducers. 
As for dtlas, we do not know whether it holds in the general case.
We note that the class of functional nondeterministic top-down tree 
transductions lies properly between the class of deterministic top-down tree transductions and 
the class of deterministic top-down tree transductions with regular look-ahead. For instance, 
the dtla $M_\textrm{ex}$ from the Introduction is equivalent to 
a nondeterministic top-down tree transducer if $\Sigma=\Delta=\{\sigma^{(1)},a^{(0)},b^{(0)}\}$
(as in Example~\ref{ex:Mex}), but not if $\Sigma=\Delta=\{\sigma^{(2)},a^{(0)},b^{(0)}\}$.

\bigskip
\noindent
{\bf Acknowledgments.} We are grateful to the referees for their detailed and constructive comments.

\bibliographystyle{plain}
\bibliography{bib}

\end{document}